\documentclass[usenatbib]{mn2e}
\usepackage{apjfonts}
\usepackage{amssymb}
\usepackage{amsmath}
\usepackage{ctable}
\usepackage{url}
\usepackage{fixltx2e} 
\newcommand{\Msun}{\mathrm{M}_{\odot}}
\newcommand{\Zsun}{\mathrm{Z}_{\odot}}

\newcommand{\Rvir}{R_{\rm vir}}
\newcommand{\Mvir}{M_{\rm vir}}
\newcommand\altaffilmark[1]{$^{#1}$}
\newcommand\altaffiltext[1]{$^{#1}$}
\newcommand{\etal}{et al.}

\title[Metal flows of the CGM]{Metal flows of the circumgalactic medium, and the metal budget in galactic halos \vspace{-0.5cm}}

\vspace{-0.2cm}
\author[Muratov \etal]{
\parbox[t]{\textwidth}{ 
Alexander L.~Muratov\thanks{E-mail:amuratov@ucsd.edu}\altaffilmark{1}, 
Du\v{s}an Kere\v{s}\thanks{E-mail:dkeres@physics.ucsd.edu}\altaffilmark{1},
Claude-Andr{\'e} Faucher-Gigu{\`e}re\altaffilmark{2},
Philip F.~Hopkins\altaffilmark{3},
Xiangcheng Ma\altaffilmark{3},
Daniel Angl{\'e}s-Alc{\'a}zar \altaffilmark{2},
T.K. Chan \altaffilmark{1},
Paul Torrey \altaffilmark{3,4},
Zachary H.~Hafen\altaffilmark{2},
Eliot Quataert\altaffilmark{5}, \&\
Norman Murray\altaffilmark{6,7}  
}
\vspace*{6pt} \\
\altaffiltext{1}{Department of Physics, Center for Astrophysics and Space Sciences, University of California at San Diego, 9500 Gilman Drive, La Jolla, CA 92093} \\
\altaffiltext{2}{Department of Physics and Astronomy and CIERA, Northwestern University, 2145 Sheridan Road, Evanston, IL 60208, USA} \\ 
\altaffiltext{3}{TAPIR, Mailcode 350-17, California Institute of Technology, Pasadena, CA 91125, USA} \\
\altaffiltext{4}{MIT Kavli Institute for Astrophysics \& Space Research, Cambridge, MA, 02139, USA}\\
\altaffiltext{5}{Department of Astronomy and Theoretical Astrophysics Center, University of California Berkeley, Berkeley, CA 94720} \\
\altaffiltext{6}{Canadian Institute for Theoretical Astrophysics, 
60 St.\ George Street, University of Toronto, ON M5S 3H8, Canada} \\
\altaffiltext{7}{Canada Research Chair in Astrophysics \vspace{-0.5cm}} \\
\vspace{-0.5cm}
}

\date{Submitted to MNRAS, June, 2016\vspace{-0.6cm}}
\begin{document}
\maketitle

\date{\today}

\begin{abstract}
We present an analysis of the flow of metals through the circumgalactic medium (CGM) in the Feedback in Realistic Environments (FIRE) simulations of galaxy formation, ranging from isolated dwarfs to $L*$ galaxies. We find that nearly all metals produced in high-redshift galaxies are carried out in winds that reach $0.25\Rvir$. When measured at $0.25\Rvir$ the metallicity of outflows is slightly higher than the interstellar medium (ISM) metallicity.  Many metals thus reside in the CGM. Cooling and recycling from this reservoir determine the metal budget in the ISM. The outflowing metal flux decreases by a factor of $\sim2-5$ between $0.25\Rvir$ and $\Rvir$. Furthermore, outflow metallicity is typically lower at $\Rvir$ owing to dilution of the remaining outflow by metal-poor material swept up from the CGM. The inflow metallicity at $\Rvir$ is generally low, but outflow and inflow metallicities are similar in the inner halo. At low redshift, massive galaxies no longer generate outflows that reach the CGM, causing a divergence in CGM and ISM metallicity. Dwarf galaxies continue to generate outflows, although they preferentially retain metal ejecta. In all but the least massive galaxy considered, a majority of the metals are within the halo at $z=0$. We measure the fraction of metals in CGM, ISM and stars, and quantify the thermal state of CGM metals in each halo. The total amount of metals in the low-redshift CGM of two simulated $L*$ galaxies is consistent with estimates from the COS halos survey, while for the other two it appears to be lower.

\end{abstract}

\begin{keywords}
galaxies: formation --- galaxies: evolution --- stars: formation --- cosmology: theory\vspace{-0.5cm}
\end{keywords}

\vspace{-1.1cm}
\section{Introduction}

Astrophysical ``metals'', i.e. elements heavier than H and He, are produced through stellar nucleosynthesis and explosions. While metals are a small fraction of the baryonic component of the universe by mass, their atomic transitions produce absorption and emission features that are in an easily accessible range of the electromagnetic spectrum. Studying the relative prevalence of these features allows one to probe the conditions of gas throughout the cosmos. In addition, the distribution and abundance of metals can be used to archaeologically infer key insights regarding when and where stars form, and how they evolve. 

Observed galaxies at $z=0$ obey well known correlations between stellar mass, $M_{*}$, and average galactic metallicity, as measured both for metals locked in stars (e.g. \citealt{gallazzi_etal05, kirby13}), and metals in the gas phase of the interstellar medium (ISM) (e.g. \citealt{tremonti04,lee_etal06}). We refer to both of these relations as MZR, and will specify if stellar or gas phase metallicity is discussed explicitly. These mass-metallicity relations provide a useful lens through which to view galaxy evolution, as they link the buildup of galactic mass to the chemical state of the galaxy. The correlations also exist at higher redshift, with some evolution in slope and normalization (e.g. \citealt{erb08,zahid_etal13}). 

It is somewhat more difficult to obtain a census of the metals in the  circumgalactic and intergalactic media (CGM and IGM). These terms both refer to gas outside of galaxies, but in this paper we loosely refer to the gas within $\Rvir$ of the host dark matter halo as CGM, and to the gas outside of $\Rvir$ as IGM. Gas in these components is far more tenuous than the ISM, so strong background sources and extreme sensitivity are required to notice absorption features. At $z=2$, only about 50\% of the metals produced by galaxies have been found in the galaxies themselves \citep{bouche_etal06}. At $z=0$, a similarly small fraction of the metals produced by stars is accounted for within the ISM and stars alone \citep{peeples_etal14} (hereafter P14). This is known as the galactic ``missing metals problem''.  
Presumably, this means a significant amount of metals has been ejected into the CGM or IGM \citep{bouche_etal07}. A large amount of ejected metals could be a natural consequence of ubiquitous metal-rich outflows observed at all cosmic epochs (e.g. \citealt{heckman00, veilleux_etal05, weiner_etal09, steidel10, martin_etal12, bouche_etal12, rubin_etal13}).

\begin{footnotesize}
\ctable[
  caption={{\normalsize Simulations}\label{tab:sims}},center,star
  ]{lcccccccccl}{
\tnote[ ]{Parameters describing the initial conditions for our simulations (units are physical): \\
{\bf (1)} Name: Simulation designation. \\
{\bf (2)} $M_{h}(z=0)$: Mass of the $z=0$ ``main'' halo (most massive halo in the high-resolution region). \\
{\bf (3)} $M_{h}(z=2)$: Mass of the $z=2$ ``main'' halo. \\
{\bf (4)} $M_{*}(z=0)$: Stellar mass of the $z=0$ ``main'' halo. \\
{\bf (5)} $M_{*}(z=2)$: Stellar mass of the $z=2$ ``main'' halo. \\
{\bf (6)} $\Rvir(z=0)$: Virial radius of the $z=0$ ``main'' halo in physical kpc. \\
{\bf (7)} $\Rvir(z=2)$: Virial radius of the $z=2$ ``main'' halo in physical kpc.  \\
{\bf (8)} $m_{b}$: Initial baryonic (gas and star) particle mass in the high-resolution region, in our highest-resolution simulations. \\ 
{\bf (9)} $\epsilon_{b}$: Minimum baryonic force softening (minimum SPH kernel smoothing lengths are comparable or smaller). Gravitational force softening lengths for the gas particles are adaptive.\\
{\bf (10)} $m_{dm}$: Dark matter particle mass in the high-resolution region, in our highest-resolution simulations. \\ 
{\bf (11)} $\epsilon_{dm}$: Minimum dark matter Plummer-equivalent force softening (fixed in physical units at all redshifts). \\
}
}{
\hline\hline
\multicolumn{1}{c}{Name} &
\multicolumn{1}{c}{$M_{h}(z=0)$} &
\multicolumn{1}{c}{$M_{h}(z=2)$} &
\multicolumn{1}{c}{$M_{*}(z=0)$} &
\multicolumn{1}{c}{$M_{*}(z=2)$} &
\multicolumn{1}{c}{$R_{\rm vir}(z=0)$} &
\multicolumn{1}{c}{$R_{\rm vir}(z=2)$} &
\multicolumn{1}{c}{$m_{b}$} & 
\multicolumn{1}{c}{$\epsilon_{b}$} & 
\multicolumn{1}{c}{$m_{dm}$} & 
\multicolumn{1}{c}{$\epsilon_{dm}$} \\ 
\multicolumn{1}{c}{\ } &
\multicolumn{1}{c}{[$M_{\sun}$]} & 
\multicolumn{1}{c}{[$M_{\sun}$]} & 
\multicolumn{1}{c}{[$M_{\sun}$]} & 
\multicolumn{1}{c}{[$M_{\sun}$]} & 
\multicolumn{1}{c}{[$kpc$]} & 
\multicolumn{1}{c}{[$kpc$]} & 
\multicolumn{1}{c}{[$M_{\sun}$]} &
\multicolumn{1}{c}{[pc]} &
\multicolumn{1}{c}{[$M_{\sun}$]} &
\multicolumn{1}{c}{[pc]} \\ 

\hline
{\bf m09} & 2.5e9 & 1.3e9 & 4.6e4 & 4.1e4 & 36 & 12 & 2.6e2 & 1.4 & 1.3e3 & 30 \\
{\bf m10} & 7.8e9 & 3.8e9 & 2.3e6 & 1.7e6 & 52 & 17 & 2.6e2 & 3.0 & 1.3e3 & 30 \\
{\bf m11} &  1.4e11 & 3.8e10 & 2.3e9 & 3.4e8 & 140 & 37 & 7.1e3 & 7.0 & 3.5e4 & 70 \\
{\bf m12v} &  6.3e11 & 2.0e11& 2.8e10 & 2.3e9 & 230 & 51&  3.9e4 & 10 & 2.0e5 & 140 \\
{\bf m12q} & 1.2e12 & 5.1e11 & 2.2e10 & 7.0e9 & 280 & 87 & 7.1e3 & 10 & 2.8e5 & 140 \\
{\bf m12i} & 1.1e12 & 2.7e11 & 6.1e10 & 3.9e9 & 270 & 71 & 5.0e4 & 14 & 2.8e5 & 140 \\

\hline \\
{\bf z2h350} & - & 7.9e11 & - & 9.0e9 & - & 98 & 5.9e4 & 9 & 2.9e5 & 143 \\
{\bf z2h400} & - & 7.9e11 & - & 7.0e9 &  - & 97 & 5.9e4 & 9 & 2.9e5 & 143 \\
{\bf z2h450} & - & 8.7e11 & - & 1.3e10 & - & 100 & 5.9e4 & 9 & 2.9e5 & 143 \\
{\bf z2h506} & - & 1.2e12 & - & 1.8e10 & - & 110 & 5.9e4 & 9 & 2.9e5 & 143 \\
{\bf z2h550} & - & 1.9e11 & - & 4.4e9 & - & 62 & 5.9e4 & 9 & 2.9e5 & 143 \\
{\bf z2h600} & - & 6.7e11 & - & 1.7e10 & - & 95  & 5.9e4 & 9 & 2.9e5 & 143 \\
{\bf z2h650} & - & 4.0e11 & - & 6.6e9 & - & 80 & 5.9e4 & 9 & 2.9e5 & 143 \\
{\bf z2h830} & - & 5.4e11 & - & 1.4e10 & - & 94 & 5.9e4 & 9 & 2.9e5 & 143 \\
\hline\hline\\
{\bf m11h383} & 1.6e11 & 4.5e10 & 4.1e9 & 5.5e8 &  140 &  39 & 1.7e4 & 10 & 8.3e4 & 100 \\
{\bf m10h573} & 3.9e10 & 1.4e10 & 3.1e8 & 3.2e7 & 89 & 26 & 2.1e3 & 10 & 1.0e4 & 100 \\
{\bf m10h1146} & 1.5e10 & 3.5e9 & 5.8e7 & 3.1e6 & 65 & 16 & 2.1e3 & 4 & 1.0e4 & 43 \\
{\bf m10h1297} & 1.3e10 & 2.5e9 & 1.7e7 & 6.5e5 & 62 & 15 & 2.1e3 & 4 & 1.0e4 & 43 \\
\hline\hline\\
{\bf m11.4a} & 2.6e11 & 8.9e10 & 6.2e9 & 5.0e8 &  170 & 46 & 3.3e4 & 9 & 1.7e5 & 140 \\
{\bf m11.9a} & 8.4e11 & 1.3e11 & 3.0e10 & 1.1e9 & 250 & 54 & 3.4e4 & 9 & 1.7e5 & 140 \\
\hline\hline\\
}
\end{footnotesize}

Absorption features in the CGM have long hinted that it may serve as the primary reservoir for the missing metals \citep{chen_etal98, chen_etal01, gauthier09, steidel10} for a broad range of redshift. The importance of the CGM as a reservoir of metals at low redshift was confirmed with the Hubble Space Telescope Cosmic Origin Spectrograph (COS) halos survey \citep{tumlinson_etal11}, which used background quasars to probe the warm component of the CGM in a large sample of $L*$ galaxies. Absorption at impact parameters up to 150 kpc from the galactic center was detected in nearly every star forming galaxy-quasar pair. In a follow up study, absorption was also found in star-forming dwarf galaxies up to 100 kpc from the galactic center \citep{bordoloi_etal13}. 

These studies provided the strongest evidence yet that the missing metals problem, as well as the missing baryons problem (e.g. \citealt{fukugita_etal98}) could both be significantly mitigated or solved by considering the content of the CGM \citep[e.g.][]{werk_etal14, peeples_etal14}. If this is so, it is now crucial for theoretical models to predict how and when these metals got there, and how to best target them with future observational studies.

Since there are many more observations of galaxies than the CGM, many theoretical studies have instead focused on reproducing the galactic MZR correlations. Present-day semi-analytical and analytical models of galaxy formation use both of these relations, in addition to other canonical correlations between the stellar and halo mass ($M_*$-$M_h$) and the stellar mass and star formation rate ($M_*$-$\dot{M}_*$), as constraints to judge each model's success. Hydrodynamical cosmological simulations of galaxy formation have recently advanced to the point where they can naturally reproduce many of the same empirical galaxy correlations. Many works have yielded insights on how the MZR arises (e.g. \citealt{brooks_etal07, finlator_dave08, dave_etal11, peeples_somerville13, torrey_etal14, ma_etal16}).

In this work, we study how metals produced in stars travel between different regions within halos and how their evolution is connected to the outflows of gas from galaxies. We aim to provide a framework to describe the metal content of stars, the ISM, the CGM, and the IGM throughout cosmic time. Previous works on the subject have used simulations where galactic winds were generated with sub-grid prescriptions for wind velocities and mass loading (e.g. \citealt{oppenheimer06,ford_etal13a, vogelsberger_etal14, angles-alcazar_etal14, ford_etal14, suresh_etal15}), with a prescription that attempts to mimic collective effect of multiple SNe \citep{oppenheimer_etal16, rahmati_etal16} or via artificially disabled cooling for a portion of wind evolution (e.g. \citealt{shen_etal13}, etc.). In our work, we utilize a suite of cosmological zoom-in galaxy simulations from the Feedback in Realistic Environments (FIRE) project\footnote{Project website: http://fire.northwestern.edu}. FIRE uses explicit stellar feedback models with energy and momentum input based on the stellar population synthesis models and simulates galaxies with resolution typically much higher than in previous models. This enables improved modeling of the evolution of metal enriched gas, from the vicinity of stars to the CGM and beyond. 

The FIRE simulations successfully reproduced  $M_{*}$-$M_h$ \citep{hopkins_etal14}, $M_*$-$\dot{M}_*$ \citep{sparre_etal15}, and both MZR (\citealt{ma_etal16}) correlations, without any fine-tuning. By satisfying these requirements, the FIRE simulations have proven themselves useful for a full accounting of the baryon cycle. FIRE also reproduced observed properties of HI absorbers of high redshift CGM (\citealt{faucher-giguere_etal14, faucher-giguere_etal16, hafen_etal16}) further showing the viability of the simulations for our present analysis. 

In \citet{muratov_etal15} (henceforth, M15), galactic and CGM outflow rates in the FIRE simulations were quantified throughout cosmic history. We demonstrated that the high redshift universe was dominated by strong outflows following bursty star formation across all mass scales, while the low redshift universe generally tended towards more quiescent gas flows: low-mass dwarf galaxies generally slowed down star-forming activity while L*-like galaxies developed stable gaseous disks with continuous star formation and have no large scale outflows. We showed that through stellar feedback alone, galaxies self-regulated their baryon budget to end up with stellar masses comparable to observations, and that typical velocities of the outflowing gas as a function of galaxy circular velocity are in good agreement with observations. The model fits presented in M15 have already been implemented into large-volume cosmological simulations, which in turn produce good agreement with observational galaxy scaling correlations \citep{dave_etal16}.

Here, we extend the analysis of M15 to consider the metal component of galactic inflows and outflows, and how they affect the metal content of the ISM, stars, and CGM throughout cosmic history. We will especially focus on the CGM to complement \citet{ma_etal16}, which focused on the metallicity of stars and the ISM. We begin by testing whether the winds transport a large amount of metals to the CGM. We determine the mass of metals ejected per unit star formation, compute the metallicity of outflows at various galactocentric distances, and compare to the metallicity of the ISM, CGM, and galactic inflows (recycling) in Section \ref{sec:winds}. We discuss the implication for $z=0$ and $z=2$ metal budgets of the CGM, stars, and ISM in section \ref{sec:budget}. We summarize our findings and comparisons with various analytical models in Section \ref{sec:discussion}. The simulation and methodology is briefly discussed in Section \ref{sec:sims}.

\begin{figure*}
\centering
\begin{minipage}{0.48\textwidth}
\centering
\includegraphics[width=\textwidth]{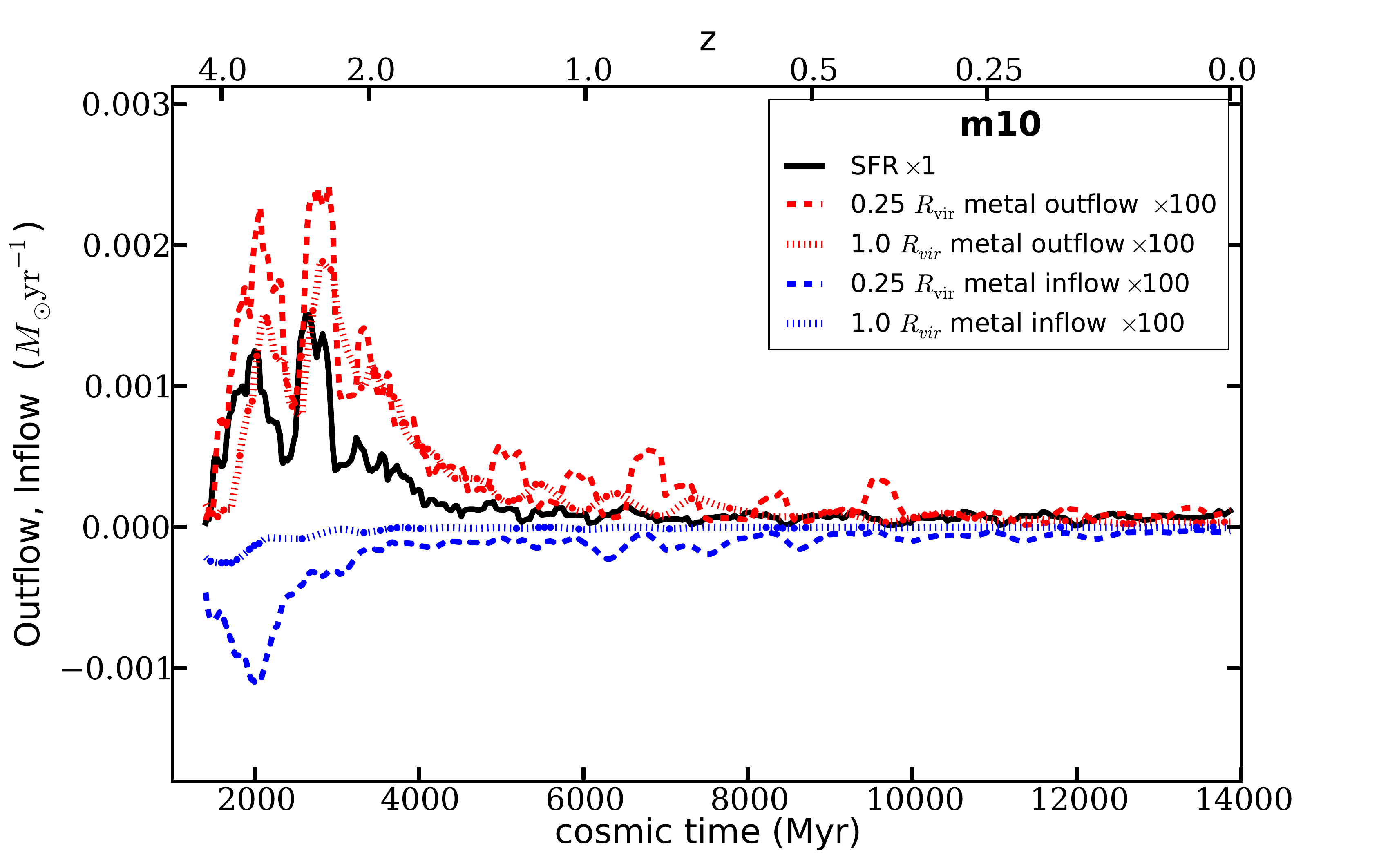}\\
\includegraphics[width=\textwidth]{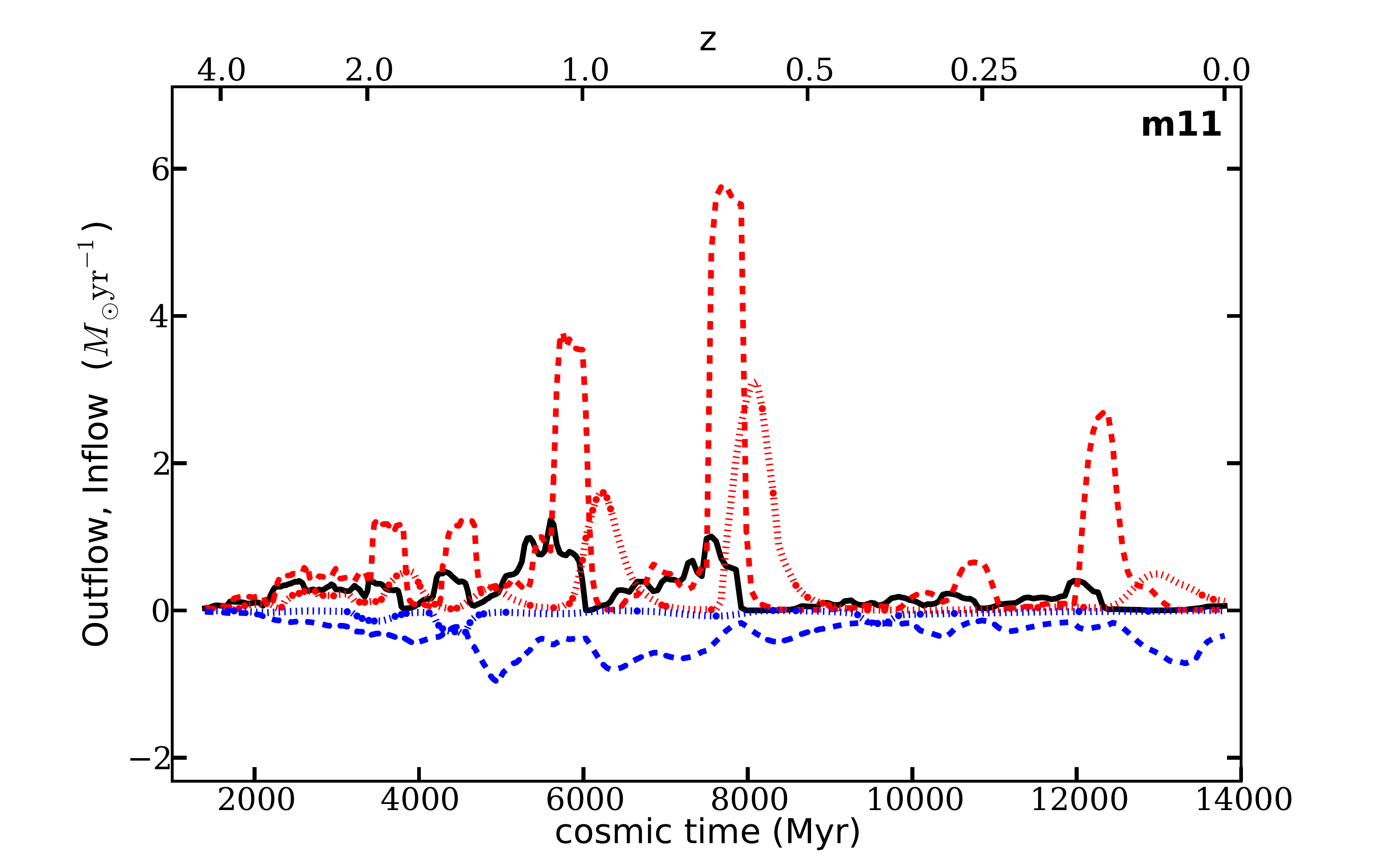}\\
\includegraphics[width=\textwidth]{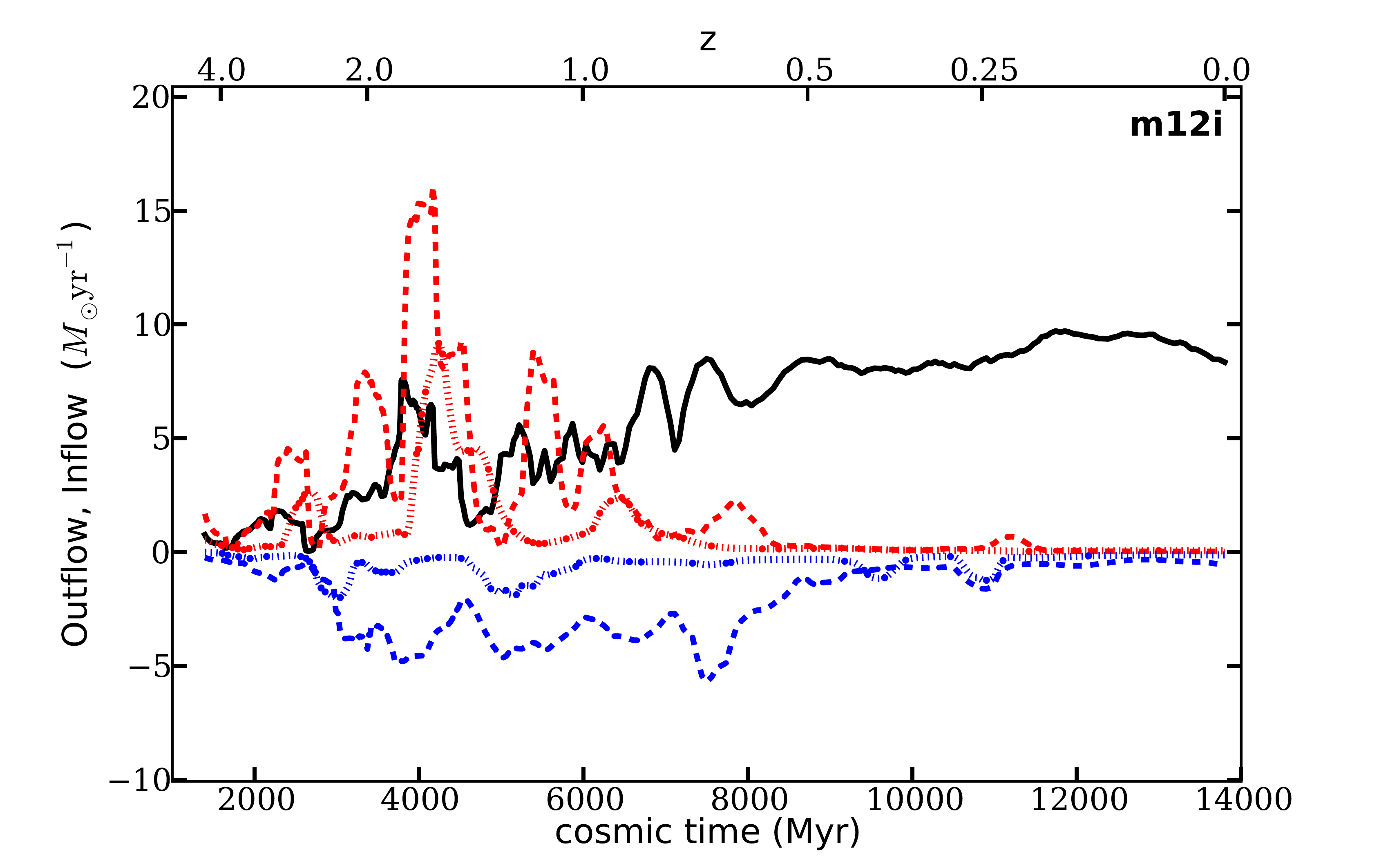}
\end{minipage}
\begin{minipage}{0.41\textwidth}
\includegraphics[width=\textwidth]{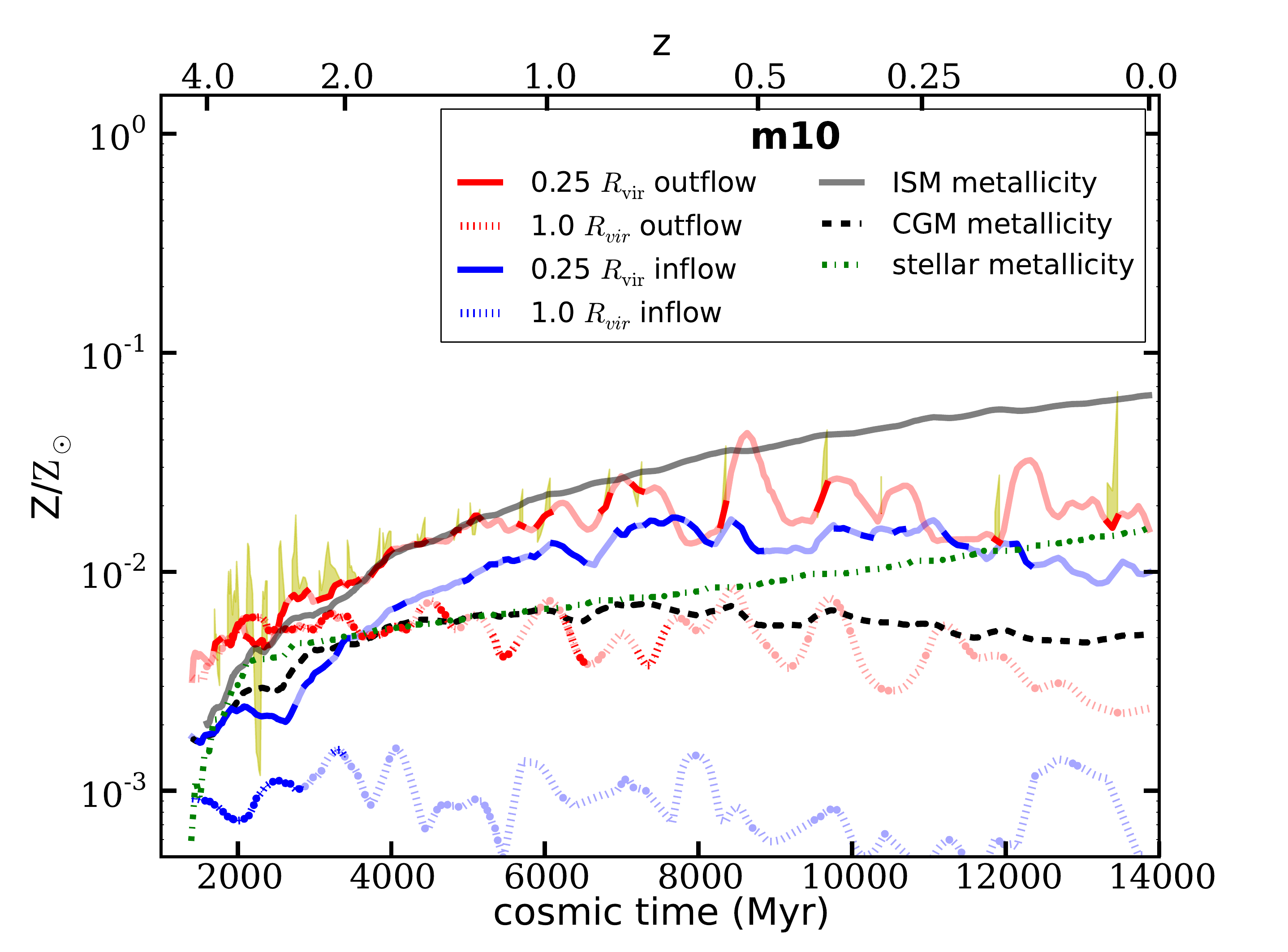}\\
\includegraphics[width=\textwidth]{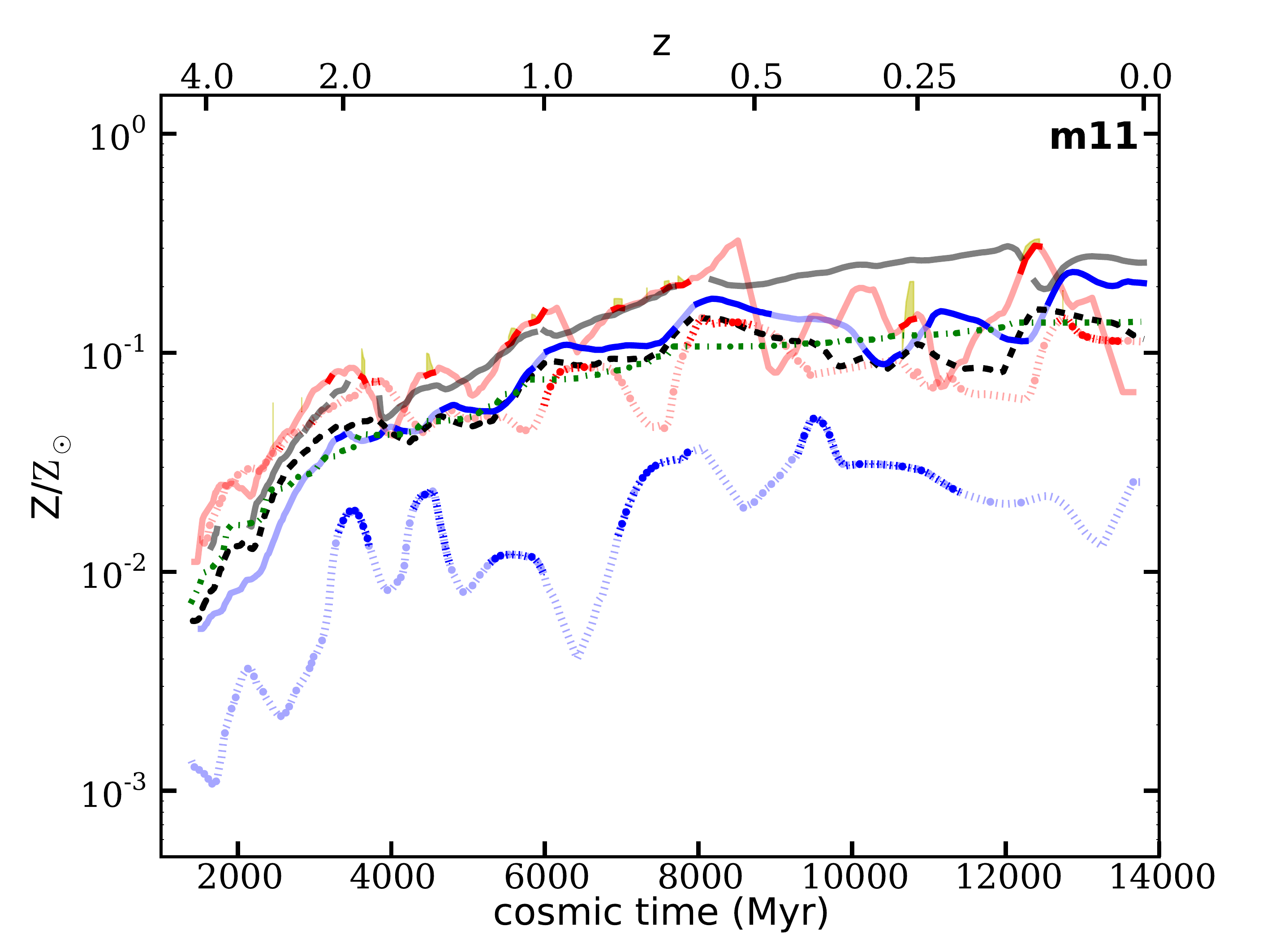}\\
\includegraphics[width=\textwidth]{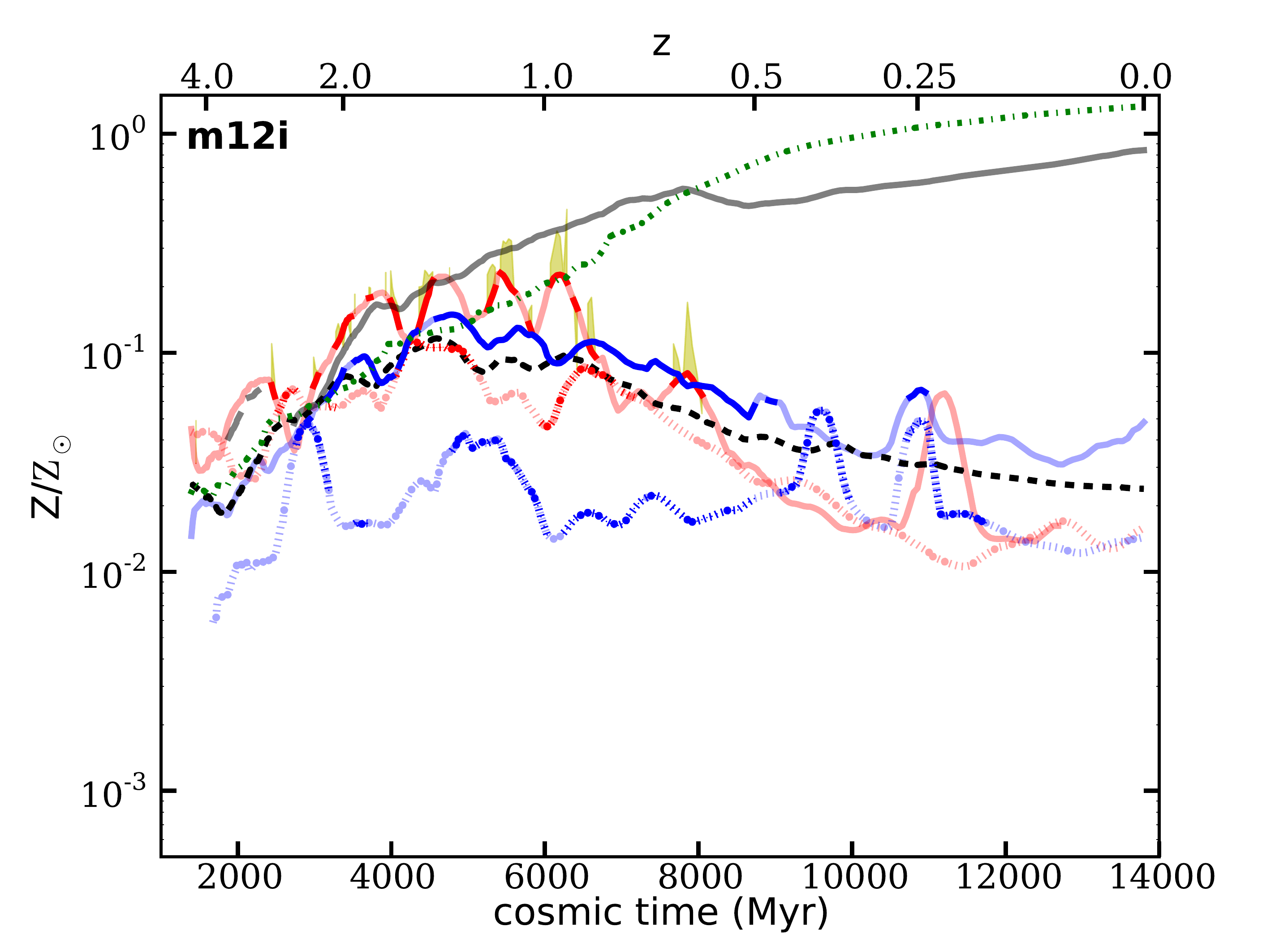}
\end{minipage}
\caption{ Left: In/outflow rates of metals at 0.25 $\Rvir$ and 1.0 $\Rvir$ for \textbf{m10} (top), \textbf{m11} (middle), \textbf{m12i} (bottom). All quantities smoothed over 400 Myr intervals. \textbf{m10} is mainly active at $z>2$. \textbf{m12i} ejects metals in bursts until $z=1$, after which enriched outflows cease, while recycled ejecta and filamentary accretion feed continuous star formation. \textbf{m11} is bursty down to z=0.  Right: Metallicity of ISM, CGM, stars (calculated as the metal mass fraction in a given component) and in/outflows (calculated as the ratio of metal flux to gas flux) at 0.25 $\Rvir$ and 1.0 $\Rvir$. ISM is defined as gas within 0.1 $\Rvir$, while CGM is the gas between 0.1 $\Rvir$ and 1 $\Rvir$.  The thick parts of lines highlight eras in cosmic history when metal in/outflow rates were highest (the smallest amount of time required to account for 80\% of the total time-integrated metal flux). The yellow spikes indicate outflows only with $v_{rad} > \sigma_{1D}$, and are smoothed over shorter timescales to demonstrate range. ISM, CGM, and outflow metallicities are close at high redshift, but can diverge at low redshift.} 
\vspace{0.3cm}
\label{fig:MetalFlowing}
\end{figure*}

\section{simulations}
\label{sec:sims}

All simulations here are previously presented FIRE simulations from \citet{hopkins_etal14, faucher-giguere_etal14, chan_etal15}, and two new runs from \citet{hafen_etal16}. They were all run with the pressure-entropy formulation of smoothed particle hydrodynamics (PSPH,  \citealt{hopkins13}), and employ the same stellar feedback algorithm described in \citet{hopkins_etal14}. Briefly, FIRE features a stellar feedback that acts locally within the multi-phase ISM as directly dictated by STARBURST99 \citep{leitherer99}. We consider feedback from stellar winds, momentum feedback due to radiation pressure from young stars, supernovae of type II and type Ia, photoelectric heating, and photoheating from ionizing radiation. The supernova feedback model ensures resolution independent momentum input: if the cooling radius of the expanding SN remnant is resolved, 100\% of available thermal energy is injected directly into local gas. If not, an appropriate amount of momentum flux is injected to account for the unresolved phase of remnant expansion. This avoids the overcooling problem \citep{katz92a, kim_ostriker14, martizzi_etal14, martizzi_etal16} by ensuring that momentum flux generated by a supernova is the same whether or not the Sedov-Taylor phase is numerically resolved. For details regarding the implementation of feedback physics, see \citet{hopkins_etal14}.

The FIRE star formation and feedback implementations require the code to resolve the multi-phase nature of the ISM. To do this we use the zoom-in technique \citep{porter85, katz93}. This enables us to reach a mass resolution of a few hundred $M_{\odot}$ in dwarfs, to $\sim 10^4 M_{\odot}$ in Milky Way mass halos and high spatial resolution in all cases; see Table \ref{tab:sims} for details. We allow stars to form only in self-gravitating gas, which is also molecular and self-shielding, and exceeds a minimum density of $n_H > 5-50 {\rm cm^{-3}}$.

We track 11 separate species (H, He, C, N, O, Ne, Mg, Si, S, Ca, and Fe) with appropriate yields from Type Ia and Type II SNe, following \citet{iwamoto_etal99} and \citet{woosley_weaver95}, and account for metal cooling separately for each species. The simulations discussed here do not employ sub-resolution metal mixing for SPH particles (we briefly discuss the potential implications of metal mixing in Section \ref{sec:discussion}). We assume solar metallicity $\Zsun = 0.02$ throughout this work. 

Table \ref{tab:sims} describes the simulations used in this paper. Our sample includes most of the initial FIRE runs \citep{hopkins_etal14} (\textbf{m09}, \textbf{m10}, \textbf{m11}, \textbf{m12i}, \textbf{m12q}, \textbf{m12v}),  as well as the \textbf{z2h} runs first introduced in \citet{faucher-giguere_etal14} which simulate low-mass Lyman-break galaxies to $z=2$, four additional dwarf galaxies run to $z=0$ first presented in \citet{chan_etal15} (\textbf{m10h1297}, \textbf{m10h1146}, \textbf{m10h573}, and \textbf{m11h383}) , and two additional L*-like galaxies (\textbf{m11.4a} and \textbf{m11.9a}) presented in \citet{hafen_etal16}. In this work, we primarily consider the central, most massive galaxy in each simulation, though we use additional galaxies that meet inclusion criteria in Section \ref{sec:average_eta}. All simulations were run with the same source code, numerical, and physical parameters for feedback, as described in \citet{hopkins_etal14}. We use adaptive gravitational softening lengths \citep{price07} for the gas, and the softening value we give for the gas are the minimum allowed. We use fixed force softening lengths for the dark matter and stars.

\begin{figure}
\includegraphics[width=\columnwidth]{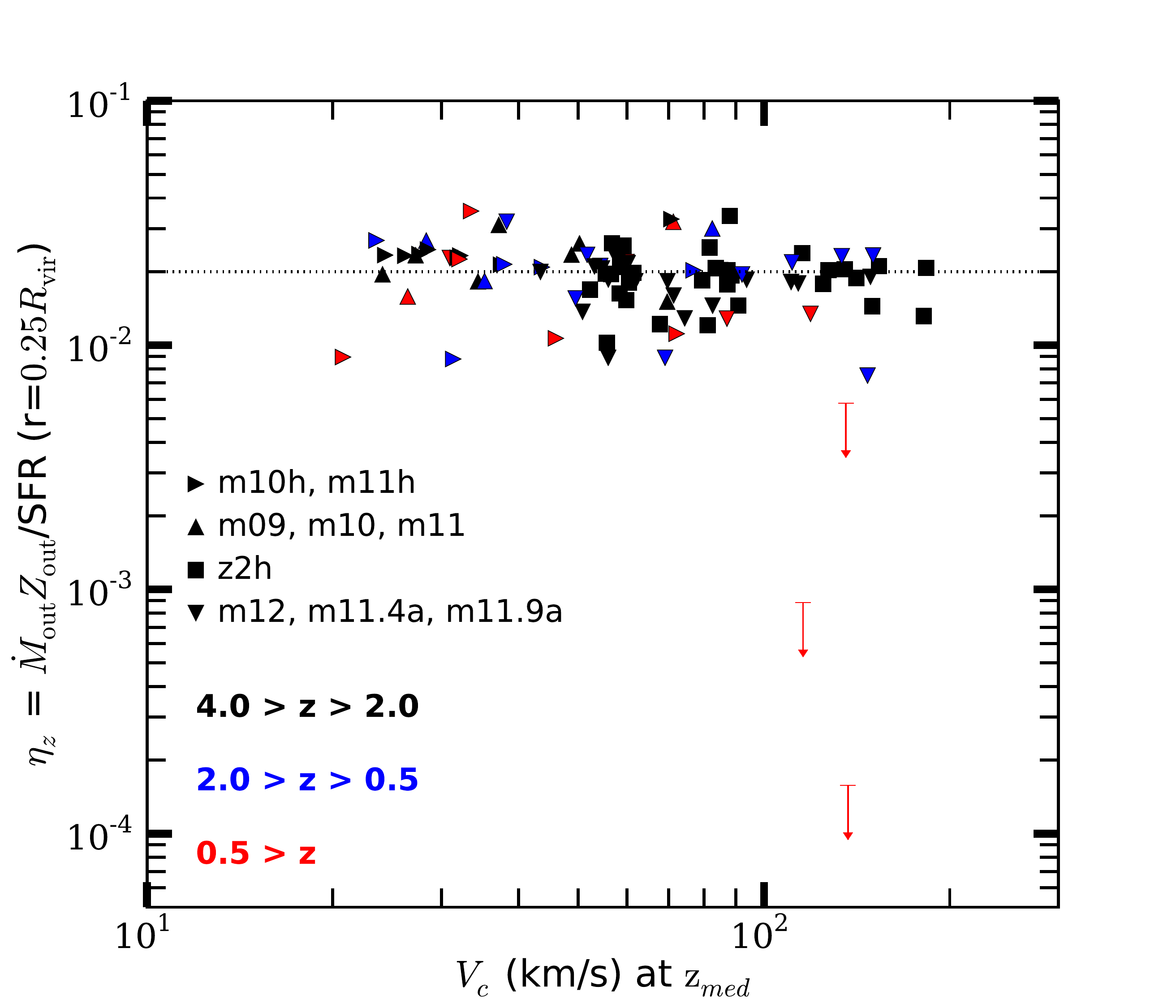}\\
\includegraphics[width=\columnwidth]{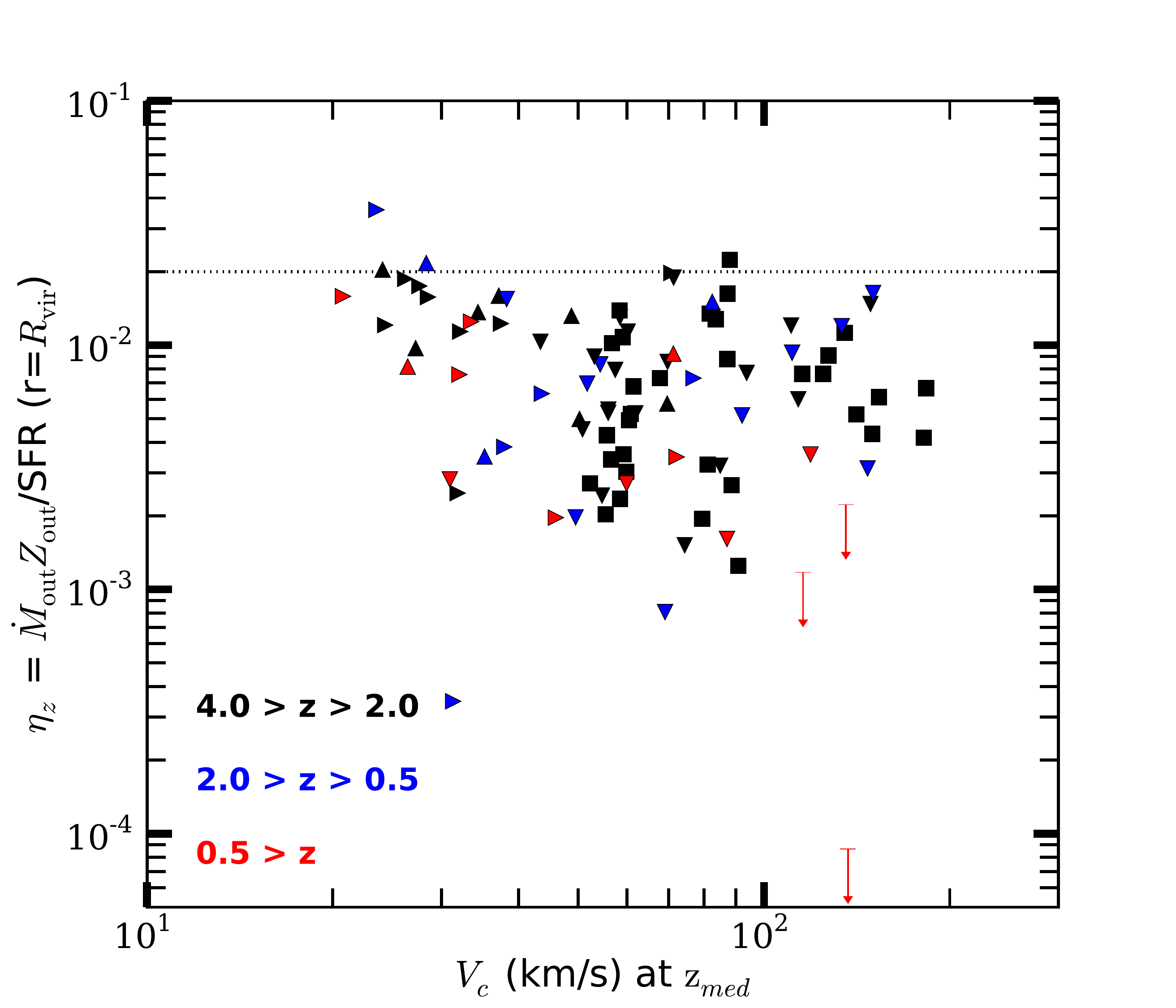}
\caption{Mass of metals ejected per unit star formation ($\eta_z$) vs. circular velocity, $v_c$. We show results for $0.25 \Rvir$ (top) and $\Rvir$ (bottom). Black, blue, red symbols refer to $z=3$, $z=1.25$, and $z=0.25$, respectively.  Right side up triangles show the halos in the zoom-in regions of \textbf{m09, m10,} and \textbf{m11}. Upside down triangles, show \textbf{m11.4a}, \textbf{m11.9a}, and the \textbf{m12} halos, but not the low-redshift \textbf{m12} ``main'' halos, which are shown as upper limits  with downward pointing arrows (see text). Squares show {\bf z2h} halos. {\bf m09} is shown only for the highest redshift bin as it has no detectable star formation and metal in/outflow at lower redshift. Sideways triangles are the \textbf{m10h} series. At $0.25 \Rvir$, galaxies have outflows with $\eta_z \approx 0.02$ independent of stellar mass, which we highlight with the black dotted line. This implies that for every star formed, about 2\% of the mass goes into metal outflows, which is similar to the yield of metals in our simulations. At $\Rvir$, $\eta_z$ has higher scatter and is generally lower than the yield. All simulated galaxies meeting the inclusion criteria outlines in Section \ref{sec:average_eta} are plotted here.}
\label{fig:etaz}
\end{figure}

\section{Metal content of winds}
\label{sec:winds}

First, we show CGM metal inflow and outflow rates in several FIRE halos (\textbf{m10, m11,} and \textbf{m12i}) from $4.5 > z > 0$ in Figure \ref{fig:MetalFlowing} (Left). We follow the methodology of M15 to compute fluxes, though in this figure, we smooth outflow rates over intervals of 400 Myr to ease visual interpretations. We note, however, that outflow and star formation rates vary on shorter timescales, particularly at high redshift. Generally, fluxes are measured in concentric shells of width $dL = 0.1 \Rvir$. For the purposes of taking measurements in the ``inner'' and ``outer'' parts of the CGM, we consider shells at $0.25 \Rvir$ and $1.0 \Rvir$ at the epoch when flux is being measured. All gas in the shell with radial velocity $v_{rad}>0$ km/s is counted as outflows, while all gas with $v_{rad}<0$ km/s is counted as inflows, which are represented as a negative flux. In each shell, metal flux is computed as: 
\begin{equation}
\frac{\partial M}{\partial t} = \sum v_{rad} Z_{SPH} m_{SPH} / dL. 
\label{eq:flux_outflow}
\end{equation}
Here, $m_{SPH}$ is the mass of each SPH particle and $Z_{SPH}$ is the particle's metal fraction. Similar representations of the metal flux of \textbf{m10} and \textbf{m11} have been previously shown in \citet{ma_etal16}.

We demonstrated in M15 that this method of computing fluxes is generally consistent with an alternative method in which particles are tracked directly as they cross spherical interfaces in the CGM. In that work, we also argued that using $v_{rad} > 0$ to define outflowing flux is a reasonable approach (compared, e.g., with only considering fast particles). Similarly, we find here that typically only about 10\% of the metal flux comes from particles with $v_{rad} < \sigma_{1D}$, the 1-D velocity dispersion of the host halo.

\begin{figure}
\includegraphics[width=\columnwidth]{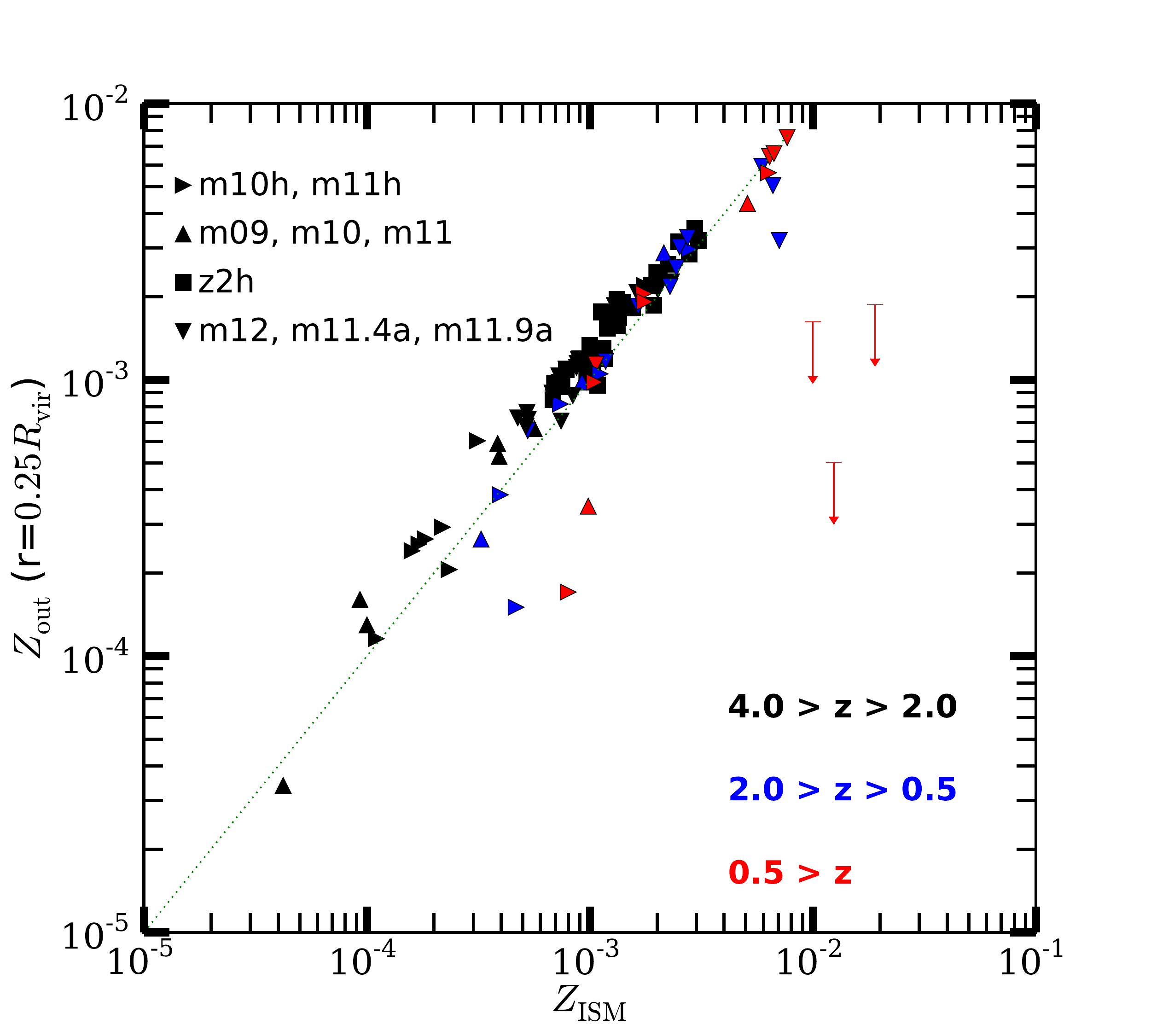}\\
\includegraphics[width=\columnwidth]{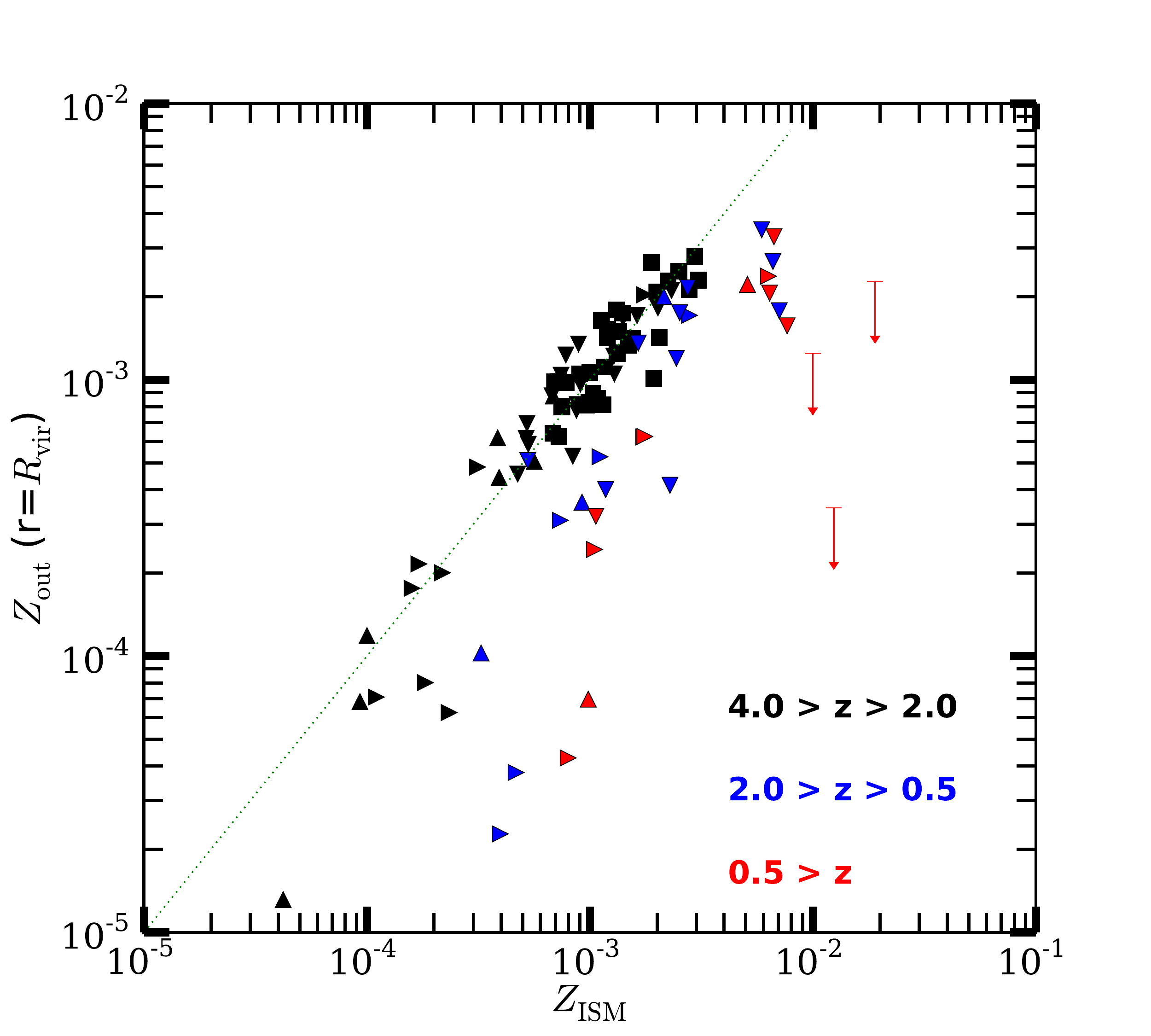}
\caption{Metallicity of outflows vs. ISM metallicity. We show results for $0.25 \Rvir$ (top) and $\Rvir$ (bottom). Outflows at $0.25 \Rvir$ are generally very similar to ISM metallicity with a small positive offset at high redshift. At low and intermediate redshift, there is a small but significant shift to lower metallicity. Outflows at $\Rvir$ have ISM metallicity at high redshift (with scatter), but have metallicity lower than ISM at low and intermediate redshifts. ISM metallicity is computed as a mass-weighted averaged over the epochs considered, while outflow metallicity is a flux-weighted average. The sample of galaxies plotted here, as well as the meaning of the color and shape of each data point is the same as in Figure \ref{fig:etaz}.}
\label{fig:ISMrats}
\end{figure}

\begin{figure}
\includegraphics[width=\columnwidth]{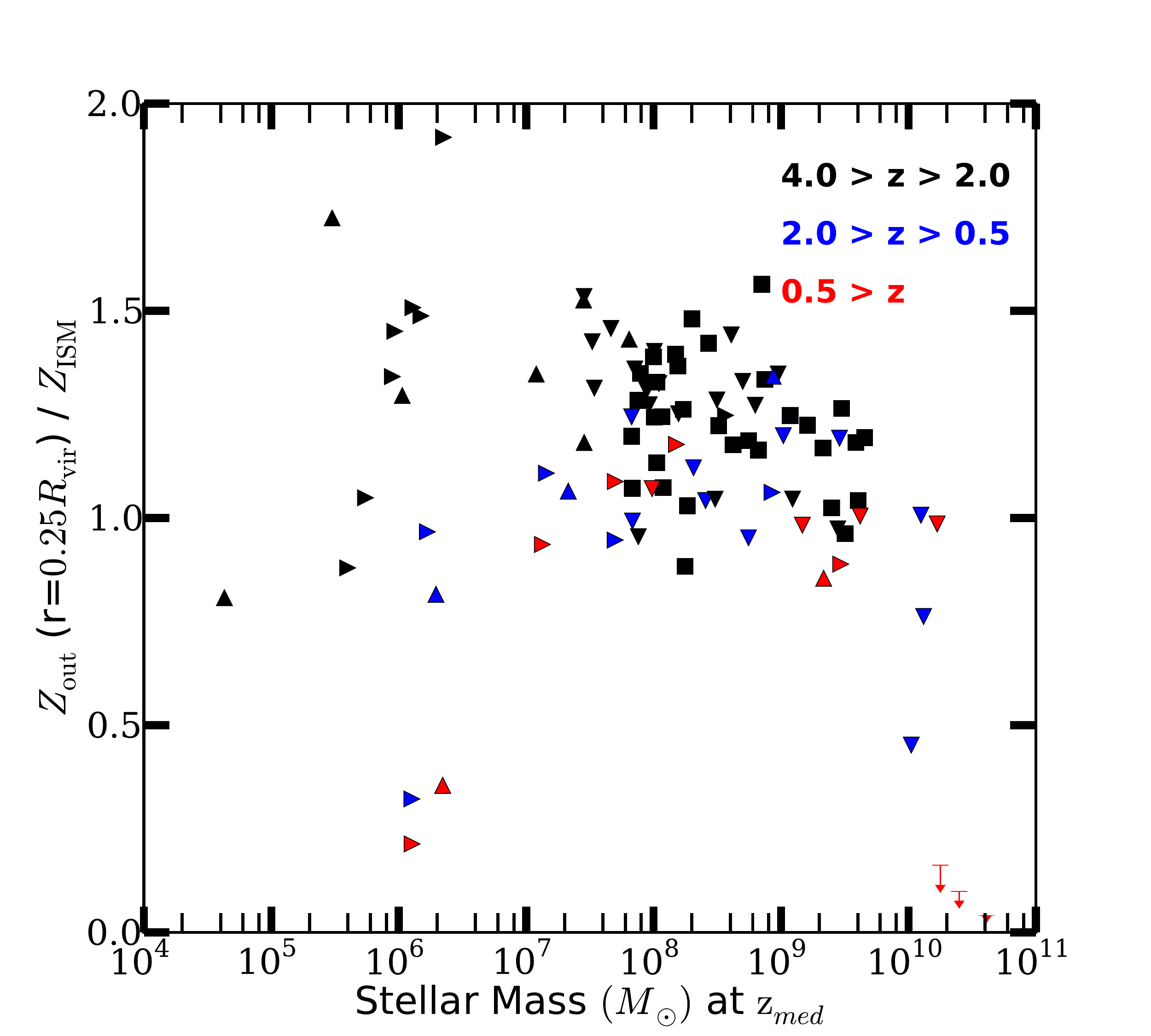}\\
\includegraphics[width=\columnwidth]{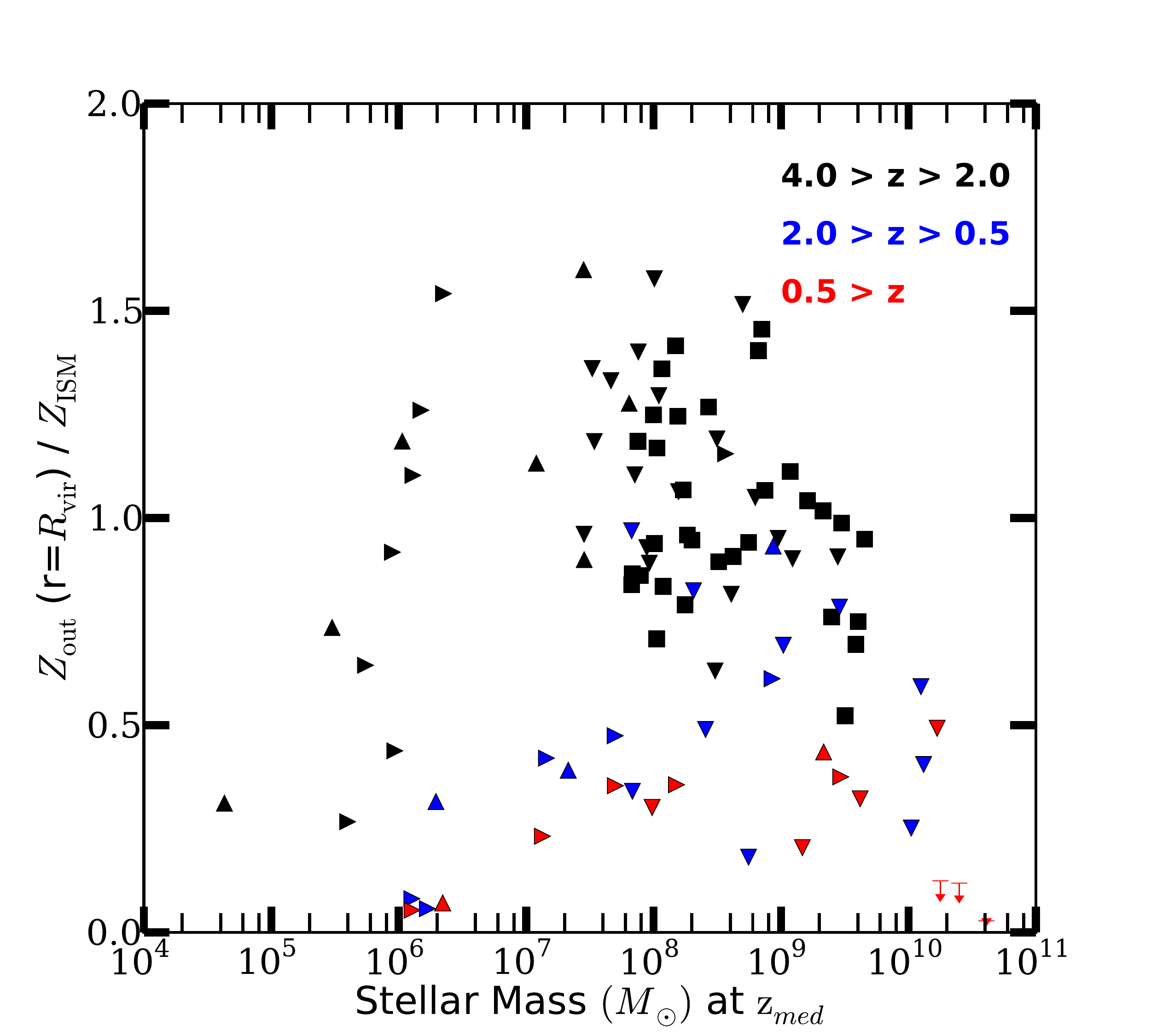}
\caption{Ratio of outflow metallicity to ISM metallicity vs. $M_*$. At $0.25 \Rvir$ (top), outflows are generally metal-rich compared to the ISM, with the exception of several low and intermediate-redshift galaxies. At $\Rvir$ (bottom), the ratio is generally more scattered, with a clear tendency for outflow metallicities to become increasingly metal poor compared to the ISM at low redshifts. The sample of galaxies plotted here, as well as the meaning of the color and shape of each data point is the same as in Figure \ref{fig:etaz}.} 
\label{fig:etaz_bymet}
\end{figure}

High redshift galaxies generally exhibit bursty star formation followed by gusts of outflows that are loaded with metals. At $z<1$, dwarf galaxies remain bursty while sufficiently massive halos in the L* range tend to have low metal outflow rates. In fact, the measured CGM outflow rates in \textbf{m12i} at $z<0.5$ are most likely not associated with stellar feedback-driven winds, but are instead merely the flux of outwardly moving gas from the CGM and gas stripped from infalling satellites.\footnote{The same applies to \textbf{m12q} and \textbf{m12v} at low redshift. For more discussion, see M15 and \citealt{hayward_hopkins15}}
Typically, the galactic metal outflow rates exceeds metal inflow rates in both the inner and outer CGM. While metals are blown out in winds over short time intervals, the inflow of metals is substantially more gradual and the rates are modest. Metal accretion appears to be dominated by contributions from recycled accretion following outflows.  A significant fraction of the metal accretion at low redshift could also come from winds launched by other galaxies via ``intergalactic transfer,'' which the particle tracking analysis of \citet{angles-alcazar_etal16} showed can be as important as fresh accretion of gas from the intergalactic medium or classical wind recycling (from the same galaxy) for $L*$ galaxies. 
Mergers can also bring some metal inwards (for example, \textbf{m12i} undergoes a major merger near $z\approx2$, while \textbf{m11} has one at $z=0.6$), but generally contribute a small fraction of the overall mass growth of galaxies below $\sim L*$. 
The contribution of metals accreted from satellite galaxies should be even smaller than their contribution to the overall mass growth of galaxies, because the MZR (satisfied by FIRE galaxies) implies that lower-mass galaxies have on average lower metallicities (see also \S~\ref{sec:budget}).

In Figure \ref{fig:MetalFlowing} (right), we show that outflow metallicities are typically higher than inflow metallicities, though in the inner CGM at $0.25 \Rvir$, they typically do not differ by more than a factor of $\sim2$. The similarity between inflow and outflow metallicity in the inner halo is broadly consistent with the prior work of \citet{oppenheimer10}, which argued that late-time gas infall is dominated by recycled wind material. At high redshift, both outflow and inflow metallicities at $0.25 \Rvir$ are close to ISM metallicities \footnote{ A brief drop in the metallicity of a fast outflow around $z\sim 3 $ apparent in the {\bf m10} panel of Figure \ref{fig:MetalFlowing} is caused by a low metallicity satellite galaxy moving outward in this snapshot, and is not associated with a feedback-driven outflow episode.}. In the outer CGM at $1.0 \Rvir$, the difference between outflow and inflow metallicity is more significant, with offsets typically at $\sim 0.5-1$ dex. 
 
In all galaxies, ISM metallicity increases substantially with time. By $z<1$, the ISM metallicity typically exceeds the outflow metallicities at both radii considered here. However, when we examine only material faster than the halo 1D velocity dispersion, $v_{rad} > \sigma_{1D}$ (yellow shaded region), the relative metal enhancement of the outflow is more pronounced at all redshifts. This suggests that during a typical gust of galactic wind, newly enriched material most affected by stellar feedback is rapidly accelerated to high velocities, and streams through the CGM. As the wind propagates, the momentum and energy of the fast, enriched wind is transferred to other gas in the ISM, as well as some material in the CGM. Subsequently, some metal-enriched material can cool and return to the ISM, while a circumgalactic outflow partially consisting of swept-up metal poor gas continues to travel outwards.  At epochs when the outflow rates are low (faint portions of the lines in Figure \ref{fig:MetalFlowing}), particularly at low redshift, our outflow metallicity measurements are primarily measuring properties of bulk-flow low-velocity CGM gas that is not associated with feedback-driven winds (see M15 for details).

The CGM metallicity is defined here as the gas mass-weighted average metallicity of gas between $0.1 \Rvir$ and $1.0 \Rvir$. This is the fiducial definition of CGM we employ in this work. ISM is gas within 0.1 $\Rvir$.  We specify whenever we consider other choices to distinguish CGM from ISM throughout the text. We note that the chosen fiducial definition of ISM  can include gas with $T>10^4 {\rm K}$, as well as gas out of the disk plane.

The metallicity of the CGM and ISM are similar at high redshift, when metals are routinely being blown out of galaxies. Unlike the ISM, which becomes increasingly metal-rich for all galaxies shown, the CGM metallicity generally stops growing at $z=1.5$. The ISM and CGM metallicities differ by an order of magnitude at $z=0$ in both \textbf{m12i} and \textbf{m10}, where outflows have ceased to be as strong as they were at high redshift. On the other hand, the \textbf{m11} low-redshift CGM metallicity is still close to the ISM metallicity, owing to the bursty outflows that still occur at low redshift.

\subsection{Time-averaged trends}
\label{sec:average_eta}
Next, we quantify the flux of metals per unit star formation. The mass-loading factor, $\eta$, is defined as the ratio of mass outflow rate to the star formation rate. We define the metal-loading factor $\eta_z$, as the ratio of the total galactic metal mass outflow rate, i.e. Equation \ref{eq:flux_outflow} for all particles with $v_{rad}>0$, to the star formation rate.

Following M15, we break down our data into three broad redshift intervals, high redshift $4 > z > 2$, intermediate redshift $2 > z > 0.5$, and low redshift $z < 0.5$.  The number of galaxies in our sample is greatest during the high-redshift range. We require that each galaxy survived for the entire duration of the interval (without being disrupted or merging into a more massive progenitor) to be included in this analysis. In addition, each included galaxy must contain at least 50,000 high-resolution dark matter particles. In order to avoid galaxies close to the edge of the zoom-in region we further require that less than 2\% of halo's dark matter mass is contributed by  dark matter particles that originate from the outside of the zoom-in region. To calculate an averaged $\eta$ and $\eta_z$ for the time interval for a given halo, we compute the total ejected mass by integrating the flux over time, and divide by the cumulative mass of formed stars. The average $\eta_z$ for each redshift interval is effectively flux-weighted, as epochs where the flux is high contribute more prominently. We report galactic properties such as stellar mass and circular velocity at the median redshift during each interval, $z_{med}$.

Figure \ref{fig:etaz} shows that $\eta_z$ does not strongly correlate with halo circular velocity $v_c = (G\Mvir/\Rvir)^{0.5}$ (a proxy for halo mass), during the high-redshift era, and is instead confined to a narrow range of values around $\eta_z \approx 0.02$. 
The yield of Type II supernova metals per star formed is $y\approx0.02$, with a slight dependence on stellar metallicity.\footnote{For older stellar particles which have had time to undergo AGB phase and Type Ia supernova, losing mass from stellar evolution, metal yield is actually closer to $y\approx0.03$. Nevertheless, the effective yield measured for simulated galaxies at $z=2$ is still $y\approx0.02$.} Since we see that $\eta_z$ is approximately equal to this yield for all galaxies, the implication is that almost all metals produced through Type II supernova stellar feedback are ejected from the galaxy, at least temporarily. This is consistent with our analysis in M15, where we argued that strong winds typically removed a substantial fraction of the dense ISM, where metals are deposited following a supernova. 

Metals that are ejected once may eventually be re-accreted. In fact, some metals can be re-accreted and then re-ejected in a second wind episode. Additionally, some amount of metals that are gained through mergers can be incorporated into the wind. On the other hand, some metals may be locked into stars without ever being ejected in the CGM. The significance of each effect will be explored in depth in a future work  (Angl{\'e}s-Alc{\'a}zar et al., in prep). For now we can say that the sum of effects from these processes leave the average normalization of $\eta_z$ to be nearly equal to the type II supernova yield as discussed above, independent of galaxy mass. However, some combination of these processes may account for the scatter of $\eta_z$. 

As we argued in M15, outflows driven by stellar feedback generally do not reach the CGM at $z<1$ for FIRE L* galaxies. This is further evidenced by the substantially lower values of metal flux for all three L* halos at $z<0.5$. Although plenty of stars are forming, the only metal flux seen at $0.25 \Rvir$ comes from random motion of metal-poor CGM gas, as well as from gas stripped from satellites. 

Next, we show the metal flux in the outer CGM, specifically at $1.0 \Rvir$, in the bottom panel of Figure \ref{fig:etaz}.\footnote{We show the total mass-loading factor $\eta$ at $\Rvir$ vs. halo circular velocity and stellar mass in Figure \ref{fig:etaRvir}. This complements similar plots for $\eta$ at $0.25 \Rvir$ shown in M15.} Here, we find $\eta_z < y$ for almost all galaxies, although the scatter is very high, with values typically ranging from  10-80\% of the Type II supernova yield, with no significant halo mass (i.e. $v_c$) dependence. This scatter suggests that galaxy mass alone cannot predict whether metals will be efficiently retained in the CGM or blown out of the virial radius.

\begin{figure*}
\centering
\begin{minipage}{0.48\textwidth}
\centering
\includegraphics[width=\textwidth]{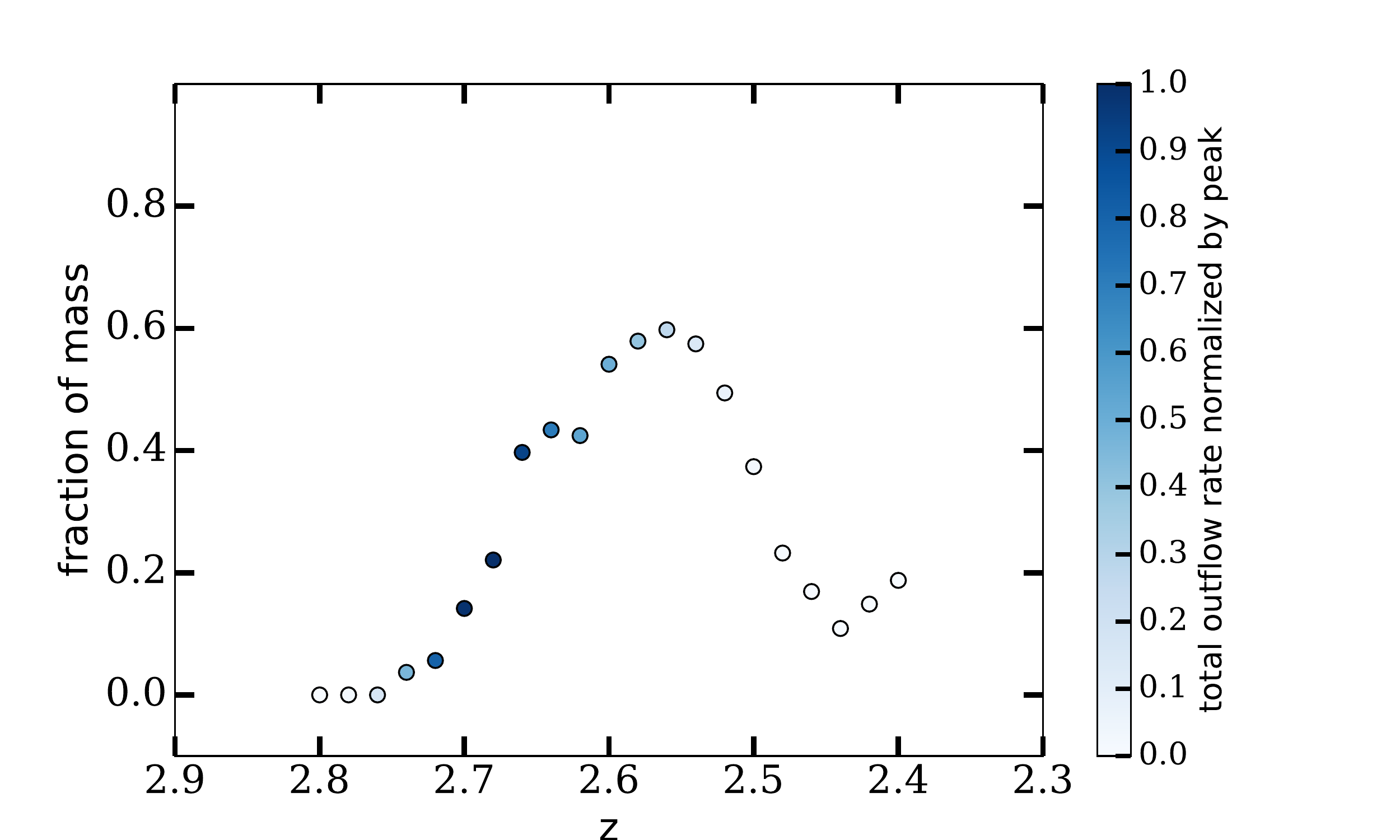}\\
\includegraphics[width=\textwidth]{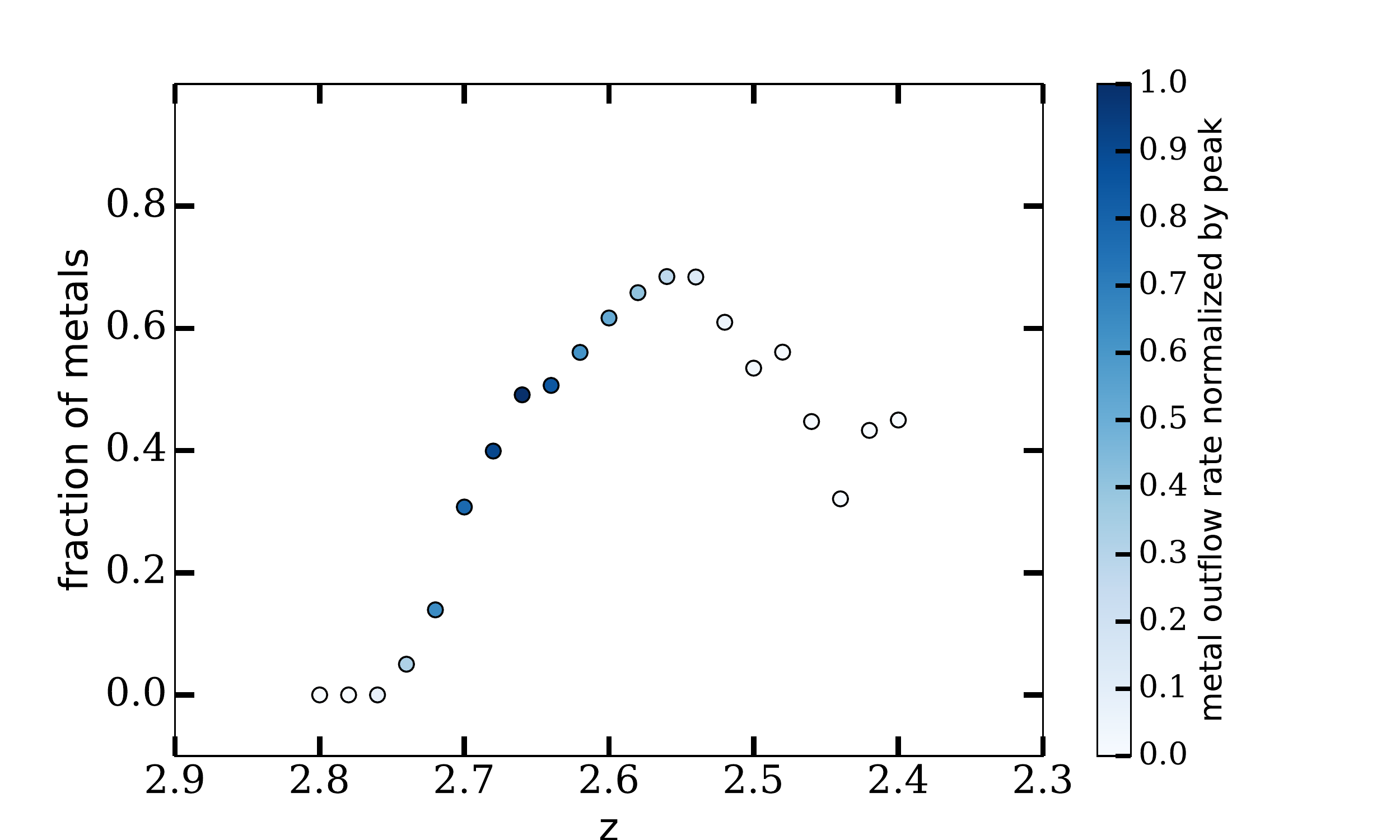}
\end{minipage}
\begin{minipage}{0.48\textwidth}
\includegraphics[width=\textwidth]{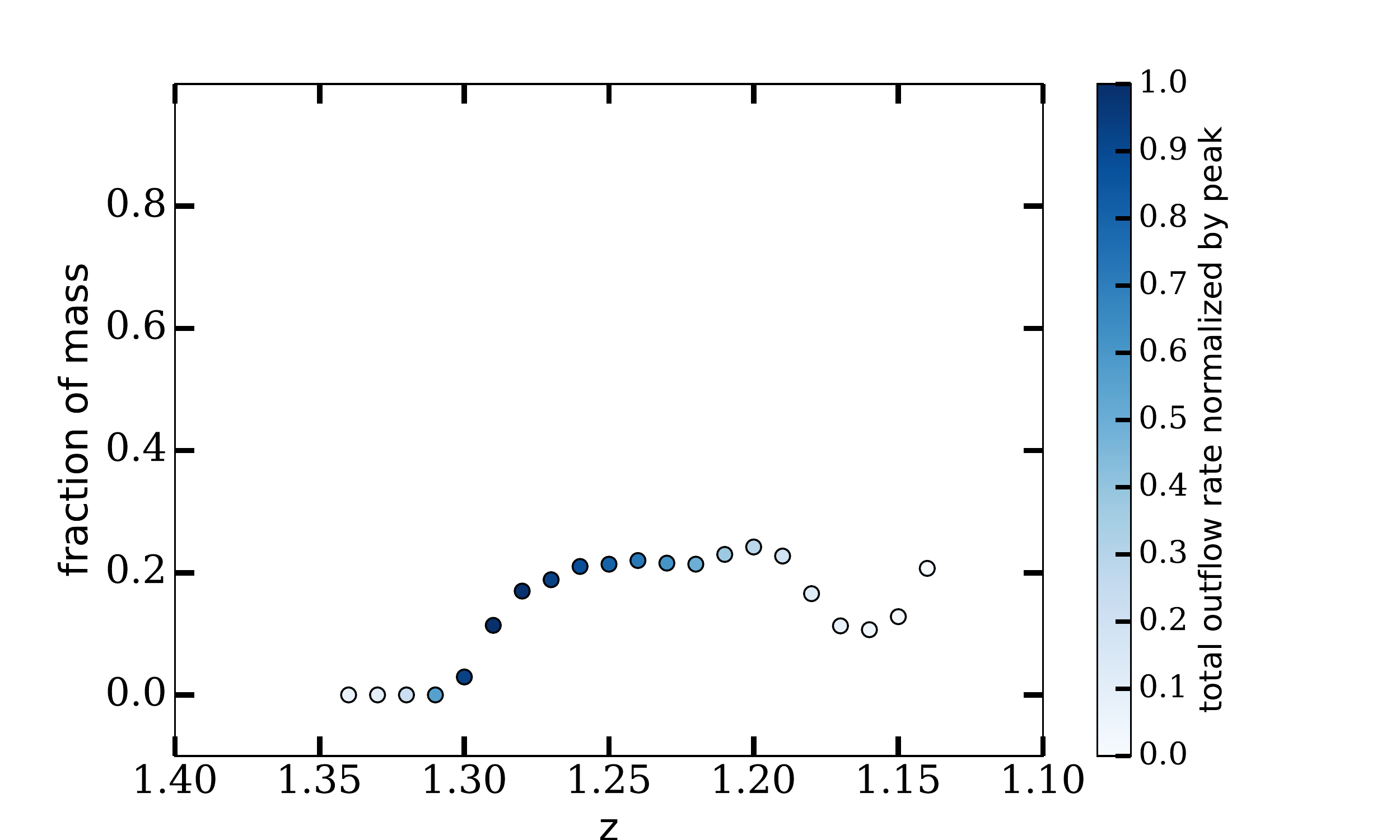}\\
\includegraphics[width=\textwidth]{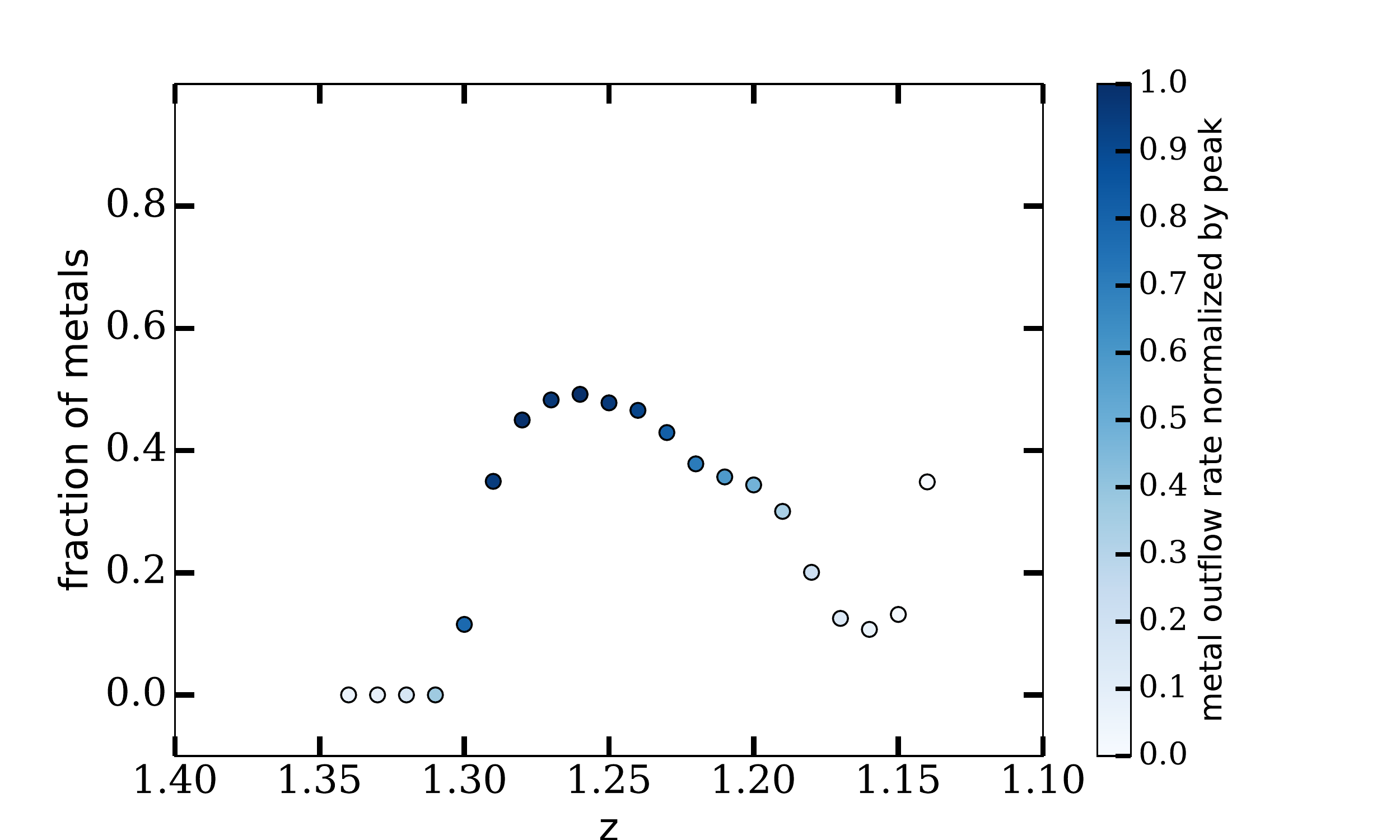}
\end{minipage}
\caption{ The fraction of total gas mass (top) and metal mass (bottom)  at $1.0 \Rvir$ that is outflowing with $v_{rad} > \sigma_{1D}$ originating from a $0.25 \Rvir$ outflow at a prior epoch ($z=2.8$ on the left and $z=1.34$ on the right) in \textbf{m12v}. A relatively low fraction of the total mass in the $1.0 \Rvir$ outflow is coming directly from the inner regions of the halo, implying that the initial ejecta exchanged momentum and energy with CGM gas as the outflow propagated. A larger fraction of the total outflowing metal mass originates in the inner halo. Combined with the decreased metallicity of the outflow at $1.0 \Rvir$ compared to $0.25 \Rvir$, this implies that swept up, metal-poor CGM gas dilutes the metallicity of the outflow at large radii. The points are shaded according to their outflow rate at $1.0 \Rvir$, normalized by the peak rate during the interval considered.}
\vspace{0.1cm}
\label{fig:TrackOutflow}
\end{figure*}

\begin{figure}
\includegraphics[width=\columnwidth]{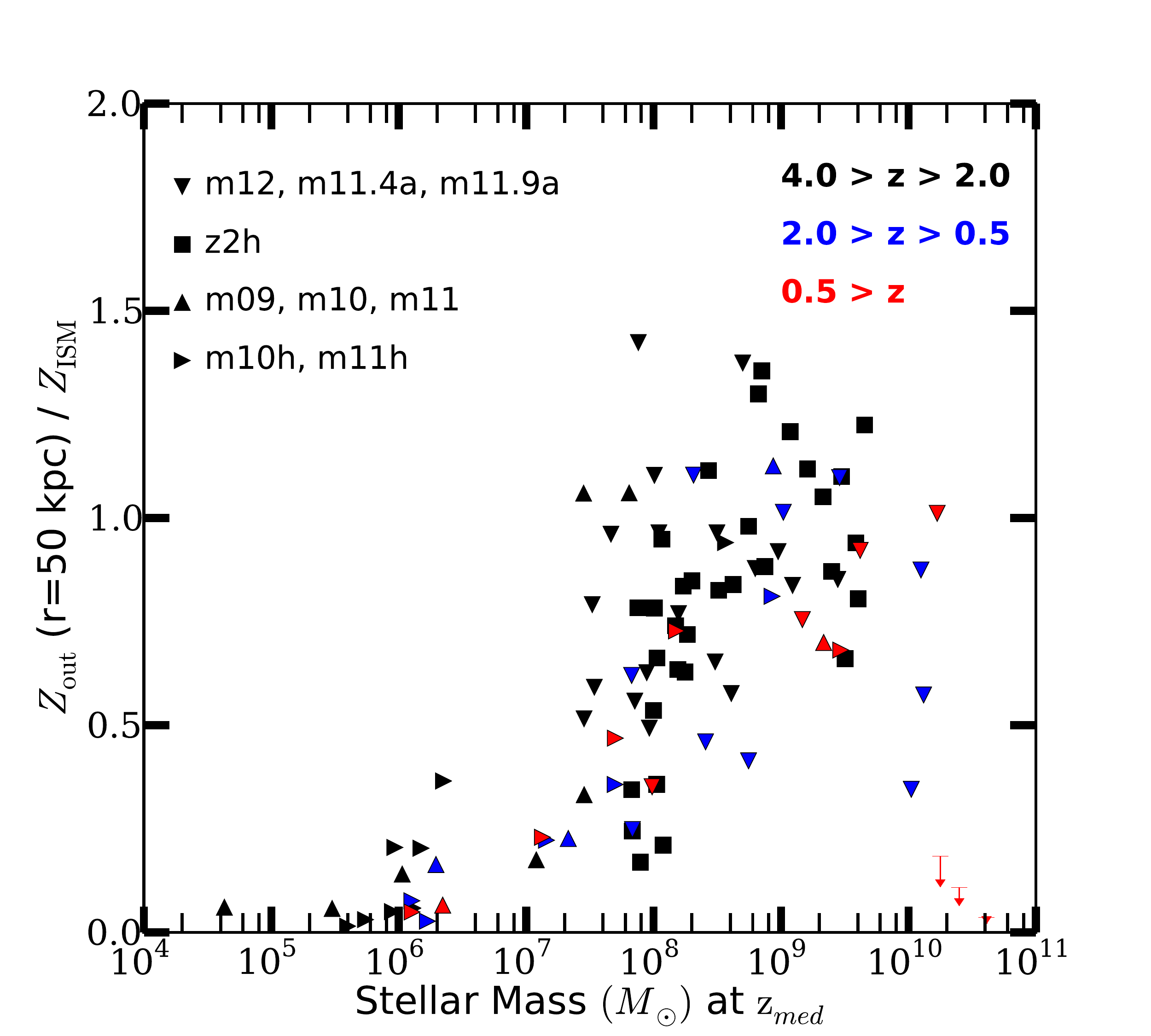}
\caption{Like Figure \ref{fig:etaz_bymet} but measuring flux in a fixed physical shell at 50 kpc from the galactic center. Redshift evolution in outflow metallicity at a given ISM metallicity is less pronounced when looking at flux through fixed physical distances.}
\label{fig:etaz_bymet_phys}
\end{figure}

\begin{figure}
\includegraphics[width=\columnwidth]{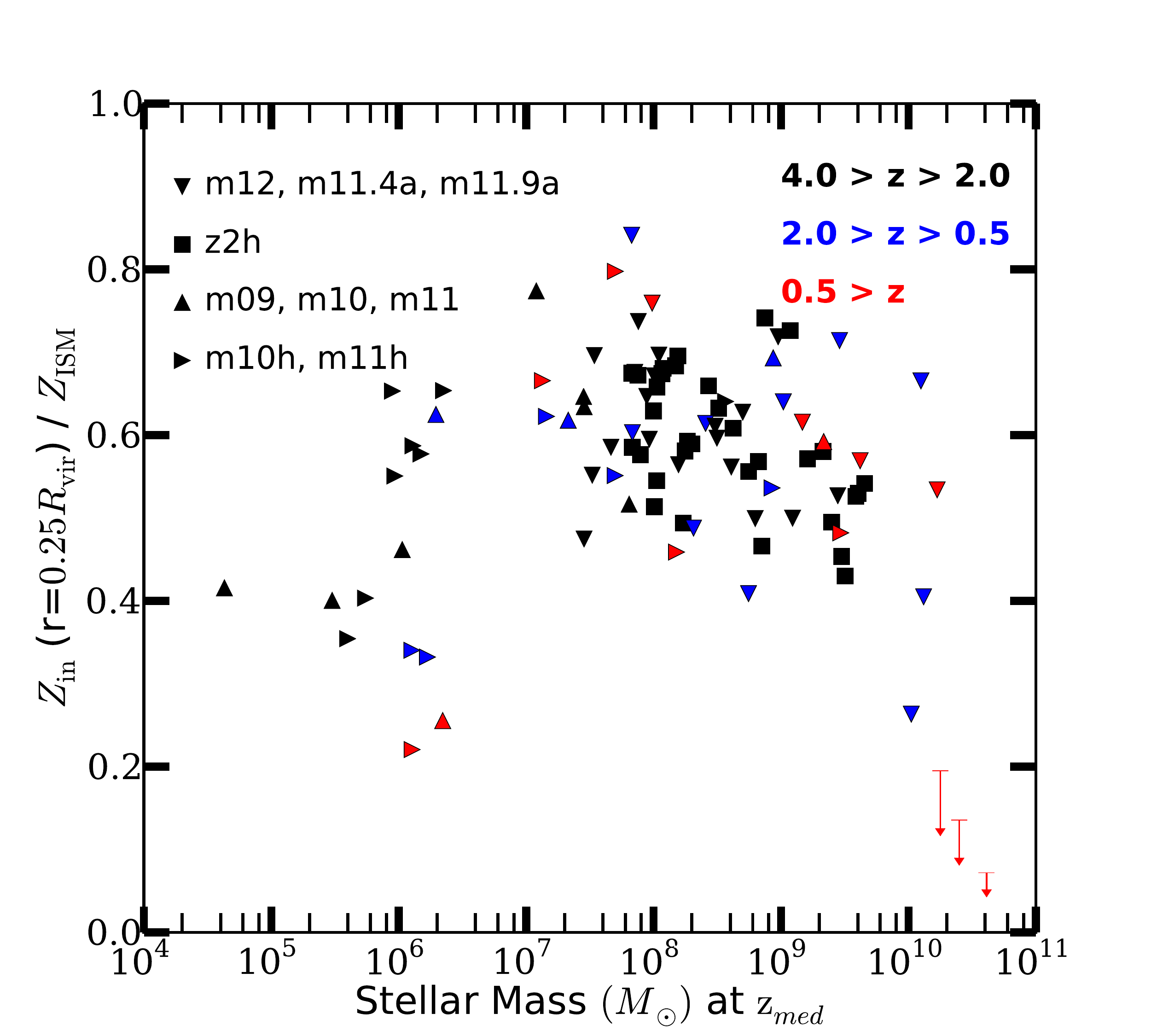}\\
\includegraphics[width=\columnwidth]{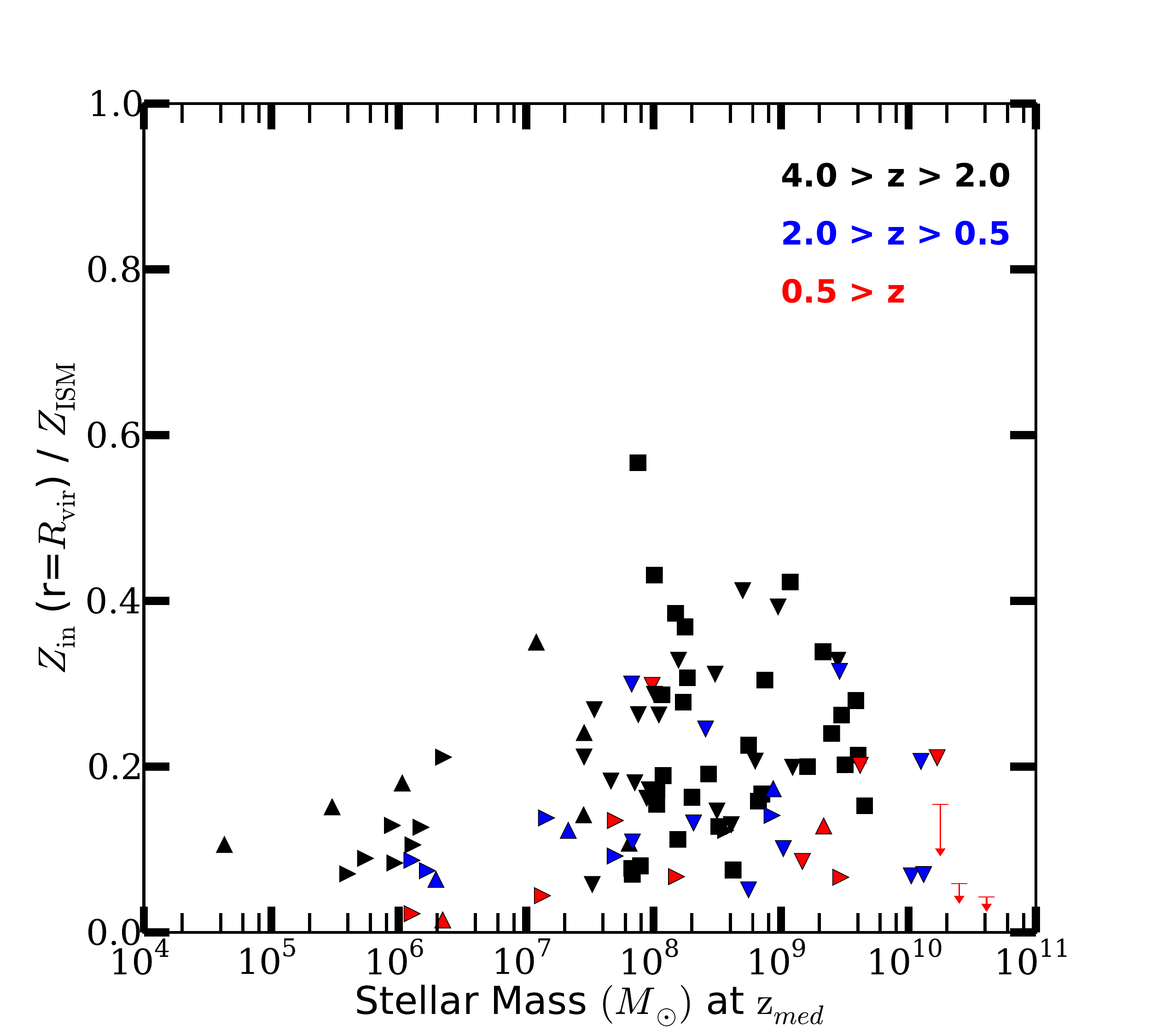}
\caption{Ratio of inflow metallicity to ISM metallicity vs. $M_*$. We show results for $0.25 \Rvir$ (top) and $\Rvir$ (bottom).  We demonstrate that inflows are generally metal poor compared to ISM at all redshifts, but only by a modest factor ($\sim$ 1.2-2) in the inner halo. ISM here is defined as gas within 0.1 $\Rvir$. The sample of galaxies plotted here, as well as the meaning of the color and shape of each data point is the same as in Figure \ref{fig:etaz}.}  
\label{fig:inflowetaz_bymet}
\end{figure}

In the upper panel of Figure \ref{fig:ISMrats}, we show that ISM metallicity is an excellent predictor of outflow metallicity (although the L*-progenitors at low redshift are again the exception because of their weak feedback-driven outflows, as discussed above). ISM metallicity shown on this figure is gas mass-averaged over the era, and includes all gas within $0.1 \Rvir$,  whether or not it is involved in star formation. We have found little difference in results using other averaging techniques (e.g. pure time average). Outflow metallicity, on the other hand, is a mass flux-weighted average, as we simply take the ratio of $\eta_z$ and $\eta$ computed for the epoch. The majority of all halos, independent of redshift, form a very tight and nearly linear correlation between ISM and outflow metallicity computed in this manner at $0.25 \Rvir$ (top panel). A least squares fit for the high and intermediate redshift data yields the following formula:

\begin{equation}
Z_{out}(r=0.25 \Rvir) = \left(0.7\pm0.1\right) \left(1 + z \right)^{0.4 \pm 0.1} Z_{ISM}^{1.0 \pm 0.03} .
\label{eq:Mstarfit}
\end{equation}

We caution that the coefficients for redshift evolution should be treated as rough estimates, as the data only consists of two possible epochs ($z=3$ and $z=1.25$ for high and intermediate redshift, respectively). In the outer CGM at $1.0 \Rvir$ (lower panel of Figure \ref{fig:ISMrats}), although there is more scatter, galaxies still essentially obey a  nearly linear correlation between ISM  and outflow metallicity. The outflow metallicities at $\Rvir$ are generally more metal-poor than they are at $0.25 \Rvir$, and are increasingly metal-poor with cosmic time for fixed stellar mass. The best-fit relation describing outflow metallicity at $r=\Rvir$ is:

\begin{equation}
Z_{out}(r=\Rvir) = \left(0.4\pm0.2\right) \left(1 + z \right)^{1.5 \pm 0.1} Z_{ISM}^{1.2 \pm 0.1} .
\label{eq:Mstarfit}
\end{equation}

We note that these fits are only crude guidance, in reality trends are more complicated and lowest and highest mass galaxies, especially at intermediate and low redshift, show significant deviations from these fits. 

\begin{figure*}
\centering
\begin{minipage}{0.48\textwidth}
\centering
\includegraphics[width=\textwidth]{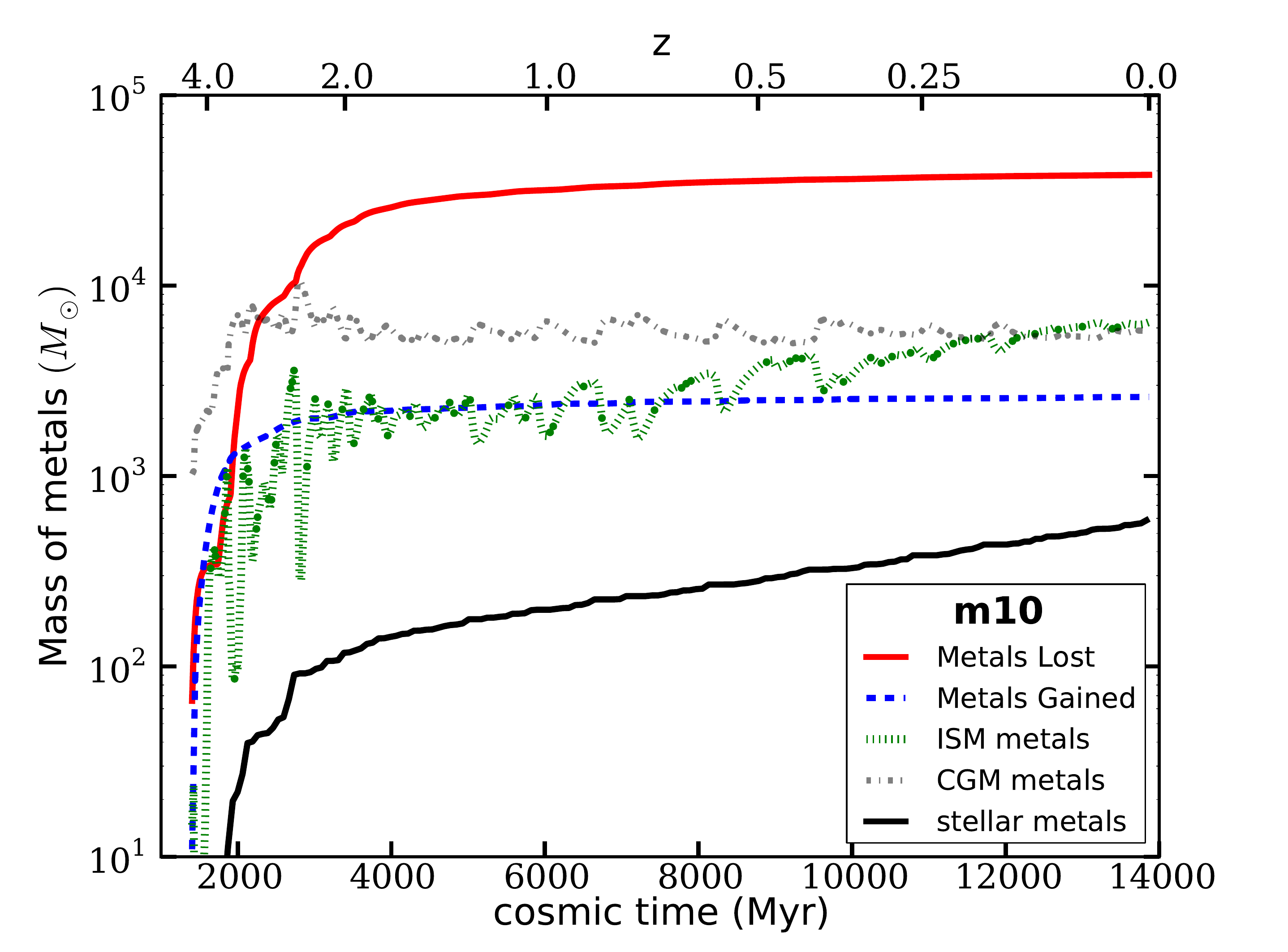}\\
\includegraphics[width=\textwidth]{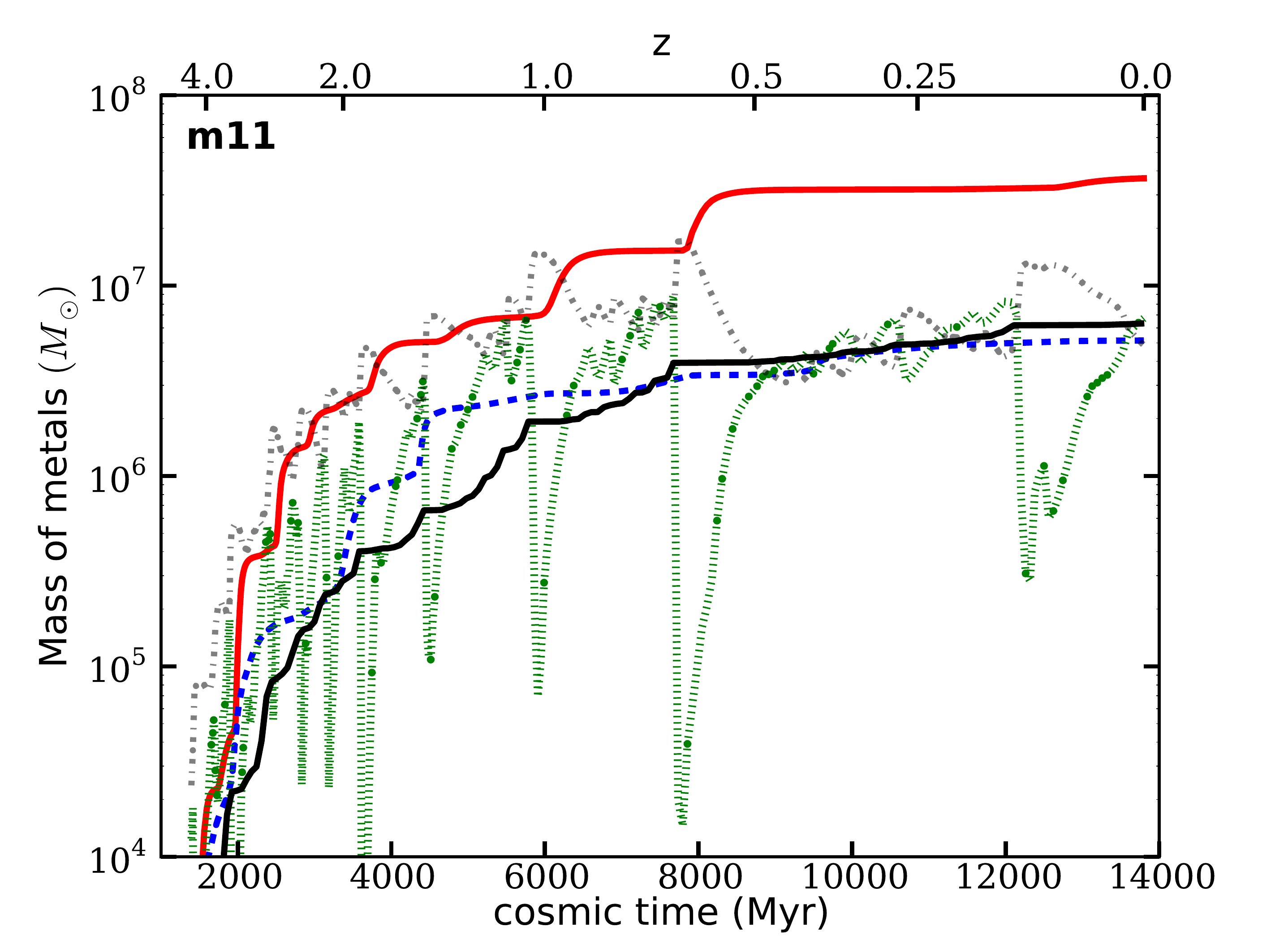}
\end{minipage}
\begin{minipage}{0.48\textwidth}
\includegraphics[width=\textwidth]{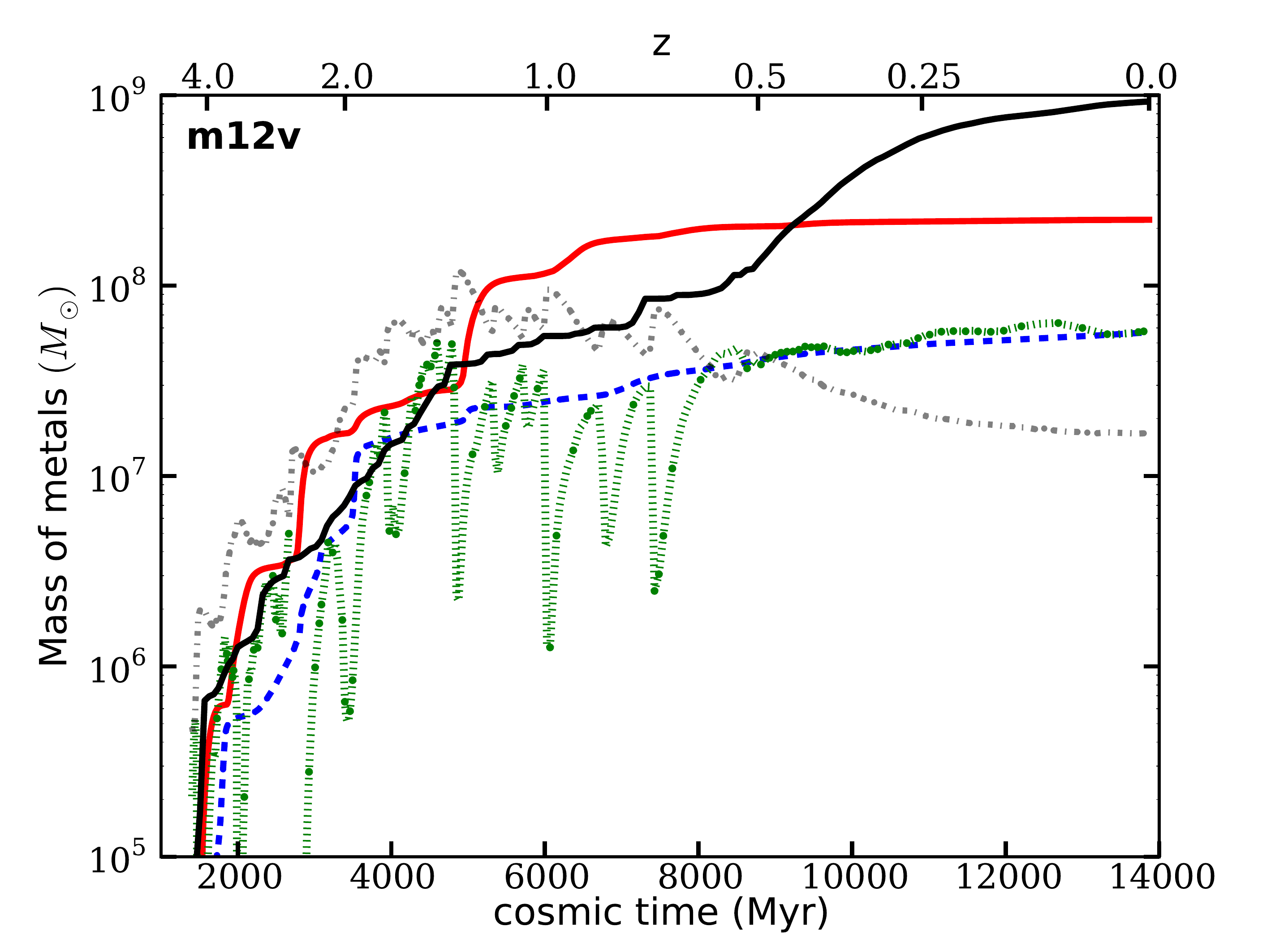}\\
\includegraphics[width=\textwidth]{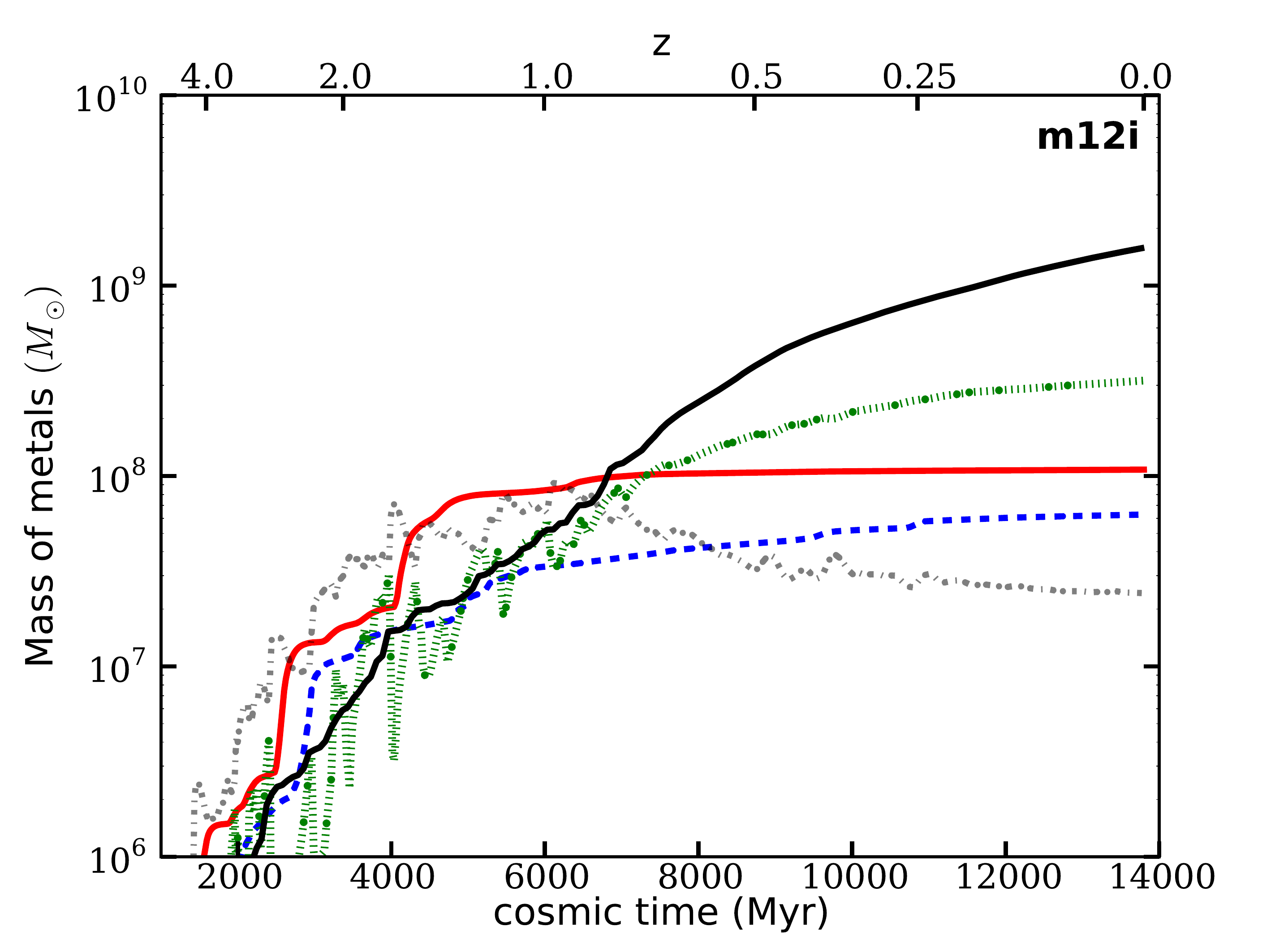}
\end{minipage}
\caption{ Time-integrated cumulative mass of material that has flown out of (red solid) and into (blue dashed) $\Rvir$ between $4.5 > z > 0$ for \textbf{m10} (upper left), \textbf{m11} (lower left),  \textbf{m12v} (upper right), \textbf{m12i} (lower right). We show the buildup of metal mass in stellar (black solid), ISM (green dotted), and CGM (black dash-dot) components. It is interesting that \textbf{m10} appears to retain a steady amount of metals in the ISM since $z=2$. The mass of ISM and CGM metals in \textbf{m11} and \textbf{m12v} fluctuate a lot and have less than they did at some higher redshift. \textbf{m12i} fluctuates until $z=1$, but then steadily grows. All galaxies have ongoing buildup of metals in their stellar component, even though star formation rate for \textbf{m10} and \textbf{m11} is low. \textbf{m12v} and \textbf{m12i} have very rapid buildup at $z<1$, when star formation is in a continuous mode.}
\vspace{0.1cm}
\label{fig:MetalFlower}
\end{figure*}

In Figure \ref{fig:etaz_bymet}, we show the ratio of outflow-to-ISM metallicity vs. stellar mass on a linear scale. Outflows in the inner halo (0.25$\Rvir$, top panel) of galaxies with $10^7 \Msun < M_* < 5\times10^9 \Msun$ are generally more metal rich than their ISM by a nearly mass-independent factor of $\sim1-1.5$. Winds with this ratio must be primarily composed of newly enriched material coming from the ISM, rather than swept up metal-poor circum-galactic gas. However, in a few galaxies, outflow metallicity is significantly less than the ISM metallicity, (i.e. $Z_{out}/Z_{ISM} \lesssim 0.8$).  These measurements correspond to the low and intermediate-redshift outflows in low-mass galaxies (such as \textbf{m10}) and intermediate-redshift outflows in the main galaxies in the original FIRE \textbf{m12} series (\textbf{m12i}, \textbf{m12q}, and \textbf{m12v}). The fraction of galaxies in the  $Z_{out}/Z_{ISM} \lesssim 0.8$ regime is much higher when considering outflow metallicity in the outer CGM at $\Rvir$ (bottom panel of Figure \ref{fig:etaz_bymet}). The metallicity of outflows at these distances also show more prominent redshift evolution.

We hypothesize that when outflows are still close to the source galaxy, they are dominated by ISM ejecta, but that the outflows interact in the CGM with metal poorer material, sweeping it up into the wind. While slower gas and metals remain within the halo, the faster material, now with additional mass loading from the swept up metal-poorer gas, can reach large distances. The additional mass loading from the CGM can then explain why the total outflow metallicity tends to decrease with increasing distance from the galaxy, especially in our intermediate and low redshift bins (see Figure \ref{fig:ISMrats}).


We tested this hypothesis by measuring the fraction of fast ($v_{rad} > \sigma_{1D}$) outflowing material at $1.0 \Rvir$ that originated from the inner CGM ($0.25 \Rvir$) at an earlier epoch when outflow rates at $0.25 \Rvir$ peaked. We considered both the total gas mass outflow rate as well as the metal mass outflow rate, and focused on \textbf{m12v} and \textbf{m12i} at high and intermediate redshifts. We found that at epochs when outflow rates at $1.0 \Rvir$ were highest, typically only $\sim 20-40$\% of the outflowing gas mass could be traced back to an earlier epoch where the same gas was outflowing at $0.25 \Rvir$. On the other hand, up to 50-60\% of outflowing metal mass at $1.0 \Rvir$ could be traced back to $0.25 \Rvir$ at the epoch when outflow rate was peaking. This is illustrated in Figure \ref{fig:TrackOutflow}, where we analyze two representative outflow episodes in {\bf m12v} and show a fraction of fast outflow material that originates in the inner halo. Therefore, while a large fraction of the metals outflowing at $1.0 \Rvir$ is associated with metal-rich ISM ejecta, a majority of the total outflowing gas mass is composed of metal-poor CGM that was instead swept up into the outflow as it traveled from $0.25 \Rvir$ to $1.0 \Rvir$. This reconciles the significant drop in the metal outflow rate seen at $\Rvir$ (Figure \ref{fig:etaz}) with the analysis of M15, who showed that the total gas mass outflow rate at $\Rvir$ is a substantial fraction of the total rate at 0.25$\Rvir$ (Figure 7 of M15), particularly at lower redshift.


\begin{figure*}
\centering
\begin{minipage}{0.48\textwidth}
\centering
\includegraphics[width=\textwidth]{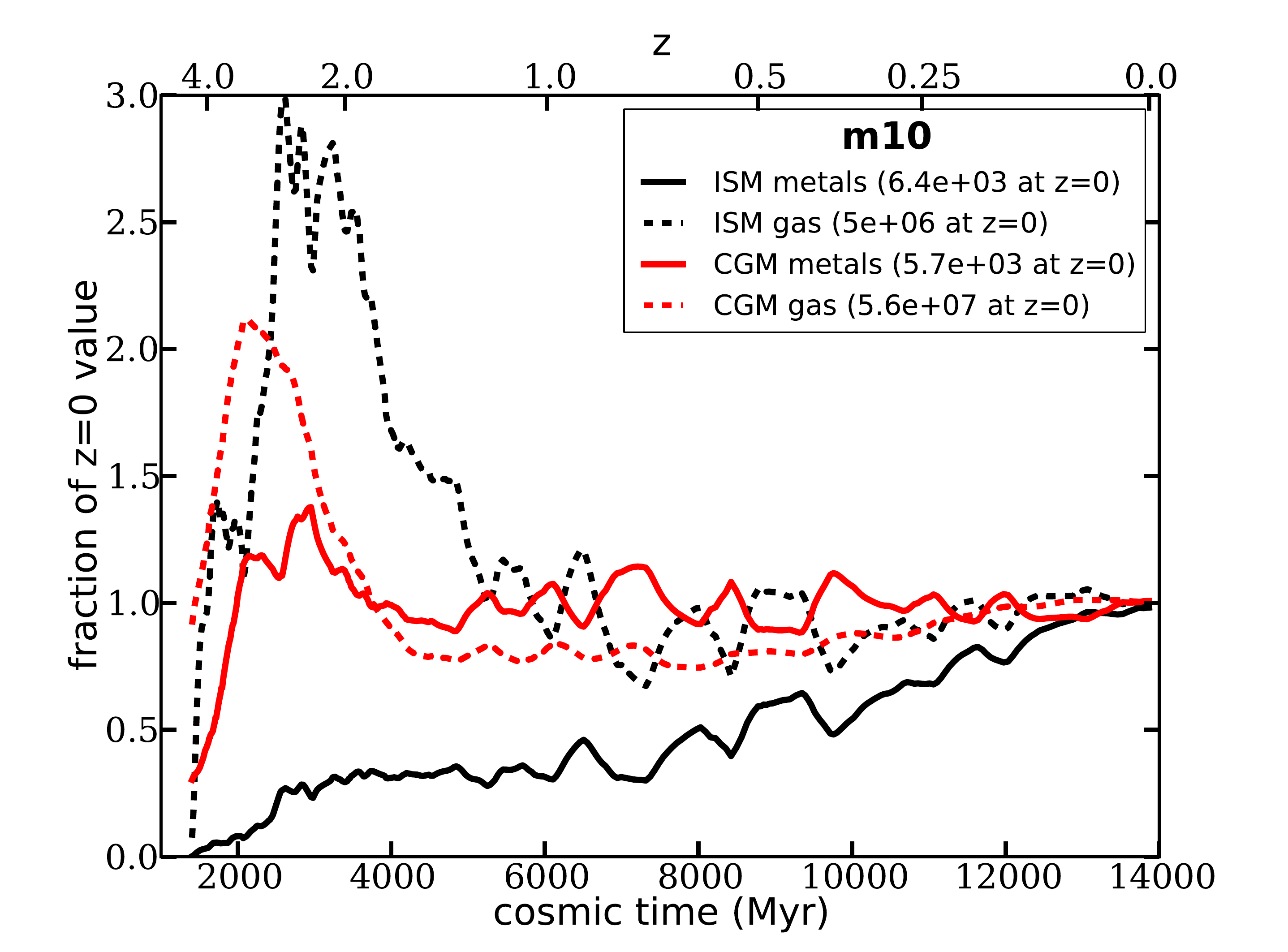}\\
\includegraphics[width=\textwidth]{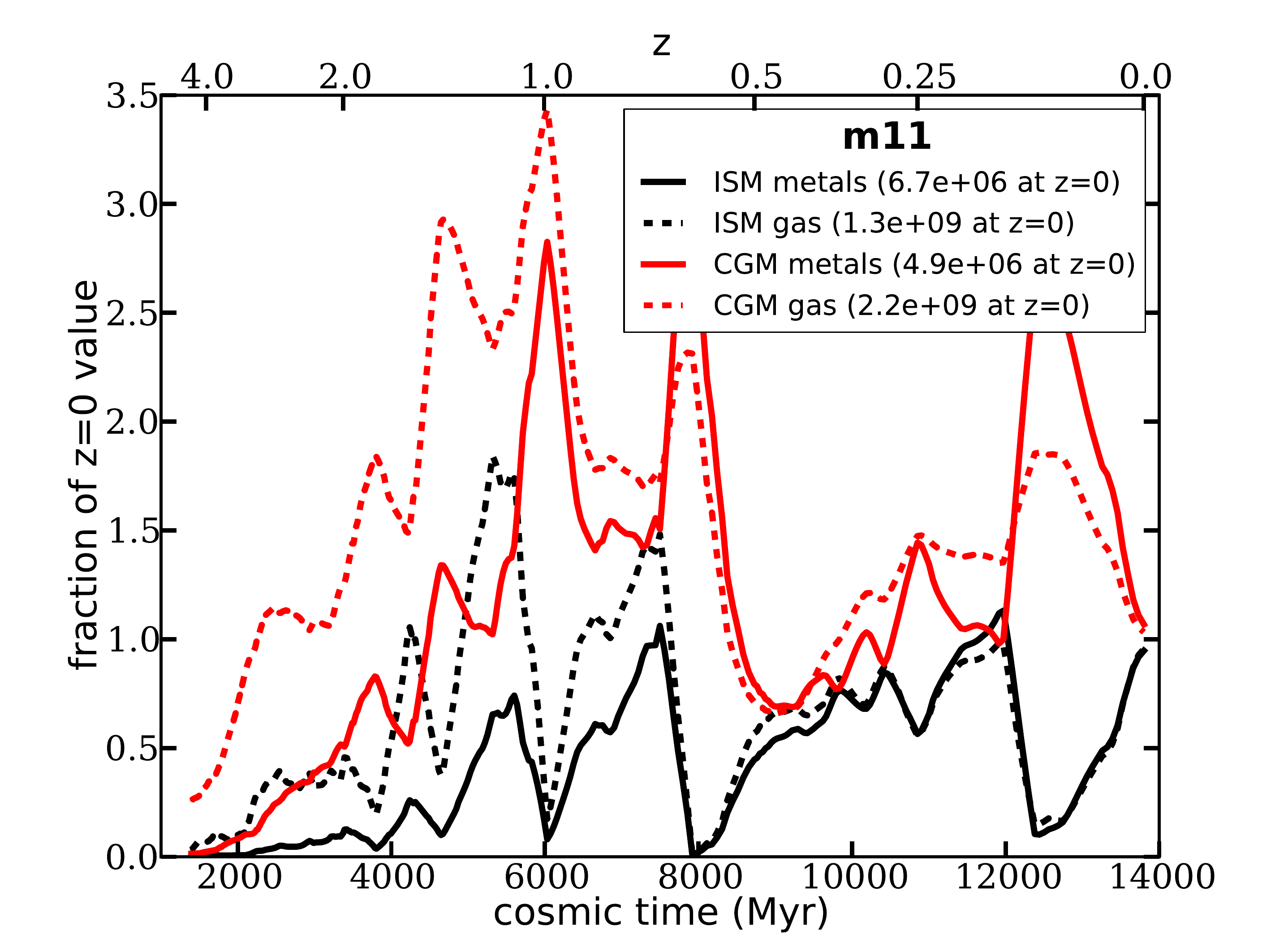}
\end{minipage}
\begin{minipage}{0.48\textwidth}
\includegraphics[width=\textwidth]{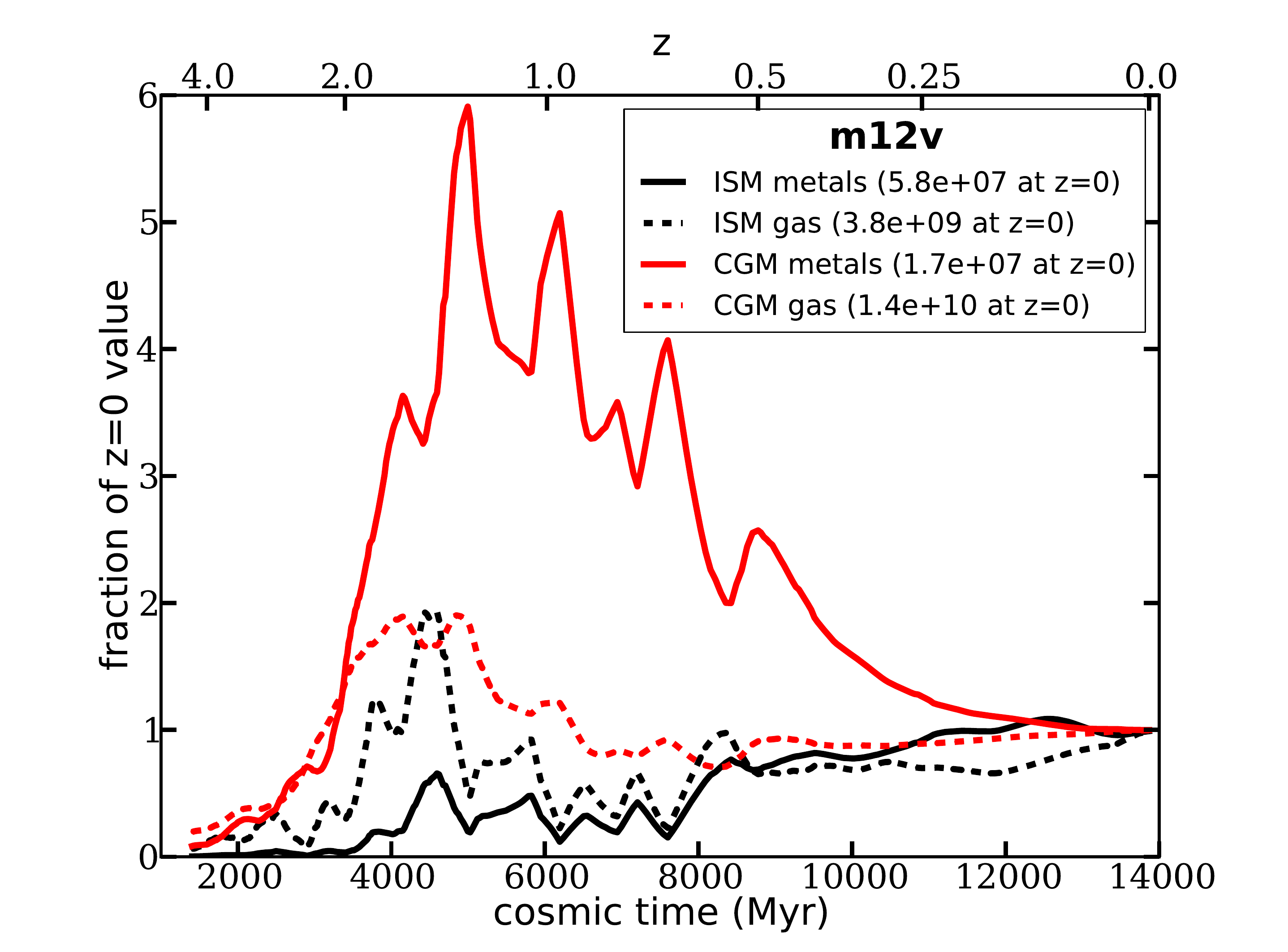}\\
\includegraphics[width=\textwidth]{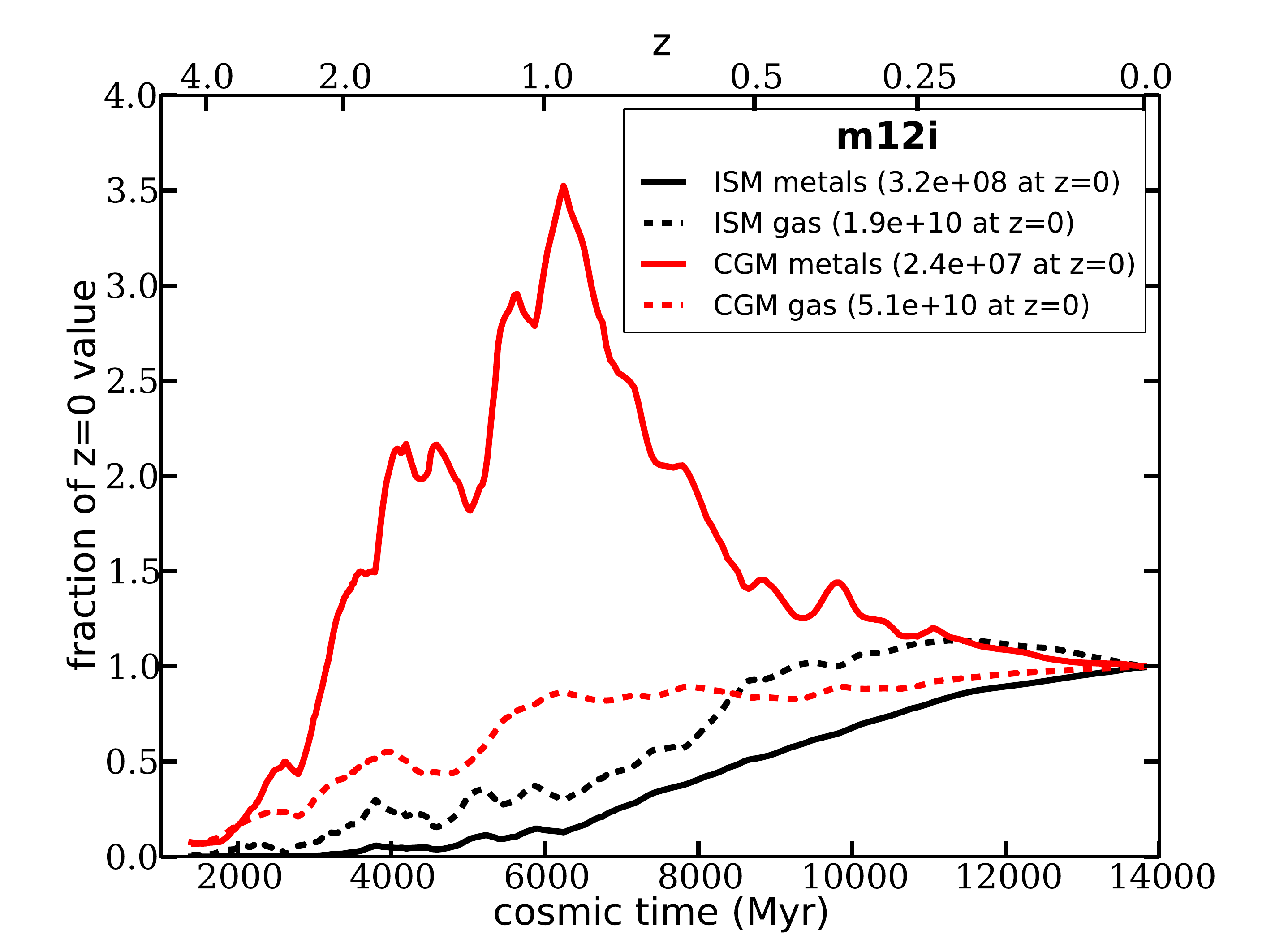}
\end{minipage}
\caption{Buildup of metal and gas mass in CGM and ISM components normalized by each component's $z=0$ value between $4.5 > z > 0$ for \textbf{m10} (upper left), \textbf{m11} (lower left), \textbf{m12v} (upper right), \textbf{m12i}( lower right). While galaxies have bursty star formation, all components fluctuate. At late times, there is a continuous buildup of metals in sufficiently massive galaxies, but also in low-mass dwarf \textbf{m10}.}
\vspace{0.3cm}
\label{fig:MetalBuildup}
\end{figure*}

We address the possibility that the redshift evolution of the metallicity of outflows seen in Figures \ref{fig:ISMrats} and \ref{fig:etaz_bymet_phys} these figures could be driven simply by the increasing physical distance corresponding to a particular fraction of comoving $\Rvir$ with decreasing redshift. Figure \ref{fig:etaz_bymet_phys}, provides a measurement of $Z_{out}/Z_{ISM}$ vs. stellar mass in a fixed physical shell at 50 kpc. In this case, we see $Z_{out}/Z_{ISM} < 0.5$ for all galaxies with $M_* < 10^7 \Msun$, confirming that these low-mass galaxies usually can't propel metal rich ejecta out very far; most of their outflows do not reach 50 kpc, though it should be noted that this distance can be well beyond the virial radii of these galaxies. Galaxies with $10^7 \Msun < M_* < 5\times10^9 \Msun$ generally have $0.5 < Z_{out}/Z_{ISM} \lesssim 1.2$, similar to the measurements at $\Rvir$, and while evolution between high and intermediate redshift epochs is statistically significant, it is less prominent than the evolution at $\Rvir$. However, we see that \textbf{m12} galaxies at intermediate redshift clearly have less metal-rich outflows than their high redshift counterparts. We found some evidence of this phenomenon through the particle tracking analysis described above, where a smaller fraction of outflowing particles at $1.0 \Rvir$ could be traced back to $0.25 \Rvir$ at lower redshifts in \textbf{m12} galaxies (see Figure \ref{fig:TrackOutflow}). We again note that in the lowest redshift bin, $z<0.5$, \textbf{m12i}, \textbf{m12q}, and \textbf{m12v} do not have any feedback-driven winds that can penetrate into the CGM.

Figure \ref{fig:inflowetaz_bymet} shows the ratio of inflow metallicity to ISM metallicity vs. stellar mass. Comparing the top panel of this figure to that of Figure \ref{fig:etaz_bymet} reveals that inflow metallicities averaged over such a long time interval are a factor of $\sim$ 2 lower than outflow metallicities in the inner CGM. Although many of the ejected metals recycle  back into the ISM following outflows, relatively pristine  gas from the IGM also contributes to the galaxy growth rate (see \citealt{keres05, dekel09, faucher-giguere11b, vandevoort_etal11b}). The bottom panel reveals that the inflow rate to $1.0 \Rvir$ is even more metal poor. Most halos are accreting gas that is a factor of $\sim$ 5 less enriched than the ISM. This provides a potential avenue for distinguishing infalling and outflowing gas in the outer halo based on its metallicity, in a statistical sense (e.g. \citealt{lehner_etal13}). 
A detailed analysis of low-redshift ($z<1$) Lyman limit systems (LLSs) in the FIRE simulations shows that such dense absorbers are confined to the inner halos of galaxies at late times \citep{hafen_etal16}. In the inner halo, inflowing and outflowing LLSs overlap significantly in metallicity, making metallicity alone an ambiguous diagnostic of inflow vs. outflow for most LLSs. However, \citet{hafen_etal16} do find that very fast outflows ($v\gtrsim2v_c$) are typically metal-rich and that very low-metallicity LLSs ($Z \lesssim 0.01 \Zsun$) are typically infalling.

\subsection{Metallicity evolution of individual galaxies}
Next, we return to the main galaxies from four individual FIRE simulations and examine their metal evolution in various components. We focus on the isolated low-mass dwarf galaxy \textbf{m10}, which has a very low baryon fraction at $z=0$; the dwarf starburst \textbf{m11} which maintains a bursty star formation history throughout its evolution; and the L*-progenitors \textbf{m12v} and \textbf{m12i}, both of which have gaseous disks and continuous star formation rates at low redshift, although the star formation rate is much higher in \textbf{m12i}. 

Figure \ref{fig:MetalFlower} shows the total mass of metals locked in stars, ISM, and CGM for the four galaxies, as well as the total amount of integrated metal flux that has gone in and out of the virial radius. While the stellar metal component grows monotonically, the ISM and CGM metals fluctuate considerably for all galaxies at $z>2$, as large amounts of metals are cycling throughout the halo and leaving the virial radius. This figure traces wind events through rapid increases in the integrated outflow flux, accompanied by re-accretion as seen through rapid increases in the integrated inflow flux. By $z=1$, \textbf{m12v} and \textbf{m10} have ejected 5-10 times more metal mass than they have re-accreted. For \textbf{m11} and \textbf{m12i} this ratio is closer to 2 or 3. We note that in addition to the metal infall (with negative radial velocities), halos can also re-accrete metals because their virial radius grows with time. In our inflow flux calculation, we do not consider the expansion velocity of the virial shell due to the growth of mass in the halo (see \citealt{faucher-giguere11b} for discussion of the effect). For this reason, the total mass of metals missing from the halo at a given redshift may differ from the difference of the integrated fluxes plotted in Figure \ref{fig:MetalFlower}. For example, this figure suggests that \textbf{m11} \textbf{m12v} have lost $\sim 3.5 \times 10^7 \Msun$ and  $\sim 1.5 \times 10^8 \Msun$ more metals than they have accreted, respectively. However, results shown later in the paper (Figure \ref{fig:MetalFinal})  suggest that these halos have all but $\sim 2.5 \times 10^7 \Msun$ and $\sim 10^8 \Msun$ of metals accounted for within $\Rvir$, respectively. In addition, we note that some inflowing flux shown in Figure \ref{fig:MetalFlower} is also due to accretion of metals generated in external galaxies (Angl{\'e}s-Alc{\'a}zar et al., in prep).

At low redshift, ISM fluctuations in metal content become gradually less drastic, first for \textbf{m12i} and \textbf{m10}, then eventually \textbf{m12v}. \textbf{m11} has a period of quiescence starting at $z=0.5$ which lasts for several Gyr, until a burst of star formation caused by a merger at $z=0.2$ once again drives out most metals from the ISM. In the L*-progenitors, the amount of metals locked in stars surpasses both the ISM and the ejected component by z=0. When it is not undergoing starbursts, \textbf{m11} maintains a comparable mass of metals in stars and ISM, while \textbf{m10} is dominated by an apparently stable reservoir of metals in the ISM. The low-mass dwarf galaxy \textbf{m10} has the highest ratio of metals ejected to the IGM to the galaxy's stellar mass, while \textbf{m12v} is the biggest IGM polluter by total ejected metal mass.

\begin{figure}
\vspace{-0.1cm}
\includegraphics[width=\columnwidth]{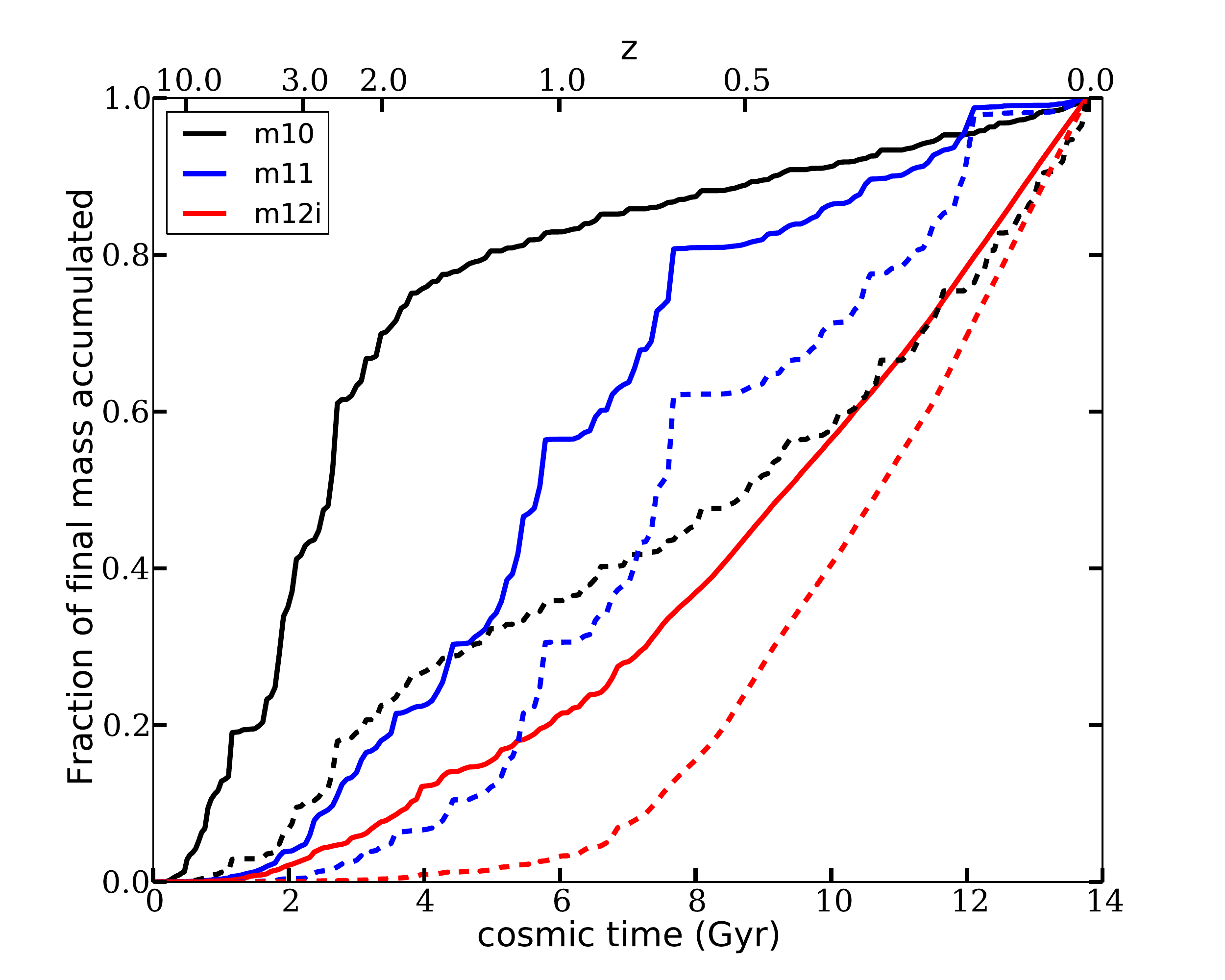}
\caption{ Cumulative buildup of stellar mass and metal content. Solid lines show the cumulative fraction of the $z=0$ stellar mass formed before cosmic time $t$ (or redshift $z$). Dashed lines show the corresponding mass of metals locked in stars compared to the $z=0$ total mass of metals in stars. \textbf{m10} builds stellar mass at relatively early times, but more than half of metals are locked in at $z<0.5$. \textbf{m11} builds  stellar and metal components gradually throughout cosmic time, while  \textbf{m12i} builds up both at late times.  }
\vspace{0.3cm}
\label{fig:Cumustellar}
\end{figure}

We examine more closely the total evolution of metals in the ISM and CGM in Figure \ref{fig:MetalBuildup}. We also show the evolution of the total gas mass of both components. As we have already demonstrated, the overall metal content of the ISM fluctuates when star formation is bursty, and grows steadily when star formation slows down. In general, the total gas mass of the ISM follows the same fluctuations as the metal content.
In the CGM, the metal content of galactic halos is high while the star formation is bursty and CGM outflows are common. However, by $z=0$, the L* progenitors have relatively metal-poor CGM compared to their high-redshift progenitors. It appears that after CGM outflows cease, a significant portion of the CGM metals have accreted into the ISM or locked into stars by z=0. We note that this conclusion is again partially determined by our definitions of CGM and ISM being divided at $0.1 \Rvir$. If we instead use a 10 kpc divide between ISM and CGM, the qualitative trends are largely similar, but the level of metal depletion in the CGM of L* galaxies at late times is less prominent (See Figure \ref{fig:radmetal}).  This has interesting consequences for the origin of currently observed metals in the CGM.

Finally, we analyze the evolution of metals in the stellar component in Figure \ref{fig:Cumustellar}. While \textbf{m10}, \textbf{m11}, and \textbf{m12i} have very different stellar mass growth histories, all three galaxies have built up more than half of their stellar metal component since $z\approx0.7$. Most strikingly, \textbf{m10} has 80\% of its final stellar mass but only 20\% of its final stellar metal mass in place by $z=2$. This has significant implications for its evolution on the stellar MZR, which we will discuss in Section \ref{sec:discussion}, and predicts a very specific distribution of stellar metallicities. \textbf{m12v} (not plotted) behaves much like \textbf{m12i}, its closest counterpart in halo mass. The cumulative star formation histories of our dwarf galaxies are consistent with observations (\citealt{weisz_etal11, weisz_etal14}, see also \citealt{chan_etal15, onorbe_etal15, wetzel_etal16}).

Since this section focused on a few specific FIRE halos, we briefly address the other main halos in our sample evolved to $z=0$. The L* progenitor \textbf{m12q} behaves much like \textbf{m12v}, though its star formation is higher at early times and lower at late times, and more of its metals are ejected into the IGM. Simulated galaxies which span the mass range between \textbf{m10} and \textbf{m11} generally behave like dwarf starbursts (i.e. \textbf{m11}). In other words, they have bursty star formation, fluctuations in ISM metal mass, and constant metal cycling through the CGM throughout all cosmic history. It appears that \textbf{m10} falls just shy of having sufficient mass to enable such prolonged burstiness, as its dramatic gas loss and relatively low densities of remaining gas only enable very low star formation rates at low redshift. We caution that this conclusion hinges on a relatively small sample of galaxies, however. The least massive FIRE galaxy \textbf{m09} was not considered in our analysis as it has almost no star formation at $z<2$. 

We also analyzed \textbf{m11.4a} and \textbf{m11.9a}, which were first introduced in \citep{hafen_etal16}, to increase galaxies in the L* mass range. However their low-redshift behavior actually more closely resembles the dwarf starburst galaxies in the \textbf{m11} range, albeit with lower $\eta$ during the outflows. The stellar mass of \textbf{m11.4a} at $z=0$ is much closer to \textbf{m11h383} than to any \textbf{m12} galaxy. It never achieves the baryon-dominated core, nor establishes an extensive stable gaseous disk, which are the conditions we hypothesize are necessary for CGM outflows to cease (M15). On the other hand \textbf{m11.9a} is considerably more massive, and even surpasses \textbf{m12v} and \textbf{m12q} in stellar mass, but the growth is more concentrated at late times, as evidenced by its lower rank in stellar and halo mass at $z=2$. More significantly, it is in the midst of a merger at $ z < 0.5$, and gains more gas mass through the virial radius than any other halo in this interval. This gas serves as fuel for star formation that \textbf{m12v} and \textbf{m12q} simply lack at late times, while \textbf{m12i} is already too stable to be triggered into a burst by accretion at late times. Currently, our sample of massive galaxies at $z=0$ is still too small to draw definitive conclusions.

\begin{figure}
\vspace{-0.2cm}
\includegraphics[width=\columnwidth]{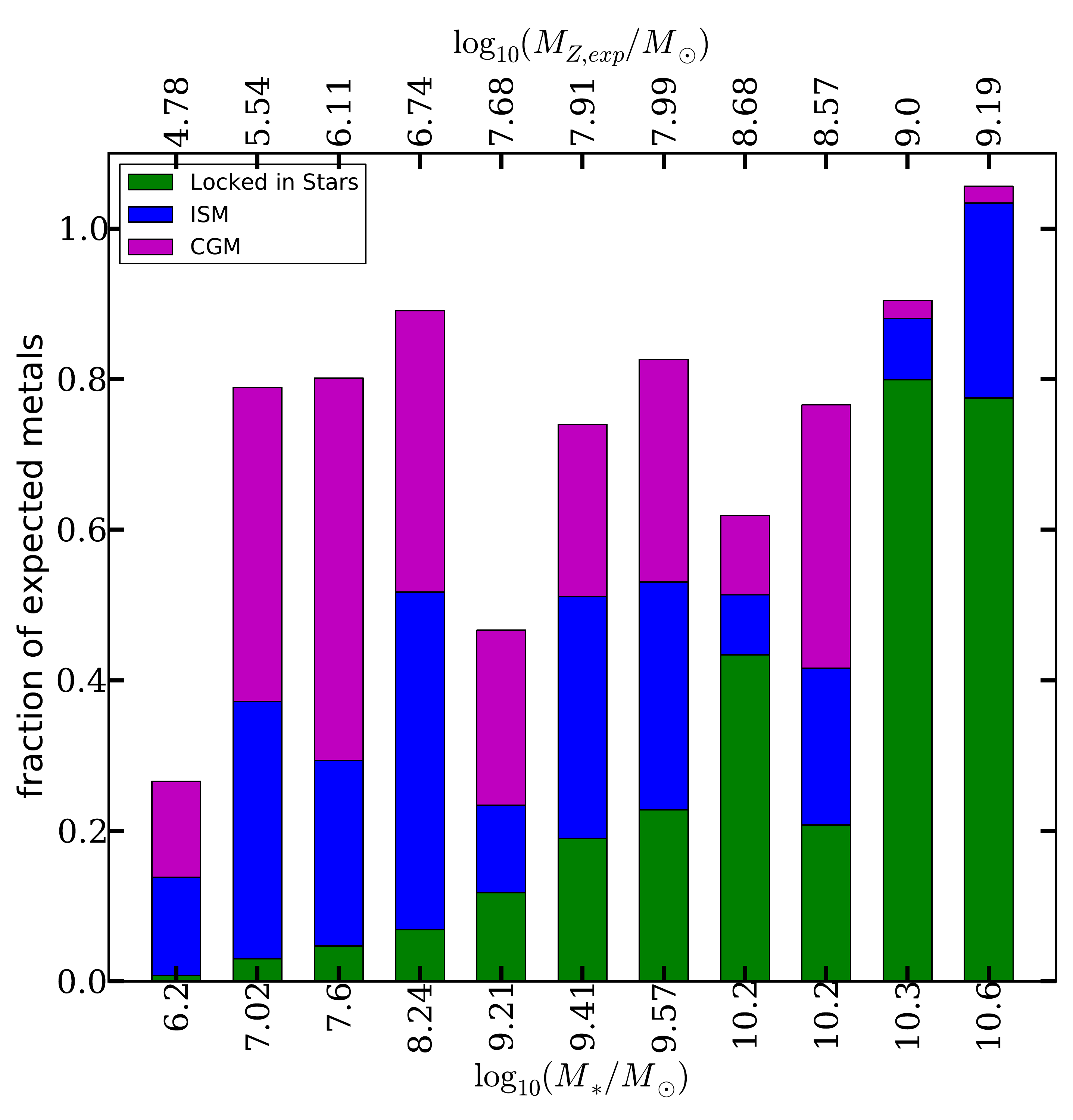}\\
\includegraphics[width=\columnwidth]{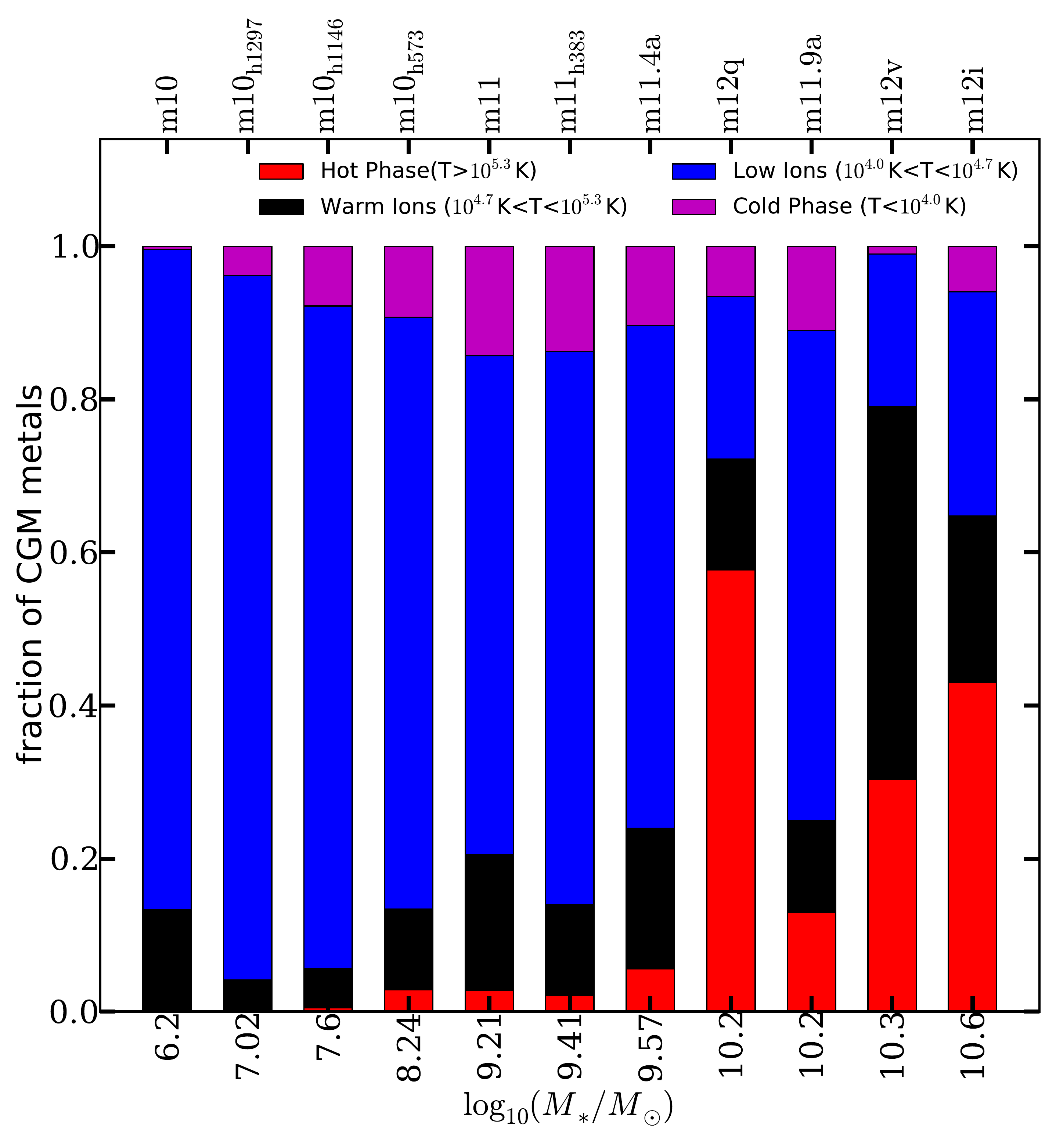}
\vspace{-0.0cm}
\caption{ Top: Amount of metals in each component within $\Rvir$ for the main galaxy in various FIRE simulations. Values are averaged over all snapshots from $0.2 > z > 0$. The dividing line between ISM and CGM is $r = 0.1 \Rvir$, the fiducial definition for the paper. The lower x-axis labels show the stellar mass of each galaxy, while the top x-axis labels show the expected total amount of metals produced by the galaxy ($M_{Z,exp}$), as computed using the empirical yield of the box (e.g. total mass of all metals in gas and stars divided by total stellar mass at $z=0$). Galaxies are sorted by stellar mass on the x-axis, but are not spaced in equal increments of stellar mass. Bottom: breakdown of what phase the CGM metals are in based on temperature. The x-axis is arranged like the top panel, with simulation names given across the top of the x-axis.   }
\vspace{0.3cm}
\label{fig:MetalFinal}
\end{figure}

\section{Metal Budget}
\label{sec:budget}

We now examine the final distribution of metals in galaxies run to $z=0$. The top panel of Figure \ref{fig:MetalFinal} shows the total amount of metals in CGM, ISM, and stars sorted by the $z\approx0$ stellar mass of each galaxy. To offset stochastic effects that could alter values for a single snapshot, we average all snapshots from $0.2 > z > 0$.  Also plotted is the expected total amount of metals produced by the stars in the galaxy at $z\approx0$ using the empirical yield described in Section \ref{sec:winds}. First we consider the dwarf galaxies, which we define broadly to be the stellar mass range $10^6 \Msun < M_* < 10^{10} \Msun$. Intriguingly, while the stellar metals increase their contribution to the metal budget in higher mass galaxies, the ISM and CGM typically each have roughly the same fractional share of metals in each of the dwarf galaxies.

The metal budgets of L* galaxies at $z\approx0$ are dominated by the stellar component. With increasing stellar mass, the ISM increasingly dominates the gas portion of the metal budget in comparison to the CGM. We note that since \textbf{m12v}, \textbf{m12i}, and \textbf{m12q} have virial radii of 230, 270, and 280 kpc, respectively, at $z=0$, our fiducial boundary between ISM and CGM of $0.1 \Rvir$ may over-emphasize the extent of the ISM. If we instead use 10 physical kpc to be the boundary between ISM and CGM, we find comparable amounts of metals in the two components of all \textbf{m12} galaxies. We show the radial distribution of metals for \textbf{m12} galaxies and \textbf{m11.9a} at $z=0.25$ in Figure \ref{fig:radmetal}. We have also verified that at $z=0.25$ these metals represent genuine CGM that is not closely associated with the extended ISM of satellite galaxies. The total contribution of metals within 5 kpc of satellite halo centers only accounts for $\sim$1\% of the total gas-phase metal content in the three \textbf{m12} halos, and 4\% in \textbf{m11.9a}.

The amount of metals sent to the IGM can be inferred by subtracting the total metal census of each halo from the expected amount. The two extremes are \textbf{m10}, which has lost over 70\% of its metals to the IGM, and \textbf{m12i} which harbors practically all metals that the stellar component has produced. We note that L* galaxies can accrete significant additional gas and metals from wind (or intergalactic) transfer from external galaxies (see \citealt{angles-alcazar_etal16}), which can explain the higher total metal content of \textbf{m12i} halo than expected from its stellar mass. The rest of the galaxies have lost an average of $\sim$30\% of their metals to the IGM with significant halo to halo variations. Our limited statistics prevent us from identifying a clear mass trend. Overall, most of the metals ever produced by the galaxies are accounted for within $\Rvir$, with a large fraction residing in the CGM for all but the most massive galaxies considered.

\begin{figure}
\vspace{-0.2cm}
\includegraphics[width=\columnwidth]{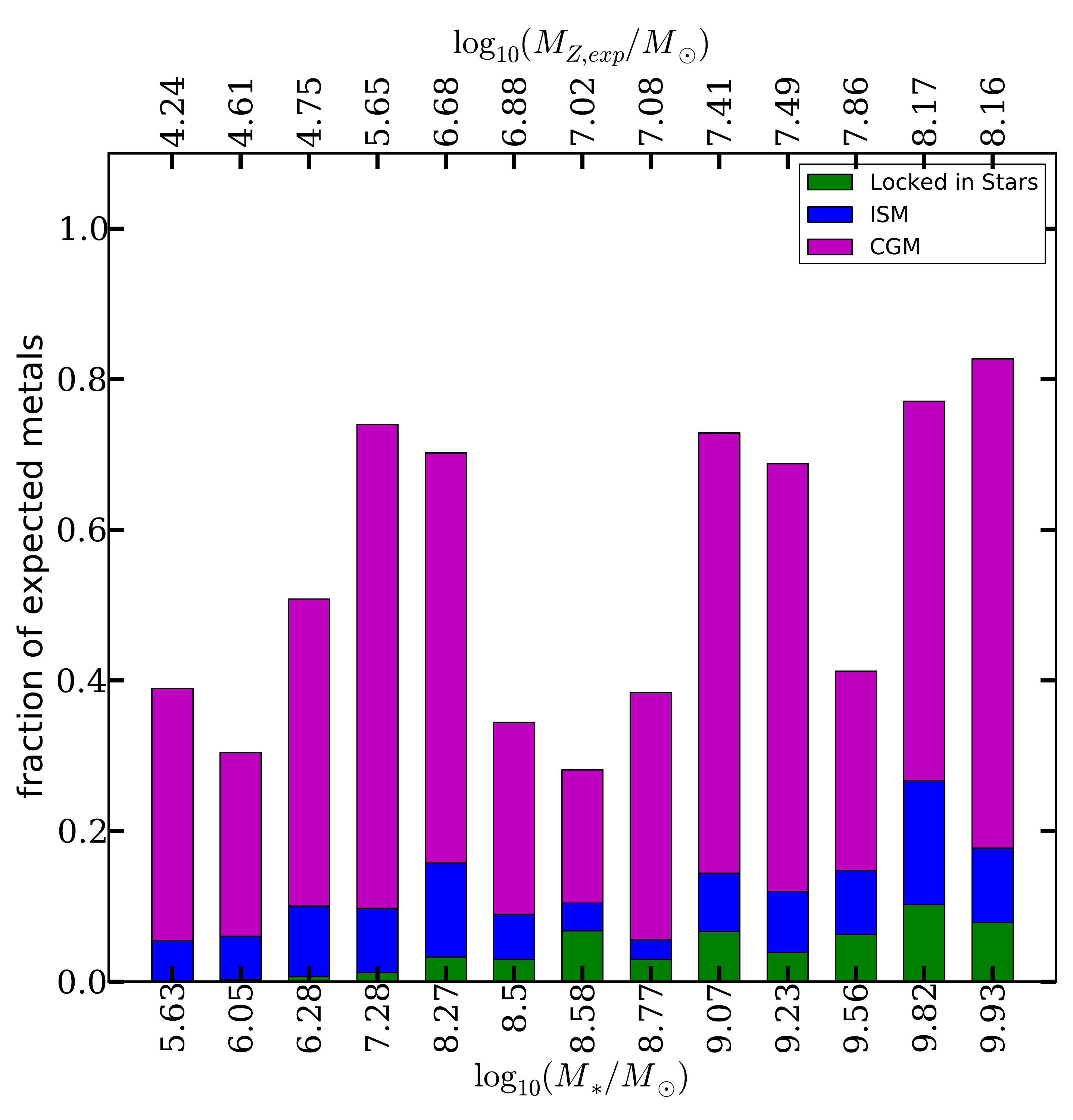}\\
\includegraphics[width=\columnwidth]{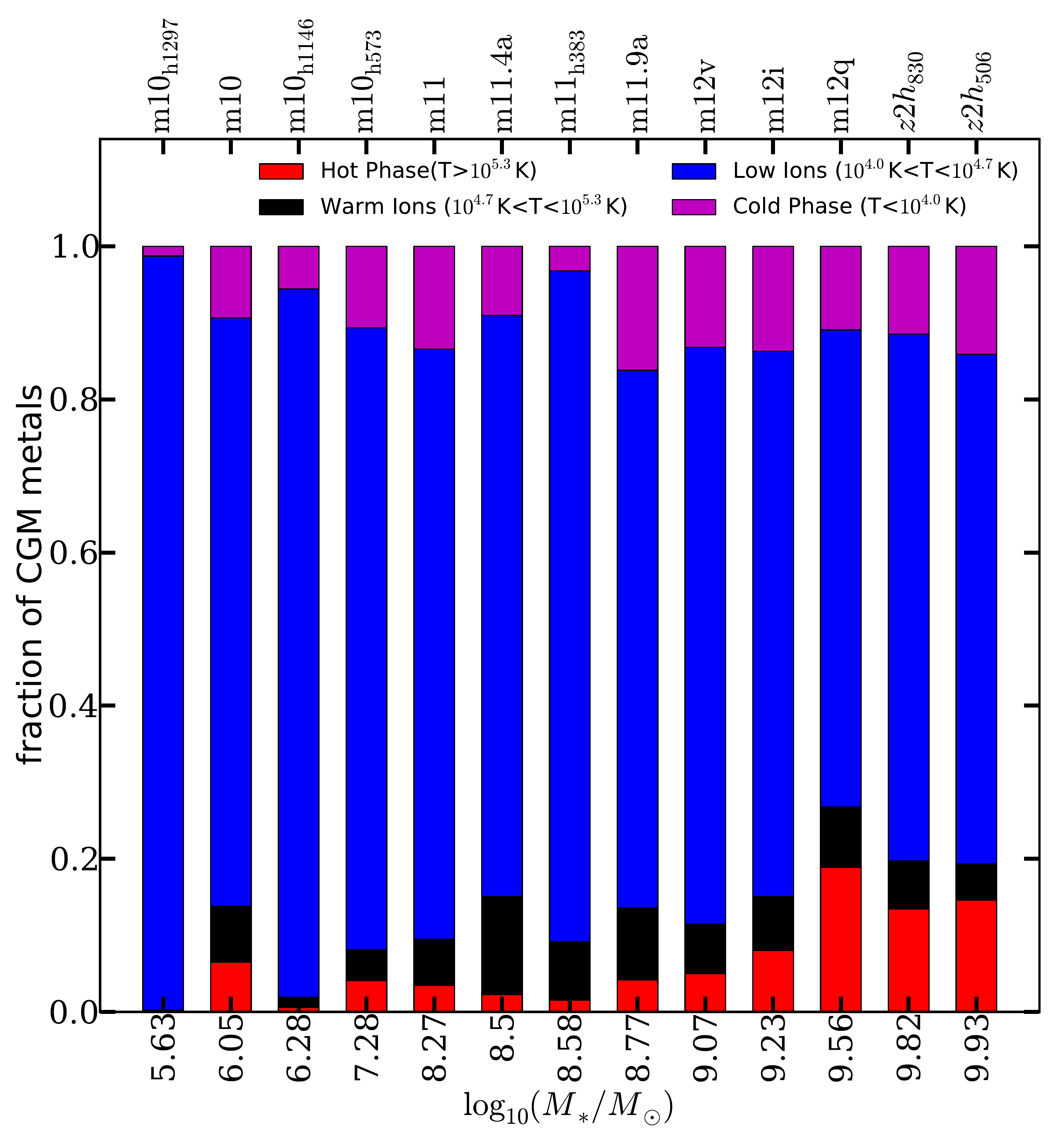}
\vspace{-0.0cm}
\caption{ Same as Figure \ref{fig:MetalFinal}, but at $z=2$ (with quantities time-averaged from redshifts $2.5 > z > 2$), with two additional galaxies from the high-redshift \textbf{z2h} sample. The dividing line between ISM and CGM is $r = 0.1 \Rvir$, the fiducial definition for the paper.}
\vspace{0.0cm}
\label{fig:MetalFinalz2}
\end{figure}

The bottom panel of this figure breaks down the temperature structure of the CGM. We use temperature ranges loosely based on  P14 to distinguish gas traced by ``warm ions'' like  photo-ionized $OVI$ apart from both hotter and colder phases, though we caution that collisionally ionized $OVI$ may still appear in the hot phase \footnote{ For reference, we summarize the virial temperatures of several halos presented in Figure \ref{fig:MetalFinal} and Figure \ref{fig:MetalFinalz2}: $T_{\rm vir} = 4\times 10^4~{\rm K},~2.6 \times 10^5~{\rm K},~1.1 \times 10^6~{\rm K}$ for {\bf m10}, {\bf m11} and {\bf m12i} at z=0 and $8.2 \times 10^4~{\rm K}, ~2.7 \times 10^5~{\rm K},~9.8\times 10^5~{\rm K}$ for their main progenitors at z=2.}. The CGM of our dwarf galaxies appears to be dominated by gas that is likely to harbor low-ions. On the other hand, L* galaxies have significant contributions from the hot phase. Warm ionized gas is present in all L* galaxies, but it only dominates the metal budget of one of them. Contribution of cold, neutral gas ($T<10^4K$) is typically small ($< 25\%$), and is dominated by contributions from satellites and subhalos. We provide additional comparisons of our metal masses to P14 in Section \ref{sec:observations}, where we take into account the specific redshift and radial range covered by the COS halos survey. 

We perform the same analysis at $z\approx2$ (averaged from $2.5 > z > 2$) in Figure \ref{fig:MetalFinalz2}, using the same galaxies plus two low-mass LBG-like galaxies (\textbf{z2h506} and \textbf{z2h830}) to slightly extend the mass range (we find that including all of our $z\approx2$ halos here does not change our conclusions).  The most notable difference between $z\approx2$ and $z\approx0$ is that all galaxies at $z\approx2$ have their metal budget dominated by the CGM, with most gas in the temperature range corresponding to photo-ionized low ions. Very few metals have been locked into stars by this epoch in halos of all masses considered. The fraction of metals lost to the IGM is somewhat higher at $z=2$ than in $z=0$, averaging about 40-50\% among all halos, with a hint of decreasing fraction at higher masses. The dominance of the CGM and IGM over stellar and ISM components at $z=2$ is consistent with the picture painted in the rest of this work: all high-redshift galaxies feature bursty star formation, which generates winds that transport large amounts of metal to the CGM and further into the IGM.

The interpretation of results is again somewhat determined by the definition of CGM. We consider two complications that arise when considering higher redshift galaxies. First, since $\Rvir$ evolves with redshift, many of the metals said to be in the IGM at $z=2$ will eventually end up within $\Rvir$ by $z=0$. Second, the CGM of each halo can include a substantial number of satellites and incoming mergers at this redshift. Within a fixed distance of 150 kpc from the galactic center, we find that the majority of halos contain more than 100\% of all metals generated by the central galaxies, with some of the extraneous metals contributed by galaxies that are incoming as mergers.

\section{Discussion}
\label{sec:discussion}

In this paper, we have explored how galactic winds in the FIRE simulations cause metals to cycle through galaxies and their surrounding environs. Here we attempt to present a unifying narrative for what role these winds have in driving metallicity evolution in the ISM, stars, and CGM for the galaxies simulated to $z=0$. The galaxies can be divided into three broad categories: low-mass dwarfs (\textbf{m10}), dwarf starbursts (\textbf{m11} and \textbf{m10h} series, as well as \textbf{m11.4a} and \textbf{m11.9a}, although they aren't dwarfs), and massive L* progenitors (\textbf{m12i}, \textbf{m12q} and \textbf{m12v}). We differentiate between ``low-mass dwarfs'' and ``dwarf starbursts'' by judging their star formation activity at low redshift, with galaxies in the latter category generally having a larger fraction of stars forming in bursts at late times, while also generally being more massive. We will address each subclass of galaxy separately to try to understand their evolution. 

\subsection{Low-mass dwarfs} We have seen that the isolated dwarf galaxy \textbf{m10} forms the majority of its stellar mass prior to $z=2$, and then lives a relatively quiet life until $z=0$. The galaxy maintains a cycle of star formation, inflows, and outflows through $z=0$, but the rates of all three processes are much lower than they had been at high redshift. However, the metallicity of the ISM grows between $z=2$ and $z=0$ by a factor of $\sim$5, while more than half of all stellar metals are locked into stars at $z<0.5$. This significant low-redshift metallicity evolution is key to the success of the FIRE halos matching observed MZR at multiple epochs \citep{ma_etal16}.

The metal content of galaxies in this mass range has been consistently difficult to reproduce in large-volume simulations where winds are prescribed a-priori. For example, in the work of \citet{torrey_etal14}, the authors point out that without higher metal retention efficiency in low mass galaxies their simulations find an MZR relation that is too steep. We find evidence for this kind of preferential retention of metals in the ISM only at low redshift. While it is a very low-mass galaxy, \textbf{m10} appears to retain a substantial fraction of metals produced by stellar evolution at $z<2$. In contrast, metal-poor gas continues to flow from the ISM into the CGM and beyond. This appears to occur due to entrainment of metal-poor gas in outflows generated by the low-intensity star formation that persists in the halo throughout its history. While the amount of stellar feedback energy generated at low redshift is low, it is enough to unbind some of the loosely bound material in the CGM. While metal loading factors are low, the overall mass-loading of the winds at $0.25 \Rvir$ is still quite high, as reported in M15. 

Even though this galaxy shows relatively efficient metal retention at late times, it does have very highly mass loaded winds at all times (M15 found $\eta \approx 30$ at $z<0.5$, and $\eta > 100$ at higher redshifts, as measured at $0.25 \Rvir$). As a consequence, \textbf{m10} is the only galaxy in our sample that lost most of the metals that were produced to the IGM, with only 29\% of produced metals residing in the halo at $z=0$.  This is likely caused by its low mass, which limits its ability to retain loosely bound baryons in its CGM, particularly at high redshift when star formation is most active. The effect is even more pronounced when considering the baryon fraction - the ratio of total baryonic mass within $\Rvir$ compared to the cosmic mean. The baryon fraction of \textbf{m10} is only 7\% at $z=0$ (see M15). We note that in spite of the high mass-loading factors seen in winds at low redshift, they do not cause large fluctuations in the total amount of gas and metals within the halo, as the reservoir of gas is still large compared to the outflow rates (see Figure \ref{fig:MetalFlower}).

\subsection{Dwarf starbursts}
\label{sec:dwarfs}
According to our simulations, galaxies with $z=0$ halo mass $10^{10} \Msun < M_h < 10^{11} \Msun$, corresponding to stellar mass $10^7 \Msun < M_* < 5 \times 10^9 \Msun$, do not maintain monotonic growth of gas or metal content in either the ISM or the CGM. Instead, metals propelled by bursts of stellar feedback cycle through the galaxy and halo throughout all of cosmic history. This behavior is critical to explain observed kinematics of stars, sizes, and dark matter cores in these galaxies \citep{onorbe_etal15, chan_etal15, el-badry_etal16}.  Once ejected from the galaxy into the CGM, the metals can cool gradually and recycle into the galaxy, or be re-heated through the next superwind outflow. Most metals produced in each burst of star formation are not immediately available to be locked up in the next generation of stars, ensuring that ISM and stellar metallicity evolve gradually rather than rapidly. 

The CGM and ISM stay closely connected through frequent, powerful outflows, and the CGM metallicity generally follows ISM metallicity. Interestingly, we have found that the $z=0$ total metal mass and total gas mass of ISM and CGM in each of these galaxies are roughly comparable. The phase structure of the CGM is dominated by low ions, which means it is not significantly hotter than ISM gas. We note that with our fiducial definition for what constitutes the CGM ($0.1 \Rvir < r < \Rvir$), the CGM volume is $1000$ times greater than the ISM volume. Although the total metal content is comparable, metal densities and column densities in CGM are much lower. Our findings for the galaxies in this mass range are broadly consistent with observations form the COS-dwarfs survey as presented in \citet{bordoloi_etal13}. We present a brief comparison in Section \ref{sec:observations}. 

While strong outflows generated by bursty star formation are promising for explaining the enrichment of the IGM, in fact all but one of these galaxies have lost less than 30\% of their total metal budget to the IGM by $z=0$. Although we have used it as the archetype, \textbf{m11} has a prominent blowout following a major merger at $z=0.6$, which ejected significantly more metals into the IGM than its counterparts of similar mass. In a time average sense, there is no strong mass trend between stellar mass and fraction of ``missing'' metals (i.e. metals in the IGM at $z=0$) for galaxies with mass $\sim 10^7-5 \times 10^{10} \Msun$. However, at higher redshift ($z \geq 2$) there is a weak trend of stronger metal loss in lower mass halos.

\subsection{L*-progenitors}

At high redshifts, the metals in L* progenitors generally behave like metals in dwarf starbursts, but at low redshift the galaxies switch from bursty to continuous star formation (see M15 and \citealt{hayward_hopkins15} for the physics of this transition). Around the same time, the galaxies of this mass no longer drive superwinds that can move metals from the ISM to the CGM. As a result, ISM metallicity begins to grow monotonically at $z<1$, surpassing the CGM as the dominant gas metal component by $z=0$. However, both are much lower than the stellar metal content, which grows rapidly at $z<0.5$. Most of the supernova ejecta created at late times fail to escape the central region and remain in the ISM, where they are locked into the next generation of stars. Metals that still remain in the more distant CGM at $z=0$ were primarily ejected at significantly earlier times. Other zoom-in simulations have reached similar conclusions \citep[e.g.][]{oppenheimer_etal16}.

\subsection{Comparison to models and observations}
\label{sec:observations}
Metals are often used as chemical tracers for the history of galactic winds. It is possible to use the MZR correlations to estimate the metal abundance for a galaxy of a given mass, and combine them with star formation histories to infer the strength and prevalence of galactic winds in the past \citep{finlator_dave08, lu_etal15, zahid_etal14}.  

We compare the metal outflow and wind outflow efficiencies measured in this work and M15 with those inferred from such \textit{a posteriori} semi-analytical models. We first note that the FIRE simulations reproduce both stellar and gas-phase MZRs \citep{ma_etal16}, which the semi-analytical works discussed here are tuned to. In \citet{lu_etal15}, metal abundances are used to constrain the maximum value of $\eta$ for galactic winds in halos of various masses. They argue that for $M_* = 10^9 \Msun $, the mass-loading factor for gas ejected from the galaxy can at most be $\eta \approx 20$, and as low as $\eta \approx 10$ if outflows are relatively metal-enriched compared to the ISM.  In \citet{zahid_etal14}, much lower values of $\eta$ are inferred. These results at first appear to conflict with M15, where higher values of $\eta$ were measured at this mass, particularly at high redshift. 

We consider the source of the disparity. One of the main parameters in the \citet{lu_etal15} model is metal mixing efficiency, and it directly factors into the computation of $\eta$. Less efficient mixing means that metals are preferentially ejected in outflows, as SN ejecta do not get fully mixed into the ISM prior to being launched into the galactic wind. When metal mixing efficiency is low, outflows are extremely enriched compared to the ISM. When $\eta$ is computed in this scenario, a given amount of metals lost from a galaxy would imply a lower value of $\eta$ compared to models where mixing is more efficient. Our result suggests that at high redshift, a fully-mixed model is a fair approximation for bulk outflow metallicity, as the metallicity of outflows is usually close to the ISM metallicity. This is most easily seen in Figure \ref{fig:MetalFlowing}, when considering the metallicity of the most prominent high-redshift bursts compared to the ISM metallicity at the same epoch. We note that the ``mixing'' seen in our simulated outflows is simply a result of the fact that large portions of the enriched ISM and CGM are disrupted and incorporated into larger scale galactic outflows following bursts of star formation, particularly at high redshift. The actual efficiency of metal diffusion within a given parcel of gas does not need to be high.

Perhaps a more relevant quantity to constrain in semi-analytical models is $\eta_{ z,{\rm eff}}$, the amount of metals permanently ejected from each galaxy as measured at $z=0$. In fact, since \citet{lu_etal15} make the explicit assumption that re-accretion is not important in their models, $\eta_{ z,{\rm eff}}$ is the quantity they are constraining rather than $\eta_z$ as defined through instantaneous flux of outflowing gas (Equation \ref{eq:flux_outflow}), which we have focused on in this work and M15. 

Large-volume simulations such as Illustris \citep{vogelsberger_etal14} are generally tuned to produce galaxies that obey particular correlations. However, an early version of these simulations failed to reproduce the MZR if metallicity of the outflows was directly linked to the ISM metallicity \citep{vogelsberger13, torrey_etal14}. They found that a model in which only 40\% of the available metals are loaded into the wind provided a much better match for the normalization of the observed MZR. Our simulations show that metals can be fully loaded into the wind, i.e. the metallicity of the wind and ISM can be very similar, and one can still match the normalization of the MZR. This likely owes to differences in wind launching velocities, structure in the ISM and the inner CGM, and recycling of the metals back into the ISM, or because L* galaxies can still drive winds at late times in Illustris, while we see significant suppression. Only in low mass halos at late times do we find lower metallicity of the wind material in the inner halos, likely caused by the entrainment of lower metallicity gas in the vicinity of the galaxy (see Section \ref{sec:dwarfs}). By construction, Illustris can only indirectly account for this by lowering the assumed metal loading, as hydrodynamic interactions are disabled near galactic disks and  entrainment and recycling in the inner CGM are not accounted for.

\begin{figure}
\vspace{-0.2cm}
\includegraphics[width=\columnwidth]{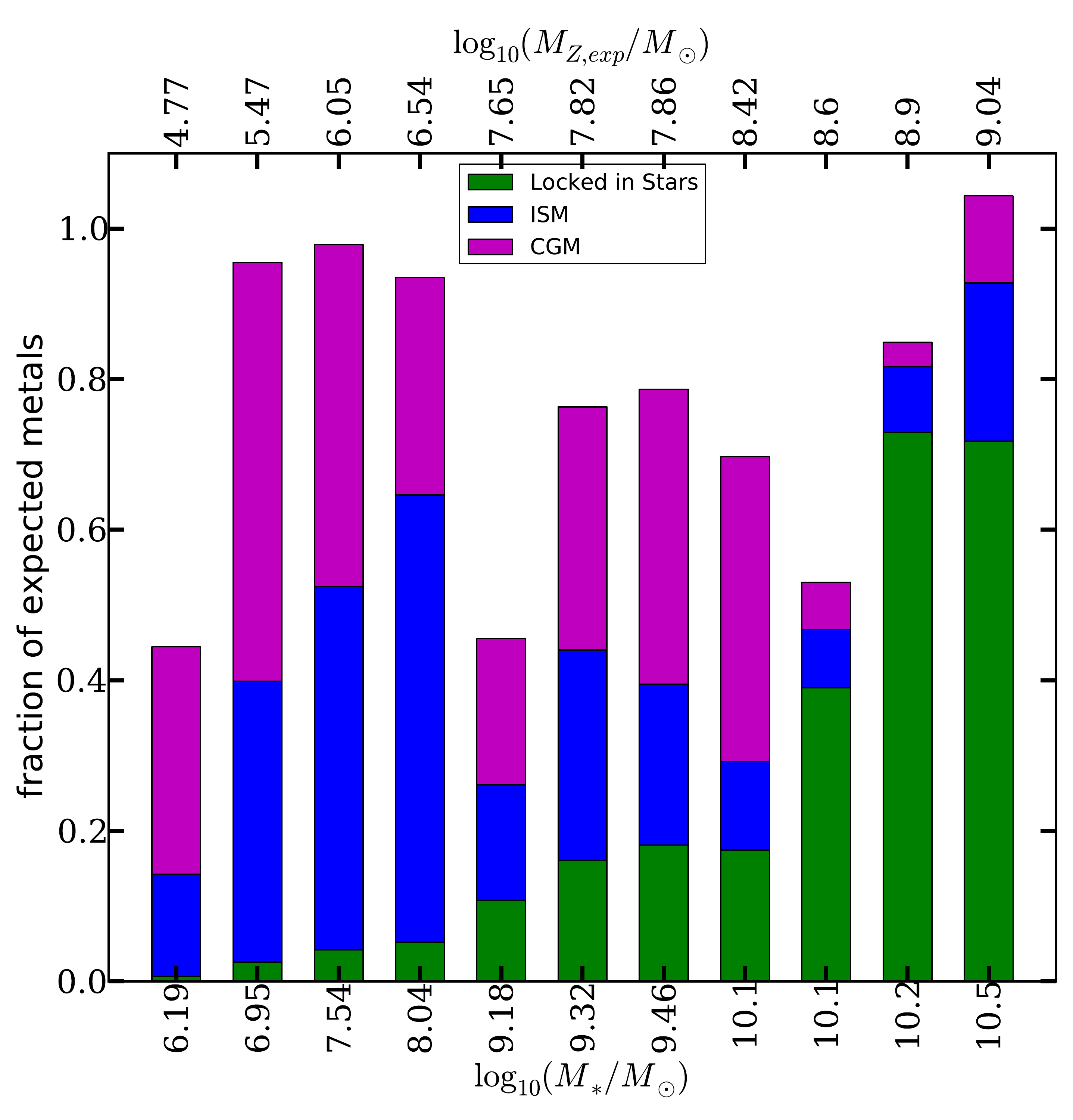}\\
\includegraphics[width=\columnwidth]{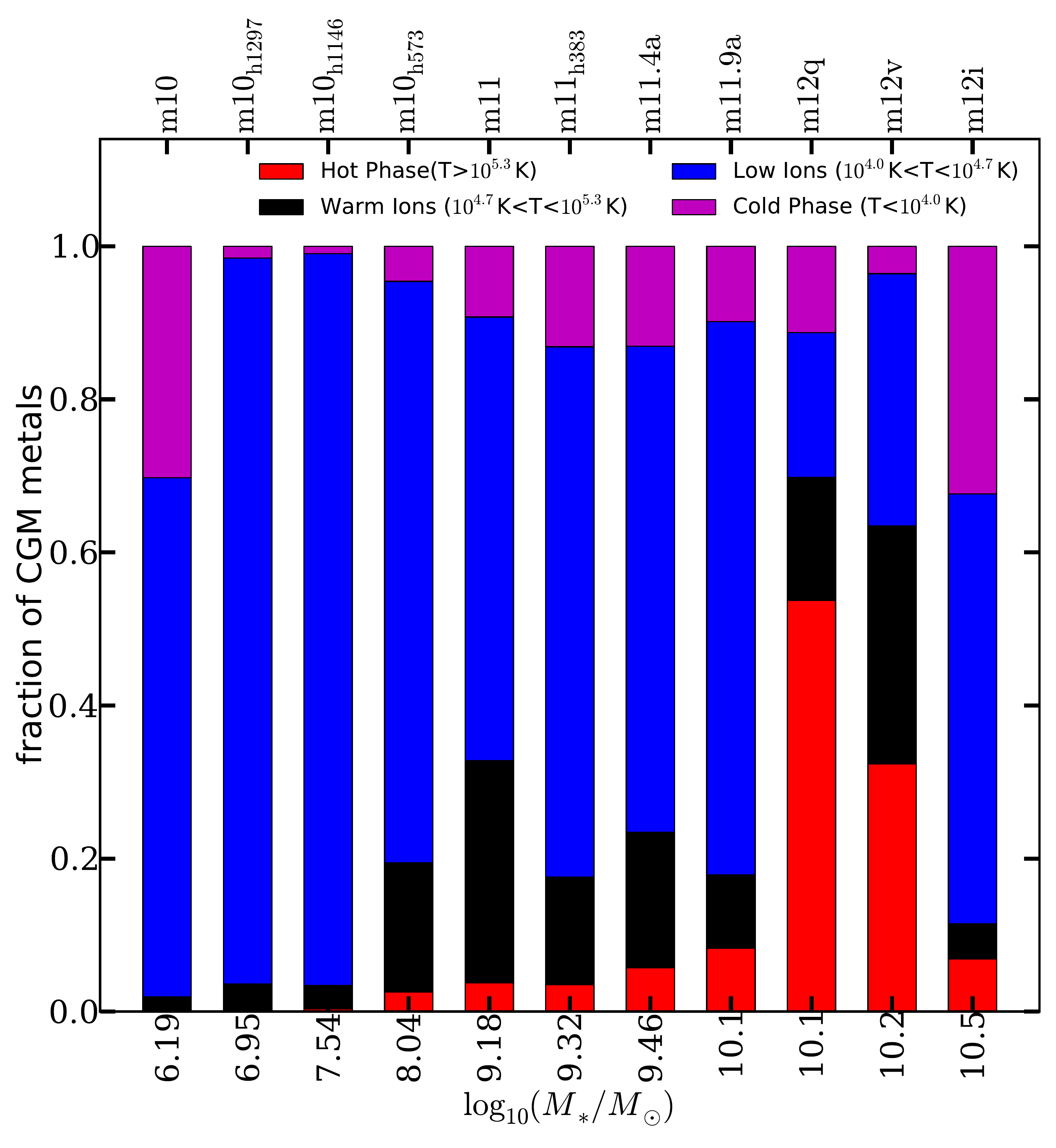}
\vspace{-0.0cm}
\caption{ Same as Figure \ref{fig:MetalFinal}, but using fixed physical distances to differentiate ISM and CGM. Gas within $r < 10 {\rm kpc}$ is considered ISM, while CGM is defined as $10 {\rm kpc} < r < 150 {\rm kpc}$. All quantities are time-averaged from $0.35 > z > 0.15$. These boundaries and epochs were chosen to facilitate a comparison with the results of the COS halos survey.}
\vspace{0.0cm}
\label{fig:MetalFinalCOS}
\end{figure}

It is also useful to consider the implication of our results for direct observational measurements of $\eta$ in galactic outflows. Some of the best constraints on galactic wind outflow rates come from instances where a quasar line-of-sight aligns with the minor axis of a star-forming galaxy \citep{bouche_etal12, kacprzak_etal14, schroetter_etal15}. The impact parameter of the quasar provides an estimate of the radial extent of the wind for the wind, while features in the quasar spectra can be used to estimate the mass flux of the outflow. \citet{schroetter_etal15} compared measurements of $\eta$ in massive low-redshift galaxies to the models of M15, among others. In short, mass-loading factors ranged from $0.5 \lesssim \eta \lesssim 2$ for galaxies between $3 \times 10^{11} \Msun \lesssim M_h \lesssim 3 \times 10^{12} \Msun$. While observations were generally found to be consistent with the M15 fits at $z\approx0$, the $z\approx1$ galaxies had values of $\eta$ lower than predicted by our simulations. However,  there are nuances that complicate the direct comparison. While measurements of $\eta$ presented in this work are often averaged over long time intervals, the authors infer their values based on a single absorption system. In addition, as the \citet{schroetter_etal15} analysis relies on a conversion between MgII equivalent width and neutral hydrogen column density $N_H$, the comparison could be further complicated by the prevalence of hydrodynamic interactions between outflows and metal-poor material described in this work, and by a multi-phase wind composition. We will analyze the phase structure of FIRE outflows in future work, but we briefly note that FIRE outflows in the CGM are multi-phase.

The analysis of the metal budget with the COS halos survey in P14 provides an estimate for the total amount of metals in the CGM. Including metals in the dust phase, this estimate suggests a  $M_* \approx 10^{10.1} \Msun$ galaxy should have a minimum CGM metal mass of $\sim 5.0 \times 10^7 \Msun$ and a fiducial CGM metal mass of $\sim 1.2 \times 10^8 \Msun$. At first glance,  Figure \ref{fig:MetalFinal} suggests our \textbf{m12} galaxies are close to this value, but generally lower than observationally inferred. To be more precise, we adopt the same definition employed in P14, which is gas from 10 to 150 kpc from the galactic center, and use a redshift range of $0.35 > z > 0.15$ to average the properties of each halo. The results are shown in Figure \ref{fig:MetalFinalCOS} (Radial profiles of the metal distribution in several \textbf{m12} halos at $z=0.25$ is shown in Figure \ref{fig:radmetal}). The total amount of metals in the CGM using this definition for \textbf{m12i}, \textbf{m12v}, and \textbf{m12q} is $1.4 \times 10^8 \Msun$, $2.7 \times 10^7 \Msun$, and $2.8 \times 10^7 \Msun$, respectively. \textbf{m11.4a} and \textbf{m11.9a}, which still drive winds at low redshift have metal masses of $3.2\times10^7 \Msun$ and  $1.3 \times 10^8 \Msun$.  We note that all of these galaxies except \textbf{m11.4a} have masses in excess of $M_* = 10^{10.1} \Msun$. The CGM metal masses of \textbf{m12i} and \textbf{m11.9a} are closest to the estimates of P14, but \textbf{m12i} is most likely dominated by extended galactic structure, as the disk extends beyond 10 kpc, and it has by far the highest stellar mass. Both \textbf{m12v} and \textbf{m12q} are marginally too metal-poor compared to the observational constraint. A considerably larger sample of simulated galaxies must be used to determine whether simulated CGM with the FIRE model is actually too metal poor compared to observations. Furthermore, metal mass inferred from the observations relies on the uncertain ionization correction and extrapolation with an uncertain radial profile for the metal distribution. A more direct way to compare our simulations to the observations is therefore to extract strengths of the specific transitions in a given ion from simulations. A rough estimate of the distribution of metals in different phases can be gleaned from the bottom panel of Figure \ref{fig:MetalFinalCOS}.

In addition to the CGM metal budget, we can compare our measurements for the total amount of metals locked in stars to the estimates from P14. We start by noting that the FIRE simulations have been shown to reproduce the \citet{gallazzi_etal05} stellar MZR \citep{ma_etal16}, which is the same data used to estimate metal mass locked in stars in P14. In Figure \ref{fig:MetalFinalCOS}, we show that at $z\approx0$, our galaxies with $M_* > 10^{10} \Msun$ have between 20\% and 70\% of the expected metal budget locked in stars. According to P14, galaxies in this same mass range should only have $\sim$20\% of their total metal budget in stars. We suggest several possible differences in the methodology to account for the discrepancy. First, the expected amount of produced metals at a given galaxy stellar mass inferred by P14 is higher than what we find in our simulation results. We believe this difference originates from assumptions about type II SNe yields and stellar mass loss rates. FIRE uses \citet{woosley_weaver95} SNe II yield tables, while P14 chose a higher yield based on other results. The stellar mass loss rates used in P14, which are resultant from choices regarding the IMF and stellar wind model, suggest that $\sim$55\% of initial stellar mass is lost from each stellar population. In the FIRE simulations, stellar mass loss rates were calculated using standard STARBURST99 models with a \citet{kroupa02} IMF, resulting in a much lower mass loss fraction per stellar population ($\sim$ 30\%). Therefore, $\sim$2 times as many metals are produced per galaxy in P14 than what was found in our simulations (see P14 Figure 1). Furthermore, P14 used a conversion factor to lower the \citet{gallazzi_etal05} metallicities from luminosity-weighted to mass-weighted metallicities, while \citet{ma_etal16} matched the metallicities to \citet{gallazzi_etal05} directly as presented. The value assumed for solar metallicity was also different, as we have used $\Zsun = 0.02$, while they adopted $\Zsun = 0.0153$. P14 would therefore also infer a lower mass of metals locked in stars at a given observed stellar metallicity compared with our work. The differing assumptions noted here create a large disparity between the fraction of expected metals in the stellar component as reported in our work and the values used in P14. 

We also consider constraints from the COS-dwarfs survey as presented in \citet{bordoloi_etal13}. In that study, CIV was detected out to 100 kpc in a large sample of dwarf galaxies in the mass range $10^8 \Msun < M_* < 10^{10} \Msun$. In particular, CIV absorption was detected in star-forming dwarf galaxies at nearly 100\% out to $0.2 \Rvir$, and 60\% to $0.4 \Rvir$. Detection around non-star forming galaxies was less common. The CGM carbon mass in star forming dwarfs was inferred to be comparable to the total carbon mass of the ISM. As mentioned previously, detailed ionization modeling is needed for a direct comparison between COS-dwarfs and FIRE. However there are clear parallels between the COS-dwarfs and the FIRE dwarf starbursts discussed in \ref{sec:dwarfs}, which occupy the same stellar mass range. We have argued that outflows following star formation supply the CGM with large quantities of metals, increasing the likelihood of detectable absorption features. The window of phase structure probed by COS-dwarfs through CIV corresponds to the ``warm ions'' temperature range that we used in Section \ref{sec:budget}, which makes up roughly 10-20\% metal budget of the CGM of the FIRE dwarf starbursts. A much larger reservoir of metals may be hidden in the cooler ``low ions'' phase, which dominates the FIRE dwarf starburst metal budget. To estimate the total CGM carbon mass, \citet{bordoloi_etal13} assumed that CIV represented about 30\% of all carbon, and found that the total mass was comparable to the ISM mass. We have also argued that the total mass of metals in the CGM and ISM is comparable in the FIRE dwarf starbursts, though our results again depend on how the ionization correction is implemented and what we consider to be the border between ISM and CGM.

The relation between the metallicity of the CGM and galaxy mass appears to be more complex in our simulations than in the simulations of \citet{suresh_etal15}, who found that CGM has a roughly constant metallicity offset from the ISM and is well correlated with galaxy mass. The main difference appears to be the increasing offset between the CGM and ISM metallicity at late times in some of our L* galaxies. We caution that our sample is small and, especially around L*, it shows a large dispersion in the amount of metals locked in the CGM.

With regards to stellar metallicity, our results predict that the MZR for both stars and gas evolves rapidly at late times. In the case of gas, this would mean that all galaxies except dwarf starbursts generally do not have winds that can escape the potential of the galaxy at $z=0$. It is difficult to say whether observations are consistent with this result, but the escape velocity of L* galaxies at $z=0$ can be quite high, and most non-starburst winds have lower velocities \citep{veilleux_etal05}. As for the stars, our simulation results suggest that more than half of all metals were locked into stars at $z<0.5$. The star formation histories of our dwarf galaxies \citep{hopkins_etal14, sparre_etal15} are generally consistent with observations \citep{weisz_etal11, weisz_etal14}. However, the metal locking history could have interesting implications for the predicted distribution of stellar metallicities and their correlation with stellar age. For low-mass dwarfs with early peaking star formation histories, a relatively small population of late-forming, metal-rich stars should contain a substantial fraction of all the metals in the galaxy. Larger statistical samples are needed for more detailed predictions, as low-mass dwarfs show a variety of SF histories \citep{weisz_etal14}.

Finally, we note that our simulations do not use a sub-resolution model of turbulent metal mixing. It has been argued that SPH simulations of galaxy formation under-mix metals unless diffusion is implemented  \citep{shen_etal10}.  
While we have not explored specifics of the metal outflows in simulations with sub-resolution metal mixing, we tested the implications of a sub-resolution model for turbulent mixing similar to that of \citet{shen_etal10} on galaxy growth and mass outflow rates, and found that subresolution metal mixing has a small effect on those bulk properties \citep{su_etal16}. This is at least partially due to the high-resolution of simulations in FIRE project, allowing for the simulations to resolve part of the turbulent cascade. 

We have also done a preliminary comparison of simulations with and without subresolution metal diffusion in FIRE-2, which uses improved hydrodynamics \citep{hopkins_etal17}. Our comparison shows that the largest effects of subgrid metal diffusion are in the metallicity distribution functions of stars and gas inside galaxies, where low and high metallicity tails are suppressed when metal diffusion is included. On the other hand, the overall galaxy averaged metallicities are relatively unaffected by subgrid metal diffusion. Given that the metal ejection episodes launch a significant fraction of the ISM into the outflow, we do not expect large changes in the global, shell averaged, metal outflow properties studied in this work.  However, the details of the distribution and properties of CGM absorbers could change and we plan to use simulations with and without metal mixing to study such effects in future work.

\section{conclusions}

The main conclusions of this paper can be summarized as follows:\\

\noindent 1. Star formation drives outflows that transport metals from the ISM to the CGM throughout cosmic history. At high redshift, CGM metallicity is only slightly lower than the ISM metallicity, as a large portion of the ISM is ejected following each burst of star formation. The difference between CGM and ISM metallicities typically increases at later times.\\

\noindent 2. Averaged over a long timescale, the average metallicity of inflows is much lower than the average metallicity of outflows in most galaxies. This is particularly true at large galactocentric distances, at the edge of the CGM (around $\Rvir$). However, in the inner halo ($0.25 \Rvir$) it is difficult to distinguish the two for any given halo and any given individual absorber. This complicates the use of metallicity as an observational diagnostic of inflow vs. outflow, especially at low redshift ($z \lesssim 1$) where Lyman limit systems that traces inflows and outflows are confined to the inner halo \citep{hafen_etal16}.\\

\noindent 3. Nearly all metals produced at high redshift are initially ejected from the galaxy, but a large fraction recycles back. Overall, galaxies retain most metals produced by stars within the virial radius at $z=0$, except in low-mass dwarfs (\textbf{m10}), which lose most of their metals to the IGM. In contrast, the baryon fraction of many of the same galaxies can be well below the universal mean by $z=0$ (see \citealt{muratov_etal15}). A full analysis of galactic recycling is presented in \citet{angles-alcazar_etal16}. \\

\noindent 4. Metallicity evolution at low redshift is driven by a variety of processes. In $L*$ galaxies, cessation of strong outflows and continuous star formation enable rapid enrichment of stars and ISM at late times, while the CGM metal content decreases. In low-mass dwarf galaxies with low star formation rates, metals are retained in the inner region of the galaxy, while metal-poor gas may continue moving through the CGM. More massive dwarf galaxies have comparable metallicities in the CGM and ISM at late times, owing to their efficient circulation of metal enriched material. Despite the variety of processes at play, all galaxies fall on the observed mass-metallicity relations for both stellar and ISM components.\\

\noindent 5. Each FIRE dwarf galaxy at $z=0$ has roughly comparable portions of metal mass in the ISM and CGM, when the dividing line between ISM and CGM is chosen to be $0.1 \Rvir$. If the ISM is defined as the inner 10 kpc the same statement applies, even for more massive galaxies. At $z=2$, a substantially larger portion of metals are in the CGM than in the ISM for all galaxies considered. \\

\noindent 6. Metals in $z=0$ CGM of dwarf galaxies mostly reside in the temperature range of $10^4 K  < T < 5 \times 10^4 K$ and are potentially observable as low-ions. In massive L* galaxies, this ``low-ion'' phase, ``warm'' phase ($5 \times 10^4 K  < T < 2 \times 10^5K$), and ``hot'' phase ($T > 2 \times 10^5 K$) contain comparable shares of CGM metals. \\\\

\noindent We would like to thank the Simons Foundation and the participants of the \textit{Galactic Super Winds II} symposium for stimulating discussions. DK was supported by NSF grant AST-1412153 and the Cottrell Scholar Award from the Research Corporation for Science Advancement. CAFG was supported by NSF through grants AST-1412836 and AST-1517491, by NASA through grant NNX15AB22G, and by STScI through grants HST-AR-14293.001-A and HST-GO-14268.022-A. Support for PFH  was  provided  by  an  Alfred  P.  Sloan  Research  Fellowship, NASA  ATP  Grant  NNX14AH35G,  and  NSF  Collaborative  Research Grant \#1411920 and CAREER grant \#1455342. DAA acknowledges support by a CIERA Postdoctoral Fellowship. EQ  was  supported  by  NASA  ATP grant 12-APT12-0183, a Simons Investigator award from the Simons Foundation, and the David and Lucile Packard Foundation. The simulation presented here used computational resources granted by the Extreme Science and Engineering Discovery Environment (XSEDE), which is supported by National Science Foundation grant no. OCI-1053575, specifically allocations TG-AST120025, TG-AST130039, and TG-AST1140023.

\bibliography{galaxies}
\begin{appendix}
\section{Total mass-loading factors at the virial radius}
\begin{figure}
\includegraphics[width=\columnwidth]{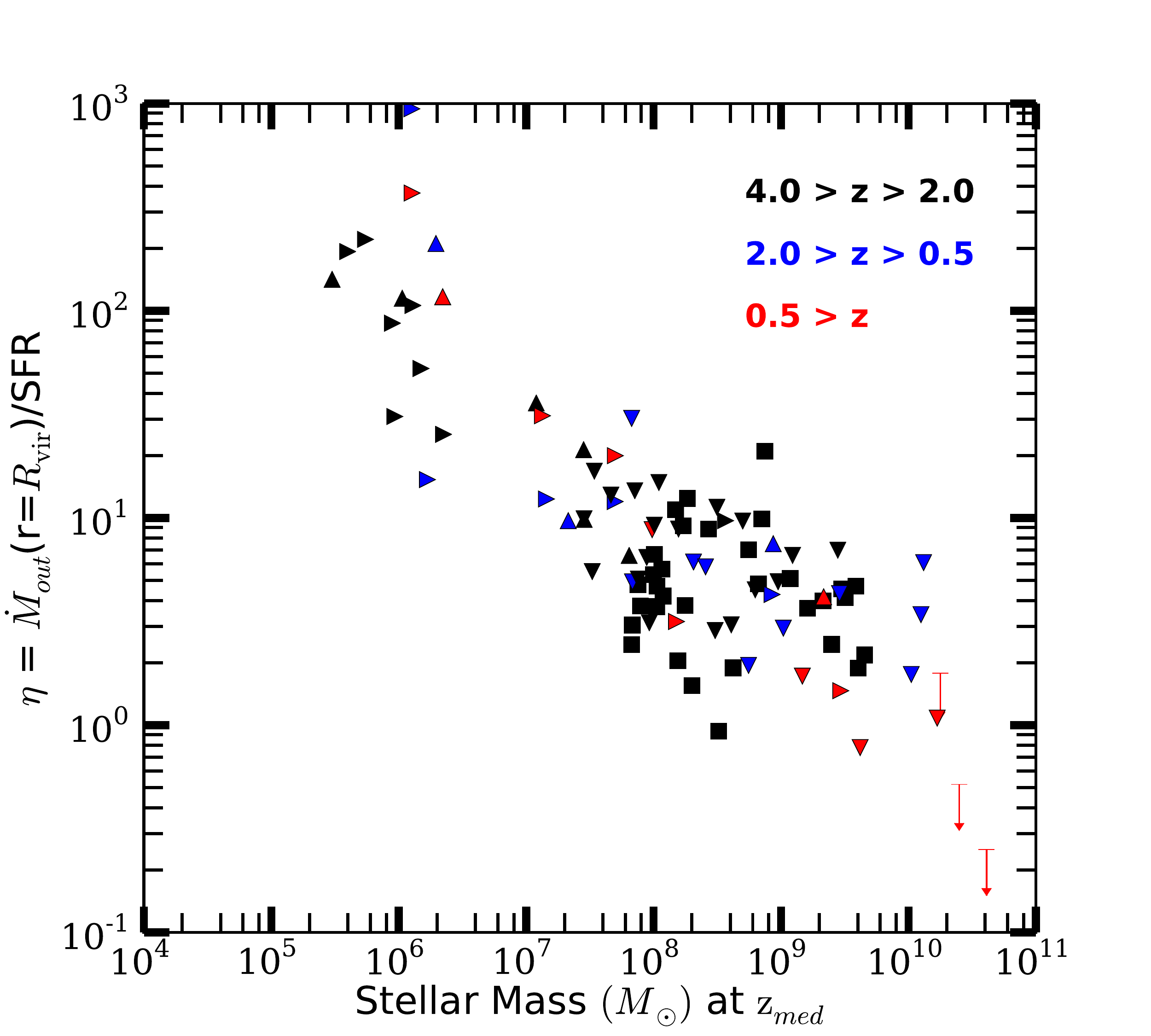}
\includegraphics[width=\columnwidth]{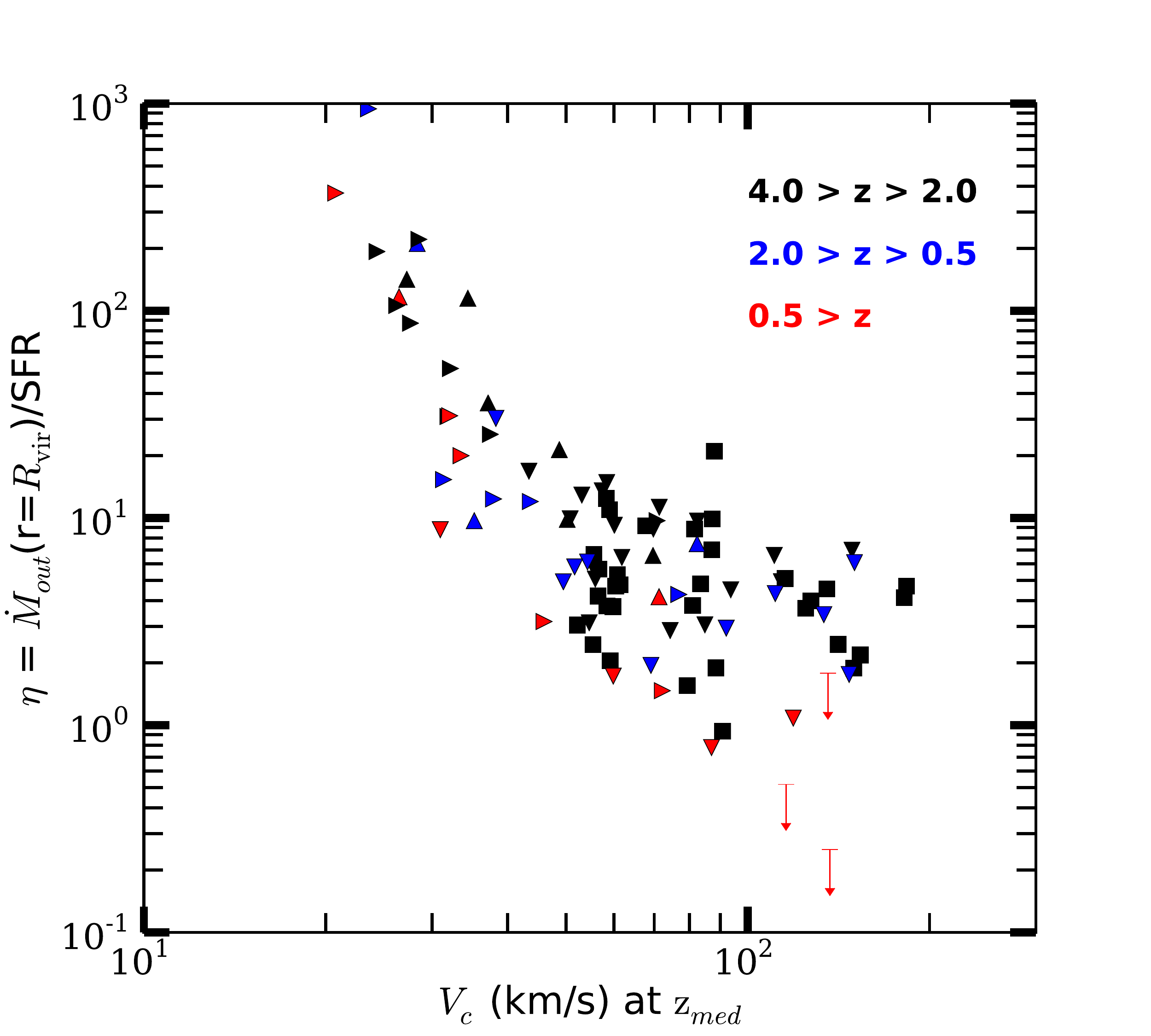}
\caption{Mass-loading factor $\eta$ as measured at $\Rvir$ vs. $M_{*}$ (top) and $v_c$ (bottom). There is a general trend of decreased $\eta$ with increasing $v_c$ and increasing $M_*$, but the trends have much more scatter than what was shown for $0.25 \Rvir$ in \citet{muratov_etal15}. The sample of galaxies plotted here, as well as the meaning of the color and shape of each data point is the same as in Figure \ref{fig:etaz}.}
\label{fig:etaRvir}
\end{figure}

In Figure \ref{fig:etaRvir}, we show the mass-loading factor $\eta$ as measured at 1.0 $\Rvir$ as a function of $v_c$ and $M_*$. The reader can find similar plots in M15 for $0.25 \Rvir$. This is the total amount of mass ejected per unit star formation, rather than only the metal component of the winds ($\eta_{z}$), which were shown in Figure \ref{fig:etaz}. While the same general trends found at $0.25 \Rvir$ in M15 apply at $\Rvir$, the overall values of $\eta$ are lower and have significantly more scatter at a given value of $v_c$ or $M_*$. 

\section{Radial distribution of metals}
\begin{figure}
\includegraphics[width=\columnwidth]{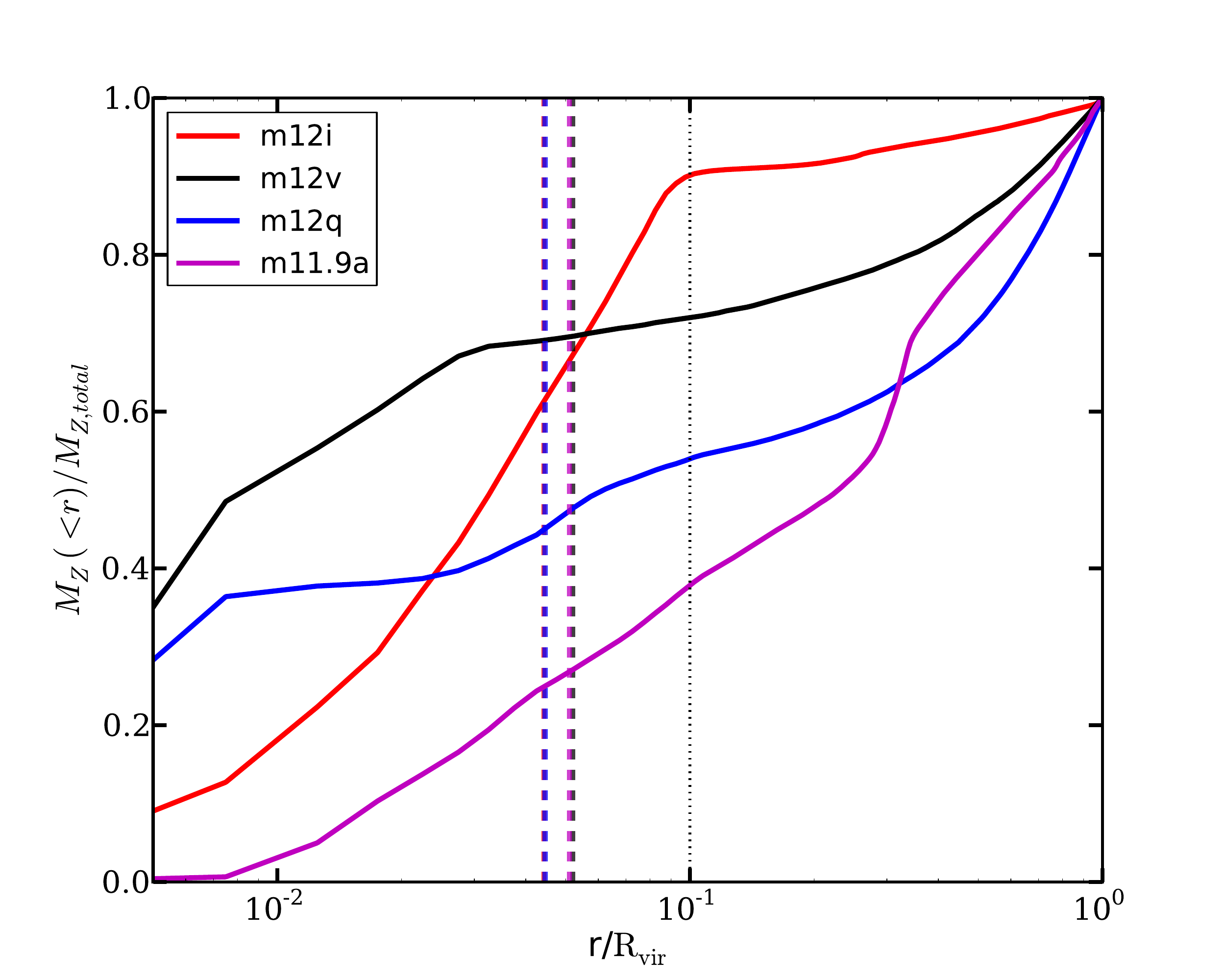}
\caption{Fraction of total gas-phase metals enclosed within radius $r$ vs. $r/\Rvir$ for all halos with $M_* > 10^{10} \Msun$. All values are computed at $z=0.25$. Dashed lines mark 10 physical kpc for each halo. Note that the dashed lines for \textbf{m12i} and \textbf{m12q} are essentially identical.  The dotted black line shows $0.1 \Rvir$. }
\label{fig:radmetal}
\end{figure}

In Figure \ref{fig:radmetal}, we show the radial distribution of metals at $z=0.25$ for all galaxies with $M_* > 10^{10} \Msun$. Our fiducial dividing line between ISM and CGM is $0.1 \Rvir$, but our comparison to the COS halos survey in Section \ref{sec:observations} instead used 10 physical kpc. This figure shows that in the case of \textbf{m12i}, a substantial amount of metals sits between 10 kpc and $0.1 \Rvir$. A similar representations of the radial profile of metals has been shown in Figure 9 of \citet{ma_etal16}.

\end{appendix}

\end{document}